\documentclass[11pt,a4paper]{article}
\usepackage{jheppub}
\allowdisplaybreaks
\usepackage{color,ulem,makecell}
\usepackage{amsmath}
\usepackage{graphicx}
\usepackage{esint}
\usepackage{slashed}
\usepackage{multirow}
\usepackage{booktabs}
\usepackage[T1]{fontenc}
\usepackage[english]{babel}

\usepackage{babel}
\begin{document}

\title{
Heavy neutrino mixing prospects at hadron colliders: a machine learning study
}

\author[a]{Si-Yu Chen,}
\author[a]{Yu-Peng Jiao,}
\author[a]{Shi-Yu Wang,}
\author[1,b,c]{Qi-Shu Yan\note{Corresponding author},}
\author[1,a]{Hong-Hao Zhang,}
\author[1,d]{Yongchao Zhang}

\affiliation[a]{School of Physics, Sun Yat-Sen University, Guangzhou 510275, China}

\affiliation[b]{School of Physics Sciences, University of Chinese Academy of Sciences, Beijing 100049, China}

\affiliation[c]{Center for Future High Energy Physics, Institute of High Energy Physics, Chinese Academy of Sciences, Beijing 100049, China}

\affiliation[d]{School of Physics, Southeast University, Nanjing 211189, China}

\emailAdd{yanqishu@ucas.ac.cn}
\emailAdd{zhh98@mail.sysu.edu.cn }
\emailAdd{zhangyongchao@seu.edu.cn}



\abstract{
We apply machine learning to the searches of heavy neutrino mixing in the inverse seesaw in the framework of
left-right symmetric model at the high-energy hadron colliders. The Majorana nature of heavy neutrinos can induce the processes $pp \to W_R^\pm \to \ell_\alpha^\pm N \to \ell_\alpha^\pm \ell_\beta^{\mp,\,\pm} jj$, with opposite-sign (OS) and same-sign (SS) dilepton and two jets in the final state. The distributions of the charged leptons $\ell = e ,\, \mu$ and jets and their correlations are utilized as input for machine learning analysis. It is found that for both the OS and SS processes, XGBoost can efficiently distinguish signals from the standard model backgrounds. We estimate the sensitivities of heavy neutrino mass $m_N$ and their mixing in the OS and SS $ee$, $\mu\mu$ and $e\mu$ final states at $\sqrt{s} = 14$ TeV, 27 TeV and 100 TeV. It turns out that the heavy neutrinos can be probed up to 17.1 TeV and 19.5 TeV in the OS and SS channels, respectively. The sine of the mixing angle of heavy neutrinos can be probed up to the maximal value of $\sqrt2/2$ and 0.69 in the OS and SS channels, respectively.




}

\maketitle


\section{Introduction}

One of the foremost objectives in particle physics is to elucidate the origin of neutrino masses and mixings. A simple and straightforward approach is to introduce the heavy right-handed neutrinos along with the associated Majorana masses $M_N$ and Dirac masses $M_D$. This addition allows for the generation of a small but nonzero neutrino masses:
\begin{eqnarray}
	m_{\nu} = - M_{D} M_{N}^{-1} M_{D}^{\sf T} \,,
\end{eqnarray}
which is known as the type-I seesaw mechanism~\cite{Minkowski:1977sc,Mohapatra:1979ia,Yanagida:1979as,Gell-Mann:1979vob,Glashow:1979nm}. In this framework, if the Dirac masses $M_D$ is at the electroweak scale, $M_{N}$ is required to be extremely large, roughly $10^{14}$ GeV. 
Such a high-energy scale is apparently inaccessible by any collider experiments. To facilitate the observation of new physics for neutrino mass generation at the TeV scale, some ingredients have to be added to the seesaw model, for instance some discrete symmetries in the lepton sector (see reviews in e.g. Refs.~\cite{Altarelli:2010gt, Ishimori:2010au, King:2013eh, Petcov:2017ggy, Feruglio:2019ybq, Xing:2020ijf, Chauhan:2023faf, Ding:2023htn, Ding:2024ozt}), or the canonical type-I seesaw mechanism has to be extended in some way. The inverse seesaw is one of such well-motivated examples, where three extra singlet fermions $S$ under the standard model (SM) gauge group are introduced~\cite{Mohapatra:1986aw,Mohapatra:1986bd}. In the inverse seesaw framework, the tiny neutrino masses are proportional to the Majorana mass parameter $\mu_{S}$, i.e.
\begin{eqnarray}
\label{eqn:inverseseesaw}
m_{\nu} = (M_D M_N^{-1}) \mu_S (M_D M_N^{-1})^{\sf T} \,.
\end{eqnarray}
The introduction of the extra singlet fermions and the new suppression parameter $\mu_{S}$ not only reduce the masses of heavy neutrinos to the accessible energy scale, but also allow for the retention of large Yukawa coupling coefficients. 


The Majorana nature of heavy neutrinos leads to distinct signatures at the high-energy colliders that are absent in the SM, e.g. the lepton number violation (LNV) signals (see reviews in e.g. Refs.~\cite{FileviezPerez:2015mlm,Cai:2017mow}). In the case of a single heavy neutrino $N$ in the framework of type-I seesaw mechanism, the ratio of LNV to lepton number conserving (LNC) signals is expected to be one. 
However, in the presence of two heavy neutrinos, the ratio can differ significantly from one, depending on the heavy neutrino masses, mixing angle, the CP violating phase as well the mixing of heavy neutrinos with the active ones~\cite{BhupalDev:2015kgw,Deppisch:2015qwa,Anamiati:2016uxp,Das:2017hmg,Antusch:2017ebe,Abada:2019bac,Drewes:2019byd,Godbole:2020jqw,Arbelaez:2021chf,Schubert:2022lcp,Drewes:2022rsk,Batra:2023ssq}. In this paper, we focus on the LNC and LNV signals induced heavy neutrinos in the inverse seesaw in the framework of left-right symmetric model (LRSM)~\cite{Pati:1974yy,Mohapatra:1974gc,Senjanovic:1975rk}. In such a framework, the heavy neutrinos can be produced at the high-energy hadron colliders from the decay of an (off-shell) $W_R$ boson, i.e. through the Keung-Senjanovi\'{c} (KS) process~\cite{Keung:1983uu} 
\begin{equation}
\label{eqn:process}
pp \to W_R^{\pm (\ast)} \to \ell_\alpha^\pm N \to \ell_\alpha^\pm \ell_\beta^{\mp,\,\pm} jj \,.
\end{equation}
The dilepton pair in the final state can be either opposite-sign (OS) or same-sign (SS), with the latter one revealing the Majorana nature of heavy neutrinos and constituting one of the smoking-gun signals of LRSMs. The extensive studies of this process can be found in Refs.~\cite{
Langacker:1984dc, Ho:1990dt, Datta:1992qw, Frere:2008ct, Schmaltz:2010xr, Nemevsek:2011hz, Chen:2011hc, Nemevsek:2011aa, Aguilar-Saavedra:2012dga, Das:2012ii, Han:2012vk, Barry:2013xxa, BhupalDev:2013ntw, Chen:2013foz, Helo:2013ika, Lee:2013htl, Deppisch:2014qpa, Aydemir:2014ama, Deppisch:2014zta, Parida:2014dla, Vasquez:2014mxa, Gluza:2015goa, Ng:2015hba, BhupalDev:2015kgw, Helo:2015ffa, Dev:2015kca, Gluza:2016qqv, Lindner:2016lpp, Mitra:2016kov, Anamiati:2016uxp, Senjanovic:2016bya, Mattelaer:2016ynf, Ruiz:2017nip, Das:2017hmg, Cottin:2018kmq, Nemevsek:2018bbt, Frank:2018ifw, Pascoli:2018heg, Cottin:2019drg, BhupalDev:2019ljp, ThomasArun:2021rwf,Nemevsek:2023hwx} (see also Refs.~\cite{Helo:2013esa,Maiezza:2015lza,Castillo-Felisola:2015bha,Mandal:2017tab, Helo:2018rll,Li:2022cuq}). 



In this paper, we apply machine learning to the search for two nearly degenerate heavy neutrinos and their mixing in the minimal LRSM with inverse seesaw, via the processes in Eq.~(\ref{eqn:process}). Some recent papers in this direction can be found in e.g. Refs.~\cite{Feng:2021eke,Liu:2023gpt,Critchley:2024wnp} (see also Refs.~\cite{CMS:2024xdq,CMS:2024hik}).  For simplicity we focus only on the charged lepton flavors $\ell = e,\, \mu$ in the final state, and require that the heavy neutrinos be lighter than the $W_R$ boson. We will consider both the OS and SS dilepton signals, and the SM backgrounds are mainly from the $W^+ W^-jj$, $W^\pm W^\pm jj$, $Z jj$, $W^\pm Z jj$ and $t \bar{t} jj$ processes, depending on the charged lepton flavors involved and whether the signal is OS or SS process. The kinetic distributions of the charged leptons, jets and the missing transverse energy for the various signals and the corresponding backgrounds are utilized as inputs for the analysis. It is found that machine learning, such as the Extreme Gradient Boosting (XGBoost), can help distinguish significantly signals from backgrounds~\cite{Chen:2016btl}. We perform the simulations for three benchmark high-energy $pp$ colliders, i.e. the 14 TeV High-Luminosity Large Hadron collider (HL-LHC) with an integrated luminosity of 3 ab$^{-1}$, the 27 TeV High-Energy LHC (HE-LHC) with the luminosity of 15 ab$^{-1}$~\cite{ATLAS:2019mfr}, and the future 100 TeV collider, such as the Future Circular Collider (FCC-hh) and the Super Proton-Proton Collider (SPPC), with the luminosity of 30 ab$^{-1}$~\cite{FCC:2018vvp,Tang:2015qga}. It turns out that the prospects of the heavy neutrino mass $m_N$ and the mixing angle at the future high-energy colliders are distinct in the OS and SS signals. In particular, the heavy neutrinos $N$ can be probed up to 3.78 (4.22) TeV, 7.9 (8.2) TeV and 17.1 (19.5) TeV at the center-of-mass energies of $\sqrt{s}=14$ TeV, 27 TeV and 100 TeV in the OS (SS) channel. The sine of the heavy neutrino mixing angle can be probed in the range of [0.30~0.707], [0, 0.707] and [0, 0.707] in the OS channels at $\sqrt{s} = 14$ TeV, 27 TeV and 100 TeV; for the SS processes the sine of the angle can be probed up to 0.49, 0.65 and 0.69, respectively.



The rest of this paper is organized as follows. In Section \ref{framework} we sketch briefly the LRSM involving the inverse seesaw mechanism. 
Section \ref{moni} is devoted to the simulations of the OS signals $e^+ e^-$, $\mu^+ \mu^-$, $e^\pm \mu^\mp$ and the SS signals $e^\pm e^\pm$, $\mu^\pm \mu^\pm$, $e^\pm \mu^\pm$ as well as the corresponding backgrounds at the hadron colliders. In this section we focus on the various kinetic distributions that can enhance the discrimination between the signal and background processes. In Section \ref{ML}, we apply the XGBoost algorithm to optimize the discrimination, for various OS and SS signal processes. We conclude in Section ~\ref{con}, with some brief discussions.


\section{Framework}
\label{framework}

\subsection{Inverse seesaw mechanism in the left-right symmetry model}

The Yukawa Lagrangian of the inverse seesaw in the framework of LRSM is based on the gauge group $SU(3)_{C}\times SU(2)_{L}\times SU(2)_{R}\times U(1)_{B-L}$, given by:
\begin{eqnarray}
{\cal L}_Y \ = \ h_\ell \bar{L}_{L} \Phi L_{R}+h_\nu \bar{L}_{R}\chi_R S 
+f \overline{L_{R}^C} \Delta_R L_{R}
+\mu_S \overline{S^C} S ~+~ {\rm H.c.} \,,
\end{eqnarray}
where $L_{L}(1,2,1,-1)$ and $L_{R}(1,1,2,-1)$ represent the lepton doublets under the $SU(2)_{L}$ and $SU(2)_{R}$ groups, respectively, and $S_{i}$ (with $i$ = 1, 2, 3) are singlets under the gauge group:
\begin{eqnarray}
	L_L=\begin{pmatrix}\nu \\ \ell_L\end{pmatrix},~~~~
	L_R=\begin{pmatrix}N \\ \ell_R\end{pmatrix},~~~
	S = S_{i,L} \,.
\end{eqnarray}
For simplicity, we have neglected the generation indices of the doublets and singlets. $h_{\nu,\,\ell}$ and $f$ are the Yukawa coupling matrices, and $\mu_S$ is the mass matrix for the $S$ fields. On the scalar fields, $\Phi(1, 2, 2, 0)$ is a bidoublet, and $\chi_R (1,1,2,0)$ and $\Delta_{R}(1,1,3,2)$ are a doublet and triplet under $SU(2)_R$, respectively:
\begin{eqnarray}
\Phi = \begin{pmatrix}\phi_1^0&\phi_2^+\\\phi_1^-&\phi_2^0\end{pmatrix},~~~
	\chi_R=\begin{pmatrix}\chi_R^0\\\chi_R^-\end{pmatrix},~~~
	\Delta_{R}=\begin{pmatrix}\frac{1}{\sqrt{2}}\Delta_{R}^{+} & \Delta_{R}^{++} \\ \Delta_{R}^{0} & -\frac{1}{\sqrt{2}}\Delta_{R}^{+} \end{pmatrix} \,.
	\label{1.1}
\end{eqnarray}

After spontaneous symmetry breaking, the vacuum expectation values (VEVs) of the neutral scalars can be represented as follows:
\begin{eqnarray}
\langle\phi\rangle=\frac{1}{\sqrt{2}}\begin{pmatrix}\kappa~~0\\0~~\kappa^{\prime}\end{pmatrix},~~~
	\langle\chi_R\rangle=\frac{1}{\sqrt{2}}\begin{pmatrix} \sigma_{R} \\ 0\end{pmatrix},~~~
	\langle \Delta_{R}^{} \rangle = \frac{1}{\sqrt{2}}\begin{pmatrix}v_{R}~~0\\0~~0\end{pmatrix} \,,
\end{eqnarray}
with the electroweak VEV $v = \sqrt{\kappa^2 + \kappa^{\prime2}}$. The resultant mass matrix for the neutral fermions can be expressed as, in the basis of $(\nu,\, N^C,\, S)$:
\begin{eqnarray}
	{\cal M} = \left(\begin{array}{ccc} 0 & M_D & 0\\ M^{\sf T}_D & \mu_R & M_N\\
		0 & M^{\sf T}_N & \mu_S \end{array}\right),
	\label{eq:mass:matrix}
\end{eqnarray}
where the mass matrices $M_D = h_\ell v$, $M_N=h_\nu \sigma_R$ and $\mu_R=fv_R$. In the limit of $||\mu_S|| \ll ||M_D|| \ll ||\mu_R|| \ll ||M_{N}||$ (with $||x|| \equiv \sqrt{{\rm tr}(x^\dagger x)}$ the
positive norm of the matrix $x$), the mass matrix ${\cal M}$ can be block diagonalized to obtain the tiny masses of the light neutrinos via the inverse seesaw mechanism in Eq.~(\ref{eqn:inverseseesaw})~\cite{Mohapatra:1986aw,Mohapatra:1986bd}.
The masses of $N_R$ and $S_L$ are predominantly governed by the matrix $M_N$, with eigenvalues at the order of $||M_N \mp \mu_S/2 \mp \mu_R/2||$. To explore the impact of interference between the heavy states on the OS and SS dilepton signals at the high-energy hadron colliders, we consider here only one generation of $N$ and $S$, and the mixing of them leads to the two real eigenstates: 
\begin{subequations}
\label{eq:N1N2}
\begin{align}
N_{1} &= c_\alpha N +s_\alpha S \,, \\
N_{2} &= i (-s_\alpha N +c_\alpha S) \,,
\end{align}
\end{subequations}
where $s_\alpha \equiv \sin\alpha$, $c_\alpha \equiv \cos_\alpha$ with $\alpha$ the mixiang angle. It is possible that the state $N$ is a linear combination of the pure flavor states $N_{e,\, \mu,\,\tau}$, i.e. $N = U_{1\beta} N_\beta$, 
with $U_{i\beta}$ the mixing matrix for $N_{e,\,\mu,\,\tau}$ ($i$ and $\beta$ are the mass and flavor indices, respectively). For simpicity, we assme $U_{1\tau} = 0$ and neglect the potential CP violating phase in the matrix $U_{}$, then $N$ can be written as
\begin{equation}
\label{eqn:Nbeta:mixing}
N = c_\theta N_e + s_\theta N_\mu \,,
\end{equation}
with $c_\theta \equiv \cos\theta$ and $s_\theta \equiv \sin\theta$. 
In the LRSM, neglecting mixing of $N$ and $S$ with the active neutrinos and the mixing of $W$ and $W_R$ bosons, the decay of heavy neutrinos occurs via the $W_{R}^{\pm}$ gauge bosons. Then the couplings of $N_{1,2}$ with the $W_R$ boson
can be expressed as~\cite{Das:2017hmg}: 
\begin{eqnarray}
\label{WL}
\mathcal{L}_{L} &=&
- \frac{g_R}{\sqrt{2}} W^-_{R,\mu} 
	\left( c_\theta \overline{e} + s_\theta \overline{\mu} \right) \gamma^\mu P_R (c_\alpha N_1 + is_\alpha N_2)   \nonumber \\
      && - \frac{g_R}{\sqrt{2}} W^+_{R,\mu} 
	 (c_\alpha N_1 - is_\alpha N_2)^{\sf T} C \gamma^\mu P_R \left( c_\theta e + s_\theta \mu \right) 
\end{eqnarray}
where $C$ is the charge conjugation operator, $P_{R} = (1+\gamma_5)/2$ denotes the right-handed chirality projection operator, and $g_R$ is the gauge coupling for the $SU(2)_R$ gauge group. Such a setup is essential for having all the flavor combinations $ee$, $\mu\mu$ and $e\mu$ for our signals in this paper. The coupling between quarks and $W_{R}$ can be expressed as:
\begin{eqnarray}
	\mathcal{L}_{Q}= - \frac{g_R}{\sqrt{2}} W_{R,\mu}^+ \sum_{i,j}\overline{u}_iV_{R,\,ij}^{\mathrm{CKM}} \gamma^\mu P_Rd_j+\mathrm{H.c.} \,,
	\label{WQ}
\end{eqnarray}
where $u_{i}$ and $d_{j}$ are the up- and down-type quarks, respectively, $i,\,j=1,\,2,\,3$ are the generation indices, and $V_{R}^{\rm CKM}$ is the right-handed Cabbibo-Kobayashi-Masakawa (CKM) matrix for quark mixing. In the minimal LRSM, the right-handed quark mixing matrix is identical to the CKM matrix in the SM, up to some quark mass signs~\cite{Zhang:2007fn,Zhang:2007da}. For the sake of simplicity and concreteness, we take  $V_{R}^{\rm CKM}$ to be the same as the left-handed CKM matrix in the SM~\cite{ParticleDataGroup:2024cfk}.

\subsection{Heavy neutrino interference}
\label{LNV on LHC}

\begin{figure}[t!]
	\centering
	\includegraphics[width=0.48\linewidth]{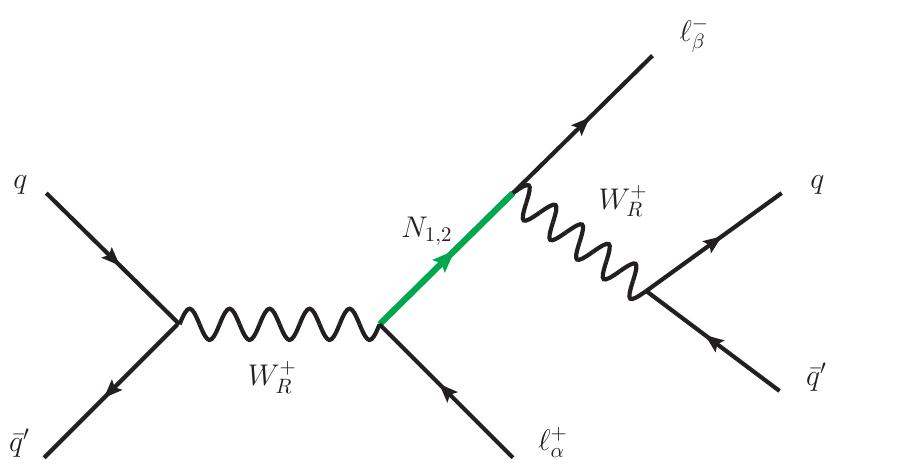}
	\includegraphics[width=0.48\linewidth]{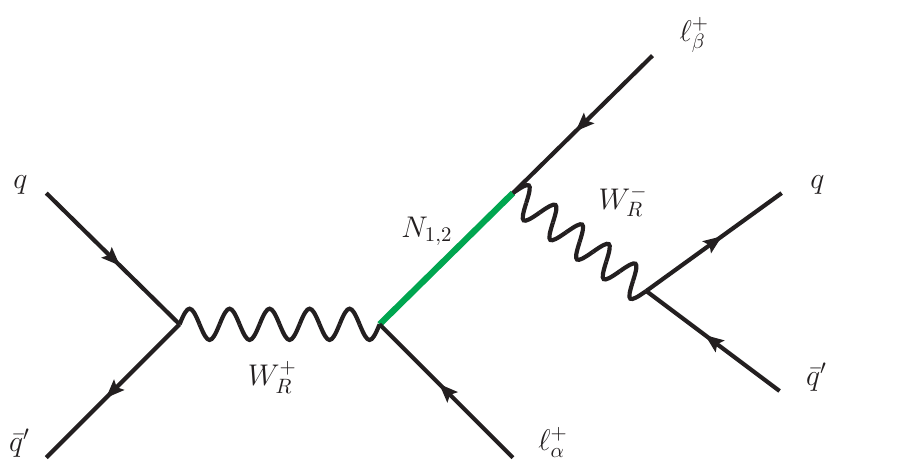}
	\caption{The representative KS signal processes of OS (left) and SS (right) dilepton pair plus two jets at the high-energy hadron colliders. 
    }
	\label{fig:diagram}
\end{figure}

The $W_R$ boson, when produced at the high-energy hadron colliders, can decay into a heavy neutrino $N_{R}$ plus a charged lepton $\ell$, eventually leading to the final state of a dilepton pair plus two jets, which is referred to as the KS process in Eq.~(\ref{eqn:process}). 
The corresponding representative Feynman diagrams are shown in  Fig.~\ref{fig:diagram}. As result of the Majorana nature of $N_{1,2}$, the heavy neutrinos can decays into either positively or negatively charged leptons, thus leading to the OS and SS dilepton signals, which is shown in presented in the left and right panels of Fig.~\ref{fig:diagram} respectively. 
In the scenario with only one single neutrino $N$, the production cross sections for the SS and OS signals are identical, i.e. $\sigma (pp \to \ell_\alpha^\pm \ell_\beta^\pm jj) = \sigma (pp \to \ell_\alpha^\pm \ell_\beta^\mp jj)$, thereby implying the same number of OS and SS signal events. 
However, with the two physical neutrinos $N_{1,2}$ in the inverse seesaw mechanism, the numbers of OS and SS dilepton events could be different, as implied by Eq.~(\ref{WL}).

Moreover, some coherence conditions must be satisfied for the interference of the two heavy neutrinos. 
In particular, if the heavy neutrinos are at the TeV range, the coherence condition of the inverse seesaw mechanism stipulates that the mass splitting between the neutrinos should not exceed the order of 100 GeV~\cite{Das:2017hmg,Akhmedov:2007fk}. 
To explore how the ratio of SS and OS events changes with the heavy neutrino parameters, we adopt the approach detailed in Refs.~\cite{Anamiati:2016uxp, Das:2017hmg}, and define the ratio ${\cal R}_{\ell\ell}$ as follows: 
\begin{eqnarray}
{\cal R}_{\ell\ell} \equiv \frac{{\cal N}_{{\rm SS},\,\ell\ell}}{{\cal N}_{{\rm OS},\,\ell\ell}}  
\ = \ \frac{\int^\infty_0 {\rm d}t\left| {\cal A}_{{\rm SS},\,\ell\ell}(t)\right|^2}{\int^\infty_0 {\rm d}t\left |{\cal A}_{{\rm OS},\,\ell\ell}(t)\right|^2}   \,,
\end{eqnarray}
where ${\cal N}_{{\rm OS},\, \ell\ell}$ and ${\cal N}_{{\rm SS},\, \ell\ell}$ are the numbers of OS and SS signal events, respectively, with the time-evolved amplitudes
\begin{subequations}
\begin{align}
{\cal A}_{\rm OS,\,\ell \ell}(t) \ &= \ C_{\ell\ell} \left[ c^2_\alpha \exp \left\{-iE_1t-\frac{1}{2} \Gamma_{N_{1}} t \right\} + s^2_\alpha \exp \left\{ -iE_2t-\frac{1}{2} \Gamma_{N_{2}} t \right\} \right] \,, \\
{\cal A}_{\rm SS,\, \ell\ell}(t) \ &= \ C_{\ell\ell} \left[ c^2_\alpha \exp \left\{ -iE_1t-\frac{1}{2} \Gamma_{N_{1}} t \right\} - s^2_\alpha \exp \left\{ -iE_2t-\frac{1}{2} \Gamma_{N_{2}} t \right\} \right] \,.
\end{align}
\end{subequations}
Here $E_{1,2}$ and $\Gamma_{N_{1,2}}$ are the energies and widths of the heavy neutrinos $N_{1,2}$, respectively, and the flavor-dependent coefficient 
\begin{equation}
C_{\ell \ell} = \begin{cases}
c_\theta^2 & \text{ for } ee \,, \\
s_\theta^2 & \text{ for } \mu\mu \,, \\
s_\theta c_\theta & \text{ for } e\mu,\; \mu e \,.
\end{cases}
\end{equation}
Under the approximation of $E_{1,2} \approx m_{1,2} \pm \Delta m /2$ with $m_{1,2}$ the masses of $N_{1,2}$ and $\Delta m$  their mass difference, we can get: 
\begin{subequations}
\label{eqn:Ncal:OS:SS}
\begin{align}
{\cal N}_{\rm OS,\, \ell\ell} ~&=~ |C_{\ell\ell}|^2 \Gamma_\mathrm{avg} \left[\frac{c_\alpha^4}{\Gamma_{N_{1}}}+\frac{s_\alpha^4}{\Gamma_{N_{2}}}+\frac{c_\alpha^2 s_\alpha^2(\Gamma_{N_{1}}+\Gamma_{N_{2}})}{\frac14 (\Gamma_{N_{1}}+\Gamma_{N_{2}})^2 + (\Delta m)^2} \right] \,, \\
{\cal N}_{\rm SS,\,\ell\ell} ~&=~ |C_{\ell\ell}|^2 \Gamma_\mathrm{avg}\left[\frac{c_\alpha^4}{\Gamma_{N_{1}}}+\frac{s_\alpha^4}{\Gamma_{N_{2}}}-\frac{c_\alpha^2 s_\alpha^2(\Gamma_{N_{1}}+\Gamma_{N_{2}})}{\frac14 (\Gamma_{N_{1}}+\Gamma_{N_{2}})^2 +(\Delta m)^2}\right] \,,
\end{align}
\end{subequations}
with $\Gamma_\mathrm{avg} \equiv (\Gamma_{N_1}+\Gamma_{N_2})/2$. The mass splitting $\Delta m$ of the heavy neutrinos originates from the Majorana mass matrices $\mu_S$ and $\mu_R$ in Eq.~(\ref{eq:mass:matrix}). The significance of LNV signals, or in other words the ratio ${\cal R}_{\ell\ell}$, depends on the parameters in the inverse seesaw framework, as detailed in Ref.~\cite{Das:2017hmg}. 

\begin{figure}[t!]
	\centering
	\includegraphics[width=0.55\linewidth]{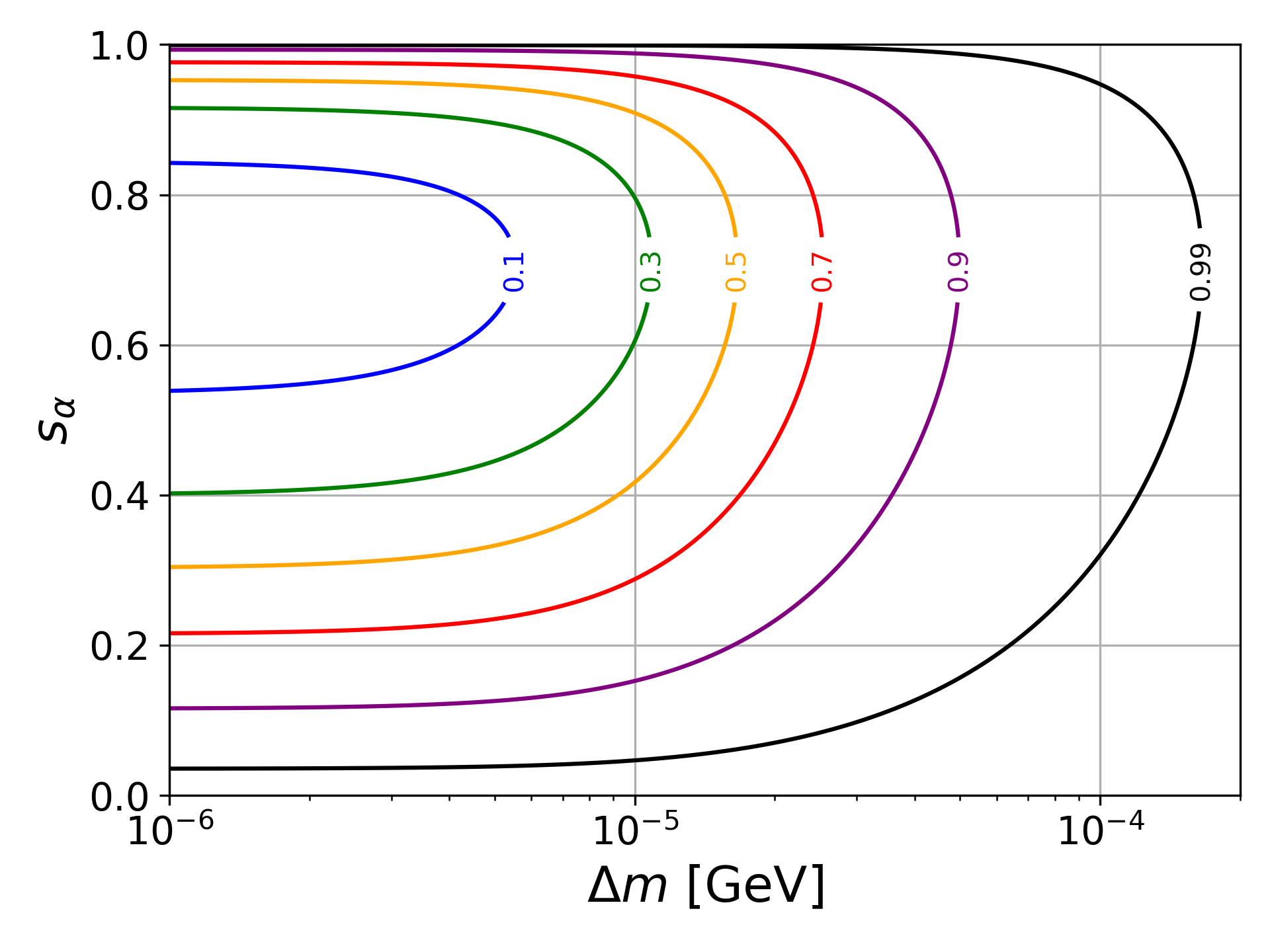}
	\caption{
 Contours of the ratio \({\cal R}_{\ell\ell} = 0.1,\, 0.3,\, 0.5,\, 0.7,\, 0.9\) and $0.99$ as functions of the heavy neutrino mass splitting \(\Delta m\) and the sine of mixing angle \(s_\alpha\), given in Eq.~(\ref{casetwo}). Other parameters are set to be: $g_R = g_L$, $m_N = 1.5$ TeV and $m_{W_R} = 6.5$ TeV. See text for more details. 
 }
\label{LNV}
\end{figure}

In the limit of $\mu_R = 0$, the splitting  arises from $\mu_{S}$, and active neutrino oscillation data require that $\Delta m \sim ||\mu_S|| \lesssim $ keV. As a result, the mixing angle $\alpha = \pi/4$, 
    which leads to $\Gamma_{N_1} = \Gamma_{N_2}$, and the ratio can be approximated as~\cite{Anamiati:2016uxp,Das:2017hmg}:
    \begin{eqnarray}
	{\cal R}_{\ell\ell}  \simeq  \frac{(\Delta m)^2}{2\Gamma_\mathrm{avg}^2+(\Delta m)^2} \,.
\end{eqnarray}
In light of the small $\Delta m$, the ratio ${\cal R}_{\ell\ell}$ is always very small, i.e. ${\cal R}_{\ell\ell} \lesssim {\cal O} (10^{-2})$, and thus less interesting for the high-energy collider searches of LNV signals. 
In this paper, we will focus on the case of $||\mu_R|| \ll ||M_N||$,
where the ratio ${\cal R}_{\ell\ell}$ can be written as~\cite{Das:2017hmg}: 
\begin{eqnarray}
\label{casetwo}
	{\cal R}_{\ell\ell} &\ \simeq &\ \frac{c_{2\alpha}^{2} + 4 \left( \frac{\Delta m}{\Gamma_0} \right)^2 }
    {1+s_{2\alpha}^{2}+ 4 \left( \frac{\Delta m}{\Gamma_0} \right)^2 } \,,
\end{eqnarray}
where $s_{2\alpha} \equiv \sin 2\alpha$ and $c_{2\alpha} \equiv \cos 2\alpha$, 
and the decay width $\Gamma_0$ is~\cite{Gluza:2015goa,Das:2016akd}:
\begin{equation}
\Gamma_0 (N \to \ell q \bar{q}') = \frac{3g_R^4 m_N}{512 \pi^3} \frac{1}{x}
\left[ 1 - \frac{x}{2} - \frac{x^2}{6} + \frac{1-x}{x} \log (1-x) \right] \,,
\end{equation}
where $x \equiv m_N^2 / m_{W_R}^2$ with $m_{W_R}$ the $W_R$ boson mass. 
In this case the LNV signals could be significant, i.e. ${\cal R}_{\ell\ell} \sim {\cal O} (1)$. For illustration purposes, the contours for the values of ${\cal R}_{\ell\ell} = 0.1,\, 0.3,\, 0.5,\, 0.7,\, 0.9$ and $0.99$ as functions of the mass splitting $\Delta m$ and the sine of mixing angle $s_\alpha$ are shown in Fig.~\ref{LNV}, where we have set the heavy neutrino masses $m_{1,2} = m_N = 1.5$ TeV and the $W_R$ boson mass $m_{W_R} = 6.5$ TeV. The right-handed gauge coupling $g_R$ for the group $SU(2)_R$ is set to be the same as $g_L$ for the SM $SU(2)_L$, i.e. $g_R = g_L$. 
It is clear in this figure that when the mass splitting $\Delta m \lesssim 10^{-4}$ GeV, the value of \({\cal R}_{\ell\ell}\) varies significantly between 0 and 1 as a function of $s_\alpha$. In scenarios with $\Delta m \gtrsim 10^{-4}$ GeV, the ratio \({\cal R}_{\ell\ell} \simeq 1\). Therefore, 
in the simulations below we will assume the the heavy neutrino mass splitting $\Delta m$ to be very small, i.e. $\Delta m \lll m_N$. In this case, the ratio ${\cal R}_{\ell \ell}$ could vary in between $0$ and $1$ while satisfying the coherence conditions for the interference of the two heavy neutrinos $N_{1,\,2}$.

\subsection{Collider constraints}



The ATLAS and CMS collaborations have searched for the heavy neutrino $N$ and the heavy $W_{R}$ boson through both the LNV and LNC processes, but no signals beyond the SM have been observed, thus imposing various constraints on $m_{N}$ and $m_{W_{R}}$. The most stringent limits are from the latest data with an integrated luminosity of 139 fb$^{-1}$ at ATLAS~\cite{ATLAS:2023cjo} and 138 fb$^{-1}$ at CMS~\cite{CMS:2021dzb}. 
For both heavy Dirac and Majorana neutrino mass $m_N < 1$ TeV, the $W_R$ boson mass $m_{W_R}$ has been excluded up to 6.4 TeV; for a heavier neutrino $N$, the $W_R$ limit get relatively weaker. 
Here for the sake of simplicity it has be assumed that the gauge coupling $g_{L}=g_{R}$, and there is no mixing of $W_R$ boson with the SM $W$ boson or the heavy-light neutrino mixing. For the more general case of $g_R \neq g_L$, see e.g. Ref.~\cite{Chauhan:2018uuy} (see also Ref.~\cite{Solera:2023kwt}). 
There are also some indirect constraints on the $W_R$ boson from the low-energy precision measurements, e.g. those from the $K_0 - \overline{K}_0$ and $B - \overline{B}$ mixings~\cite{Beall:1981ze,Mohapatra:1983ae,Ecker:1985vv,Zhang:2007da} and the neutrinoless double-beta decays~\cite{Mohapatra:1981pm,Hirsch:1996qw,Tello:2010am,Chakrabortty:2012mh,Barry:2013xxa,BhupalDev:2013ntw,Huang:2013kma, BhupalDev:2014qbx,Bambhaniya:2015ipg, Li:2020flq,Li:2022cuq,deVries:2022nyh,Li:2024djp}. However, these limits are relatively weaker, or depend largely on other parameters in the LRSM. Therefore, we will not consider these low-energy constraints.  

\subsection{Parameter setups}

Taking into account the high-energy constraints from the LHC, the parameter setups of $m_N$ and $m_{W_R}$ at the center-of-mass energies of $\sqrt{s} = 14$ TeV, 27 TeV and 100 TeV for the collider simulations below are collected in Table \ref{tab:parameter}. In this table, we also specify our choice for the mixing angles which are relevant to the heavy neutrino induced signals. For simplicity we set $\theta = \pi/4$ in Eq.~(\ref{eqn:Nbeta:mixing}), which implies that the couplings of $N$ to the charged leptons $e,\,\mu$ are the same. In light of Eq.~(\ref{casetwo}), we will restrict the mixing angle $\alpha$ in the range of $[0,\,\pi/4]$.

It should be noted that in the simplified assumptions in our paper (neglecting the mixing of $N_{e,\,\mu}$ with $N_\tau$ and the CP phase), the cross sections for the $ee$, $\mu\mu$ and $e\mu$ channels are proportional to $c_\theta^4$, $s_\theta^4$ and $s_\theta^2 c_\theta^2$, respectively (cf. Eq.~\ref{eqn:Ncal:OS:SS}). For the more general case of $\theta \neq \pi/4$, the sensitivities in these signals will be very different. In particular, in the limit of $\theta$ approaching $0$ or $\pi/2$, the $e\mu$ signals will disappear in the setup of this paper. 


\begin{table}[t!]
	\centering
 	\caption{The benchmark values of \(m_{W_{R}}\) and ranges of \(m_{N_{}}\) at $\sqrt{s} = 14$ TeV, 27 TeV and 100 TeV, the value of the mixing angle $\theta$ and the range of the mixing angle $\alpha$. See text for more details. 
    }
    \vspace{5pt}
	\label{tab:parameter}
	\vspace{5pt}
	\begin{tabular}{|c|c|c|c|}
	\hline
	$\sqrt{s}$ [TeV] & 14 & 27 & 100 \\ \hline
        $m_{W_{R}}$ [TeV] & 6.5 & 9 & 20 \\ \hline
        $m_{N_{}}$ [TeV] & [1, 6.5] & [1, 9] & [1, 20] \\ \hline
        $\theta$ & \multicolumn{3}{c|}{$\pi/4$} \\ \hline
        $\alpha$ & \multicolumn{3}{c|}{$[0,\, \pi/4]$} \\ \hline
	\end{tabular}
\end{table}

\section{Simulation details}
\label{moni}

\subsection{Monte Carlo simulations}

To investigate ${\cal R}_{\ell\ell}$ at the LHC and future higher energy colliders, the following Monte Carlo simulation chain is implemnted. The initial parton level events are generated using {\tt MadGraph5}~\cite{Alwall:2014hca}. These events are then processed for hadronization and parton showering using {\tt PITHIA 8.2}~\cite{Sjostrand:2014zea}. Detailed detector effects are simulated using {\tt DELPHES 3.5.0} with the CMS detector card~\cite{deFavereau:2013fsa}. Jet reconstruction is performed using the anti-$k_t$ algorithm, with the radius parameter \( R \) set to be 0.4 and the minimum jet transverse momentum \( p_{T}^{\rm min}({\rm jet}) = 20 \) GeV~\cite{Cacciari:2008gp}. For all simulations, the {\tt NN23LO1} set is chosen for the parton distribution functions (PDFs)~\cite{Ball:2012cx,Ball:2013hta}. 

\begin{table}[t!]
	\centering
\caption{The production and decay modes of the OS and SS dilepton signals at the high-energy hadron colliders, and the corresponding dominant SM backgrounds, with the charged lepton flavors $\ell_{\alpha,\,\beta} = e,\,\mu$. The $K$-factors for the SM backgrounds are listed in the last column~\cite{Alwall:2014hca}. 
}
\vspace{5pt}
\label{tab:process}	
\begin{tabular}{|c|c|cc|c|}
\hline
\multicolumn{2}{|c|}{} & production & decay & $K$-factor \\ \hline\hline
\multirow{9}{*}{OS} & signal & $pp \to W_{R}^{\pm}$& $W_{R}^{\pm} \to \ell_\alpha^{\pm} \ell_\alpha^{\mp}jj$ &  \\ \cline{2-5}
& \multirow{2}{*}{background} & $pp \to W^{+}W^{-}j j$ & $W^{\pm}\to \ell_\alpha^{\pm} \nu_{}$ & $1.2$ \\ \cline{3-5}
&& $pp \to Z j j$ &$Z \to  \ell_\alpha^{+} \ell_\alpha^{-}$ & 1.13 \\ \cline{2-5}
& signal & $pp \to W_{R}^{\pm}$& $W_{R}^{\pm} \to \ell_\alpha^{\pm} \ell_\beta^{\mp}jj$ ($\alpha\neq\beta$) &  \\ \cline{2-5}
& \multirow{3}{*}{background} & $pp \to W^{+}W^{-}j j$ & $W^{\pm}\to \ell_\alpha^{\pm} \nu_{},\, W^{\mp} \to \ell_\beta^{\mp} \nu$ & $1.2$ \\ \cline{3-5}
&& $pp \to  W^{\pm} Z j j$ & $W^{\pm}\to \ell_\alpha^{\pm} \nu_{},\, Z \to  \ell_\beta^{+} \ell_\beta^{-}$ & 1.13 \\ \cline{3-5}
&& $pp \to t\bar{t} j j$ &$ t \to b \ell_\alpha^{+} \nu_{},\, \bar{t} \to \bar{b} \ell_\beta^{-} \bar\nu$ & 1.3 \\ \hline\hline 
\multirow{6}{*}{SS} & signal & $pp \to W_{R}^{\pm}$ & $W_{R}^{\pm} \to \ell_\alpha^{\pm} \ell_\alpha^{\pm} jj$ & \\ \cline{2-5}
& \multirow{2}{*}{background} & $pp \to W^{\pm} W^{\pm} j j$ & $W^{\pm}\to \ell_\alpha^{\pm} \nu$ & 1.52 \\ \cline{3-5}
&& $pp \to W^{\pm} Z j j$ & $W^{\pm}\to \ell_\alpha^{\pm} \nu,~ Z \to \ell_\alpha^{+} \ell_\alpha^{-}$ & 1.13 \\ 
\cline{2-5}
& signal & $pp \to W_{R}^{\pm}$ & $W_{R}^{\pm} \to \ell_\alpha^{\pm} \ell_\beta^{\pm} jj$ ($\alpha\neq\beta$) & \\ \cline{2-5}
& \multirow{2}{*}{background} & $pp \to W^{\pm} W^{\pm} j j$ & $W^{\pm}\to \ell_\alpha^{\pm} \nu,\, W^{\mp} \to \ell_\beta^{\mp} \nu$ & 1.52 \\ \cline{3-5}
&& $pp \to W^{\pm} Z j j$ & $W^{\pm}\to \ell_\alpha^{\pm} \nu,~ Z \to \ell_\beta^{+} \ell_\beta^{-}$ & 1.13 \\ 
\hline
\end{tabular}
\end{table}

The tau events behave much like hadron states at the high-enregy colliders, therefore we consider only the electron and muon flavors in this paper. Then we can have the following flavor combinations for the dilepton signals: $\ell_\alpha \ell_\beta = ee,\; \mu\mu,\; e\mu$. The various OS and SS dilepton signals and the corresponding SM backgrounds are collected in Table~\ref{tab:process}. For the OS $e^+ e^-$ and $\mu^+ \mu^-$ dilepton signals, the dominant SM backgrounds are the following processes: 
\begin{itemize}
    \item The $W^+ W^- jj$ process with the $W$ boson decaying leptonically, which has the same final state as the signal events.
    \item The $Zjj$ process, with the $Z$ boson decaying into charged leptons, i.e. $Z \to \ell^+ \ell^-$. Here we have included the contribution from the process $WZ \to Zjj$ with the two jets from the $W$ boson decay. 
\end{itemize}
The $t\bar{t} jj$ process could also contribute the backgrounds, with the (anti)top quark decaying into a $b$($\bar{b}$) quark plus leptonic states, i.e. $t \to b W \to b\ell \nu$. The $b$-jet identification rate is 75\% for the HL-LHC and HE-LHC cards, and reach up to 82\% for the FCC-hh card~\cite{deFavereau:2013fsa}. 
At the parton level, taking into account the $b$-jet identification efficiency and imposing the requirement that the leading and next-to-leading jets do not contain $b$-jets, the cross section for \( t\bar{t} \) production is found to be smaller than those for the background processes \( W^{+}W^{-} \) and \( Zjj \) above. Therefore, we neglect the \( t\bar{t} \) background for the $e^+ e^-$ and $\mu^+ \mu^-$ signals in this analysis. 
For the $e^\pm \mu^\mp$ signals, the corresponding SM backgrounds are different. As collected in Table~\ref{tab:process}, the $W^+ W^- jj$, $W^\pm Z jj$ and $t\bar{t} jj$ processes are relevant, with $W^\pm$, $Z$ and $t \to b W$ decaying into electrons and muons in the final state. 
For the SS signals of $\ell^\pm \ell^\pm = e^\pm e^\pm,\; \mu^\pm \mu^\pm,\; e^\pm \mu^\pm$, the backgrounds are mainly the $W^\pm W^\pm jj$ and $W^\pm Z jj$ processes, with the leptonic decays $W \to \ell \nu$ and $Z \to \ell^+ \ell^-$, and the charged lepton flavors matching that for the signals. For the process of $W^\pm Z \to \ell_\alpha^\pm \nu \ell_\beta^+ \ell_\beta^-$, one of the charged leptons is missed by detectors, leading to the SS dileptons $\ell_\alpha^\pm \ell_\beta^\pm$. The $K$-factors corresponding to the SM background processes above are given in the last column of Table~\ref{tab:process}~\cite{Alwall:2014hca}.

\begin{figure}[t!]
	\centering
	\includegraphics[width=0.48\linewidth]{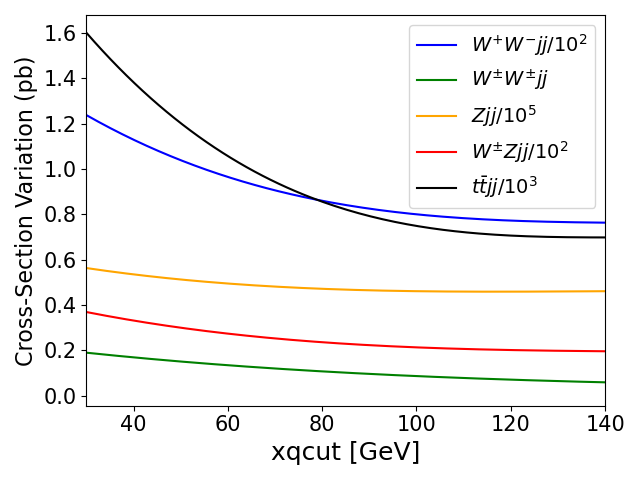}
	\caption{Production cross sections for the SM background processes $W^+ W^-$, $W^\pm W^\pm$, $Z$, $W^\pm Z$ and $t\bar{t}$ as functions of the value of {\tt xqcut} at $\sqrt{s}=14$ TeV, without considering the decays of $W$, $Z$ and $t$. 
    }
	\label{xqcut}
\end{figure}

For the SM backgrounds, the MLM matching scheme is utilized for the scenarios involving at least two jets matched to initial state radiation~\cite{Mangano:2001xp,Mangano:2006rw}. To optimize the value of {\tt xqcut}, we have conducted parameter scans for the background processes $W^+ W^- jj$, $W^\pm W^\pm jj$, $Z jj$, $W^\pm Z jj$ and $t\bar{t} jj$ 
These scans are used to observe the behavior of the production cross sections as {\tt xqcut} changes, with the results displayed in Fig.~\ref{xqcut}. All the SM processes in this figure indicate that setting ${\tt xqcut} = 100$ GeV minimizes the rate of change in cross sections for the SM backgrounds.  

The parton level cuts below are adopted to match the detector's geometric acceptance and detection capabilities:
\begin{eqnarray}
	&& p_{T} (j) > 100 ~\text{GeV} \,, ~~~~|\eta (j)| < 4.5 \,, \quad
	p_{T} (\ell) > 25 ~\text{GeV} \,, ~~~~|\eta (\ell)| < 2.5 \,, \nonumber \\ 
	&& H_{T} (j) > 1 ~\text{TeV} \,, ~~~~\Delta R(x,y) > 0.4 \,,
\end{eqnarray}
where $H_{T}(j)$ represents the scalar sum of the transverse momenta of the final state jets, and $\Delta R(x, y)$ denotes the angular separation angle between two objects $x$ and $y$. The total energy of the final state particles in the signals is related to the $W_R$ boson mass $m_{W_R}$. To suppress effectively the SM backgrounds that closely resemble the kinematic distributions of signals, we require that $H_{T}^{j} > 1$ TeV. Additionally, for the $Z$ boson processes, we impose an extra requirement that the invariant mass of the same-flavor OS dileptons $\ell^+ \ell^-$ exceeds 100 GeV to suppress the backgrounds.



\begin{table}[t!]
\centering
\caption{The parton level cross sections for the OS dilepton signals and the corresponding backgrounds at the center-of-mass energies $\sqrt{s} = 14$, 27 and 100 TeV. The cross sections are the same for the $ee$, $\mu\mu$ and $e\mu$ signals at the parton level. 
}
	\vspace{5pt}
    \label{table:cs:OS}
	\renewcommand\arraystretch{1.2} 
	\begin{tabular}{|c|c|ccc|}
	\hline
\multicolumn{2}{|c|}{$\sqrt{s}$ [TeV]}  & 14 & 27 & 100 \\ 	\hline\hline
\multirow{3}{*}{\makecell[c]{signal \\ $\text{[pb]}$}} & $m_{W_R}$ [TeV] & 6.5 & 9 & 20 \\ \cline{2-5}
& $m_N = $ 1.5 TeV  & $1.6\times 10^{-6}$ & 1.0$\times$10$^{-5}$ & $7.5\times$10$^{-6}$ \\ \cline{2-5}
& $m_N = $ 2 TeV & 1.6$\times$10$^{-6}$ & 1.2$\times$10$^{-5}$ & 1.5$\times$10$^{-5}$ \\  \hline\hline
\multirow{2}{*}{\makecell[c]{bkg for \\ $ee,\,\mu\mu$ [pb]}}	& $W^{+}W^{-}jj$   & 5.8$\times$10$^{-4}$ & 3.7$\times$10$^{-3}$ & 5.2$\times$10$^{-2}$ \\ \cline{2-5}
& $Zjj$   & 1.6$\times$10$^{-3}$ &  7.8$\times$10$^{-3}$  & 8.3$\times$10$^{-2}$ \\ 		\hline\hline
\multirow{3}{*}{\makecell[c]{bkg for \\ $e\mu$ [pb]}} & $W^{+}W^{-}jj$    & 1.1$\times$10$^{-3}$     & 7.4$\times$10$^{-3}$  & 1.0$\times$10$^{-1}$ \\		\cline{2-5}
 & $W^{\pm}Zjj$ & 8.8$\times$10$^{-6}$ & 5.4$\times$10$^{-5}$ & 6.8$\times$10$^{-4}$  \\ 	\cline{2-5}
 & $t\bar{t}jj$  & 2.3$\times$10$^{-2}$  & 2.3$\times$10$^{-1}$    & 3.9   \\ \hline

\end{tabular}
\end{table}

\begin{table}[t!]
	\centering
 \caption{The same as Table~\ref{table:cs:OS}, but for the SS dilepton signals.
 }
 \vspace{5pt}
 \label{table:cs:SS}
	\renewcommand\arraystretch{1.2} 
	\begin{tabular}{|c|c|ccc|}
\hline
\multicolumn{2}{|c|}{$\sqrt{s}$ [TeV]}  & 14 & 27 & 100 \\ 		\hline\hline
\multirow{3}{*}{\makecell[c]{signal \\ $\text{[pb]}$}} & $m_{W_{R}}$ [TeV] & 6.5  & 9 & 20  \\ \cline{2-5}
& $m_N = $ 1.5 TeV & 4.3$\times$10$^{-7}$ & $2.5\times$10$^{-6}$ & 4.0$\times$10$^{-7}$ \\ \cline{2-5}
& $m_N = $ 2 TeV & 4.0$\times$10$^{-7}$  & 2.8$\times$10$^{-6}$  & 3.0$\times$10$^{-6}$ \\ \hline\hline
\multirow{2}{*}{\makecell[c]{bkg for \\ $ee,\,\mu\mu$ [pb]}} &	$W^{\pm}W^{\pm}jj$         & 1.1$\times$10$^{-4}$    & 4.8$\times$10$^{-4}$ & 4.6$\times$10$^{-3}$ \\ \cline{2-5}
& $W^{\pm}Zjj$ &   3.4$\times$10$^{-6}$ &  2.2$\times$10$^{-5}$  & 2.8$\times$10$^{-4}$ \\ \hline\hline
\multirow{2}{*}{\makecell[c]{bkg for \\ $e\mu$ [pb]}} &  $W^{\pm}W^{\pm}jj$ &  2.2$\times$10$^{-4}$ & 9.6$\times$10$^{-4}$ & 9.3$\times$10$^{-3}$ \\		\cline{2-5}
& $W^{\pm}Zjj$ & 8.8$\times$10$^{-6}$ & 5.4$\times$10$^{-5}$ &   6.8$\times$10$^{-4}$  \\ \hline
\end{tabular}
\end{table}

Signal events are generated based on the Lagrangian in Eqs.~(\ref{WL}) and (\ref{WQ}). In the model we are considering, there are some key parameters affecting the ratio ${\cal R}_{\ell\ell}$, such as the mass $m_{N}$ of the two heavy neutrinos, the mass $m_{W_R}$ of the right-handed $W_R$ boson, and the mixing angle $\alpha$ between the heavy neutrinos. The coupling of $W_R$ to the quarks depend on the right-handed mixing matrix $V_R^{\rm CKM}$. For simplicity we set $V_R^{\rm CKM}$ to be an identity matrix, i.e. $V_R^{\rm CKM} = I_{3\times3}$, which is a good approximation. 
For simulations in this paper, we assume the neutrino masses are below the $W_R$ mass, i.e. $m_{N} < m_{W_R}$. The mixing angle of the two heavy neutrinos is within the range of ${\alpha}\in[0,\pi/4]$ (cf. Eq.~(\ref{casetwo})).
For illustration purposes, the cross sections for the OS dilepton signals and the corresponding SM background processes are shown in Table~\ref{table:cs:OS}, and that for the SS signals are collected in Table~\ref{table:cs:SS}. In both tables, we have chosen two benchmark values for the heavy neutrino mass $m_N = 1.5$ TeV and $2$ TeV. The corresponding values of $W_R$ mass at the center-of-mass energies of 14 TeV, 27 TeV and 100 TeV are taken to be 6.5 TeV, 9 TeV and 20 TeV, respectively. 
Our parton level cross sections have also confirmed that other SM processes such as $ZZjj$ contribute negligibly to the backgrounds.

\subsection{ Reconstruction of physical objects}

At the detector level, observable physical objects include jets, electrons and muons, which can be reconstructed through the transverse momentum $p_T$, pseudorapidity $\eta$, and the azimuthal angle $\phi$. The missing transverse energy $E_T^{\rm miss}$ can be reconstructed as $E_T^{\rm miss} = -\sum^{v_i} \vec{p}_T(v_i)$, where $\vec{p}_T(v_i)$ represents the transverse momentum of the $i$th visible object $v_i$. Another useful quantity is the total transverse energy $H_T = \sum_i p_T(v_i)$. 
For our analysis, we will use the reconstructed jets, charged leptons, $E_T^{\rm miss}$ and $H_T$ for our analysis.  
To enhance the distinction between signals and backgrounds, we require that the leading jet $j_1$ (charged lepton $\ell_1$) and the next-to-leading jet $j_2$ (charged lepton $\ell_2$) in each event meet the following pre-selection cuts:
\begin{align}
p_{T} (j_1,\,j_2) &> 100 ~\mathrm{GeV} \,, & |\eta (j_1,\, j_2)| &< 4.5 \,, \nonumber \\
p_{T} (\ell_1,\, \ell_2) &> 25 ~\mathrm{GeV} \,, & |\eta (\ell_1,\, \ell_2)| &< 2.5 \,.
\end{align}

\begin{figure}[t!]
	\centering
	\includegraphics[width=0.48\linewidth]{"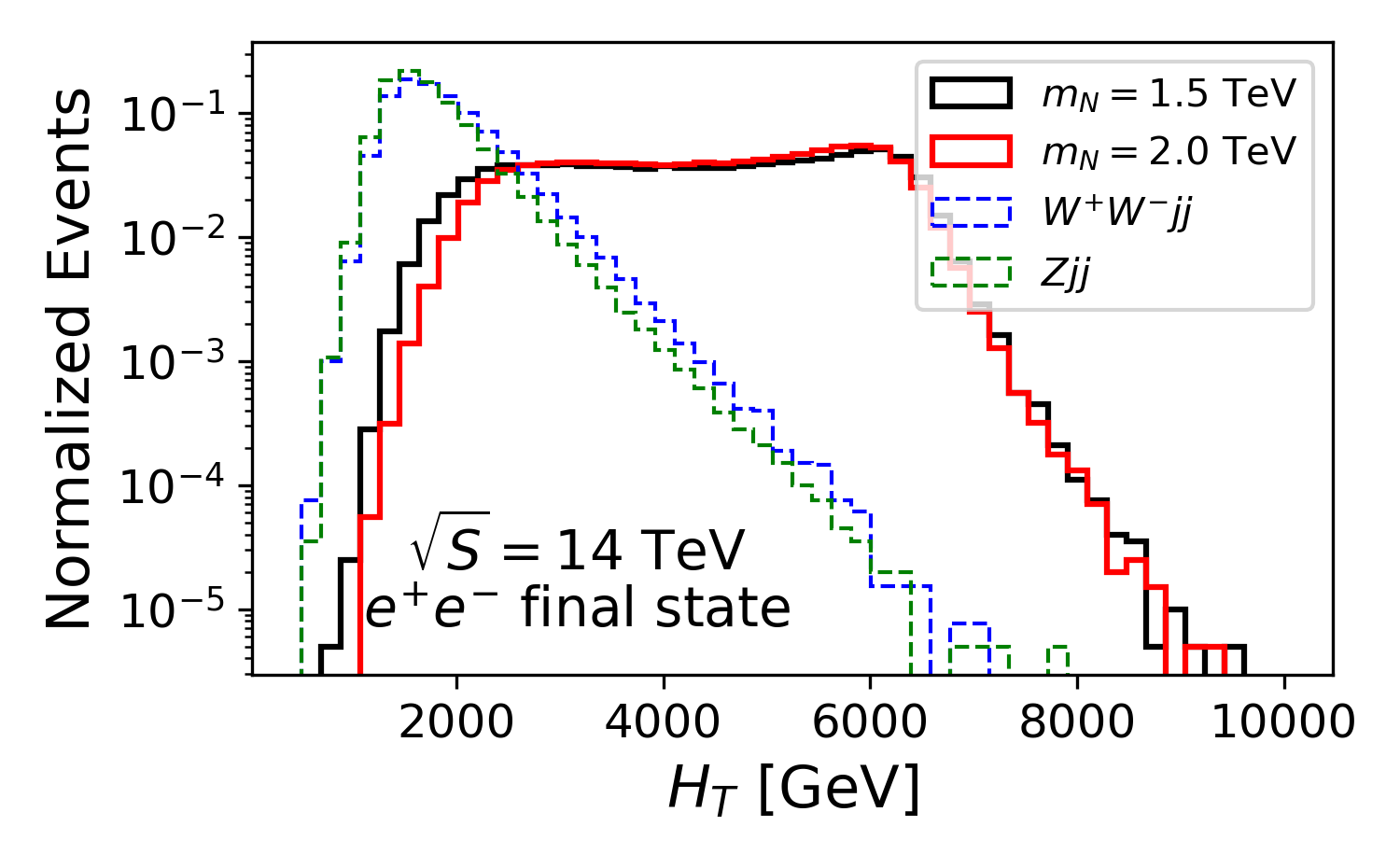"}
	\includegraphics[width=0.48\linewidth]{"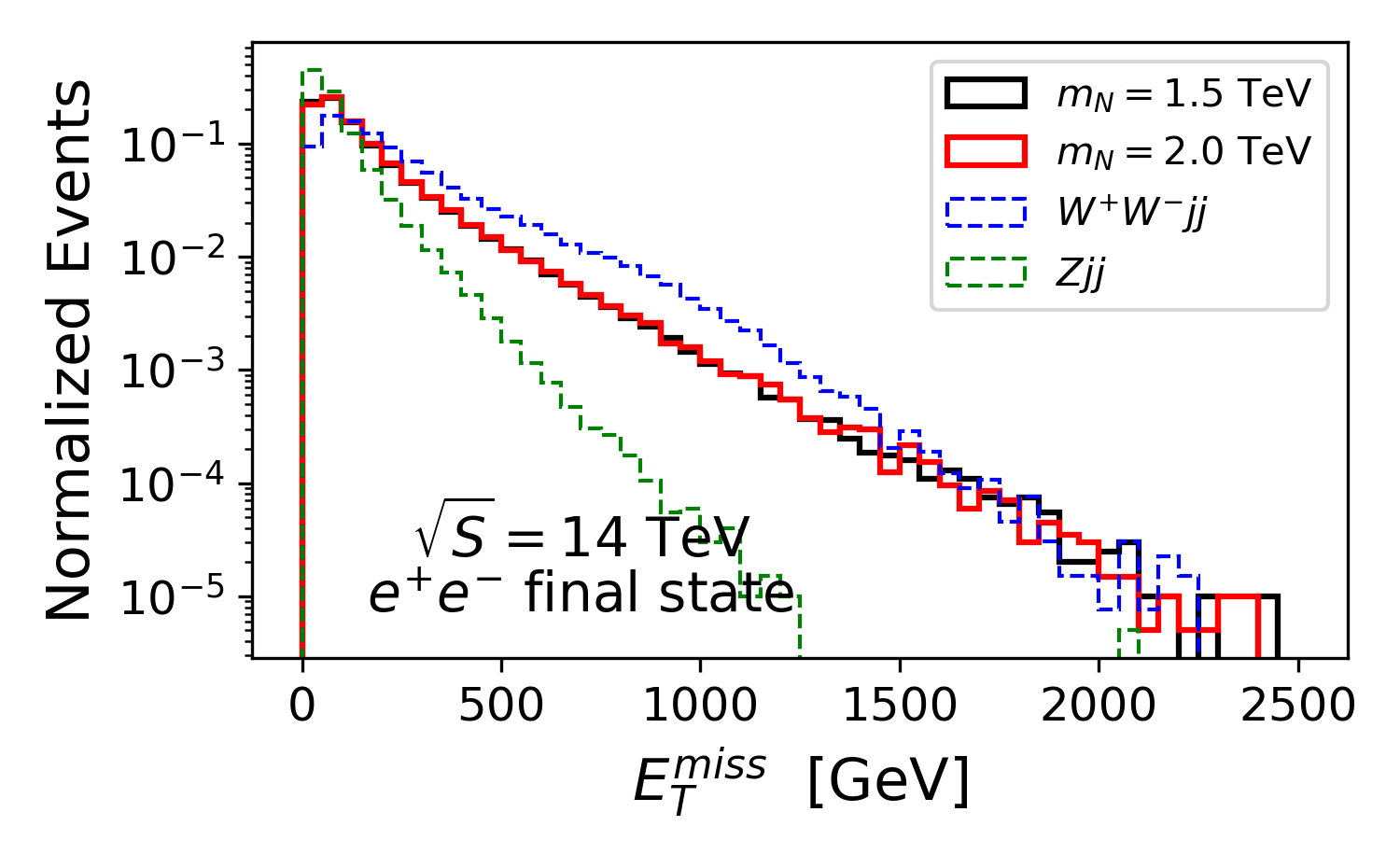"}\\
 \includegraphics[width=0.48\linewidth]{"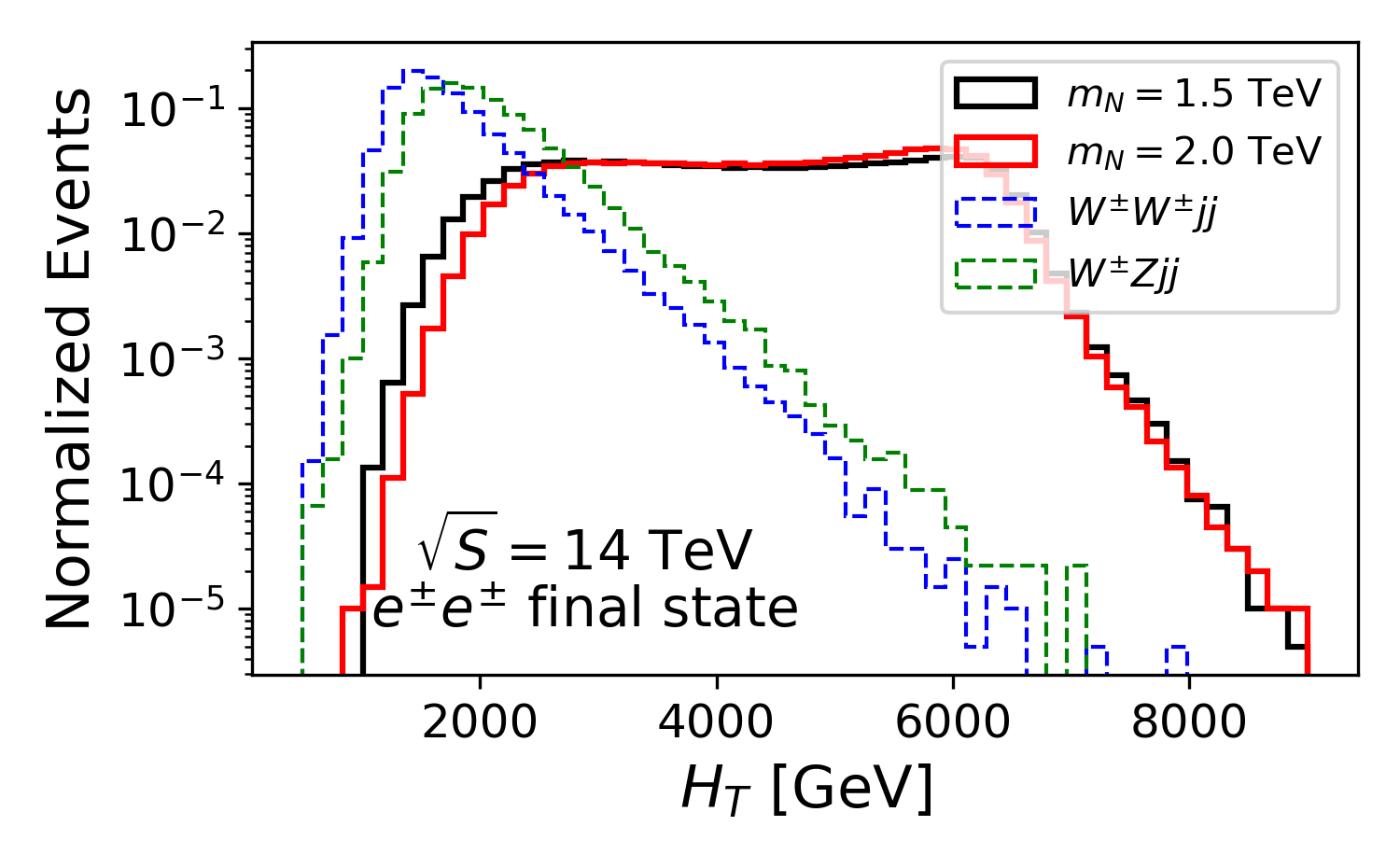"}
	\includegraphics[width=0.48\linewidth]{"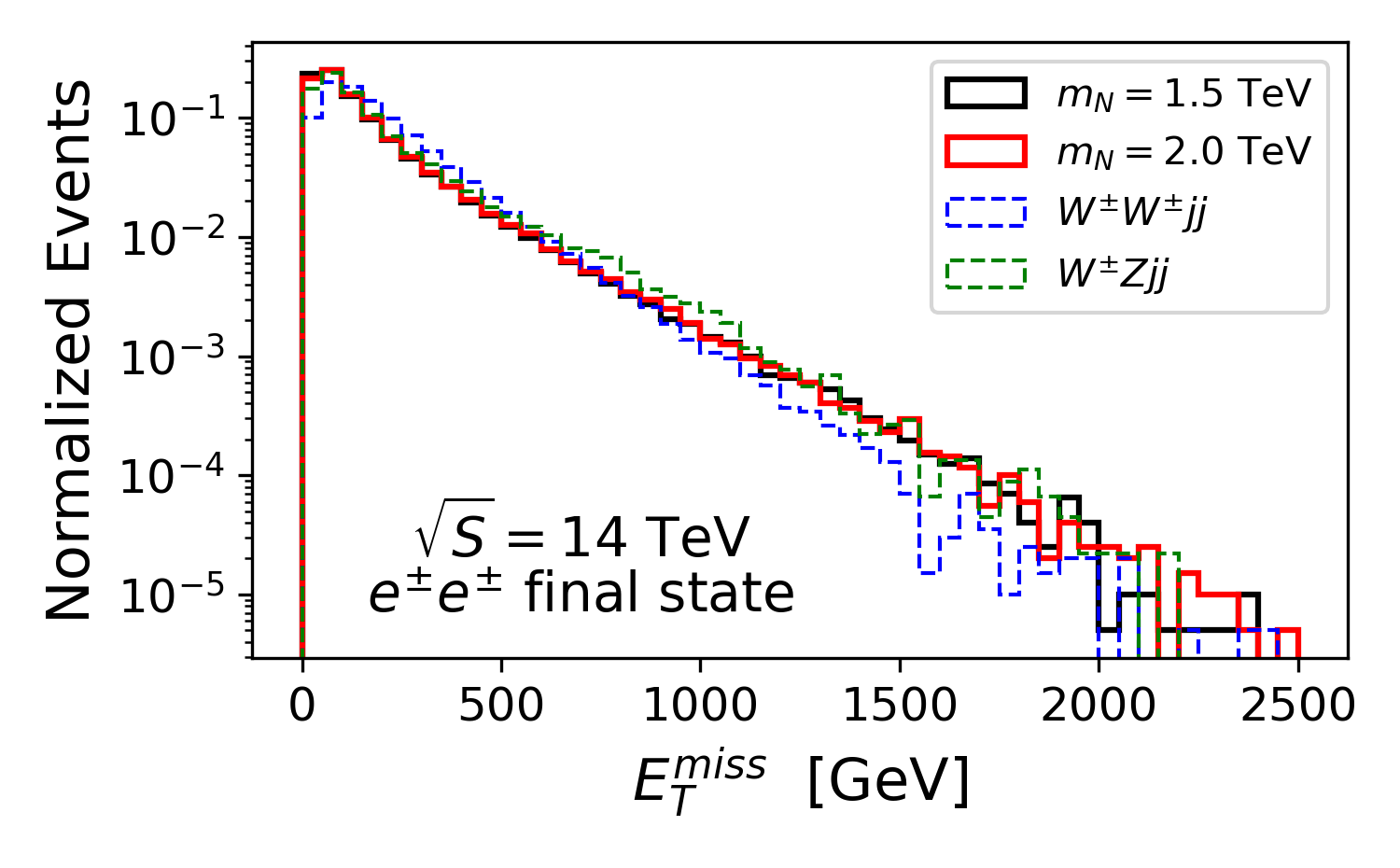"}
\caption{Distributions of the heavy neutrino $e e$ signals with $m_N = 1.5$ TeV (solid black line) or $m_N=2.0$ TeV (solid red line) and the corresponding SM backgrounds (dashed lines), for the OS (upper panels) and SS (lower panels) dilepton events at \(\sqrt{s} = 14\) TeV. The left and right panels are for the sum $H_T$ of the transverse momenta of all leptons and jets and the missing transverse momentum $E_T^{\rm miss}$, respectively.
The distributions have been normalized. 
}
\label{fig:HTandMET}
\end{figure}

\begin{figure}[t!]
	\centering
	\includegraphics[width=0.48\linewidth]{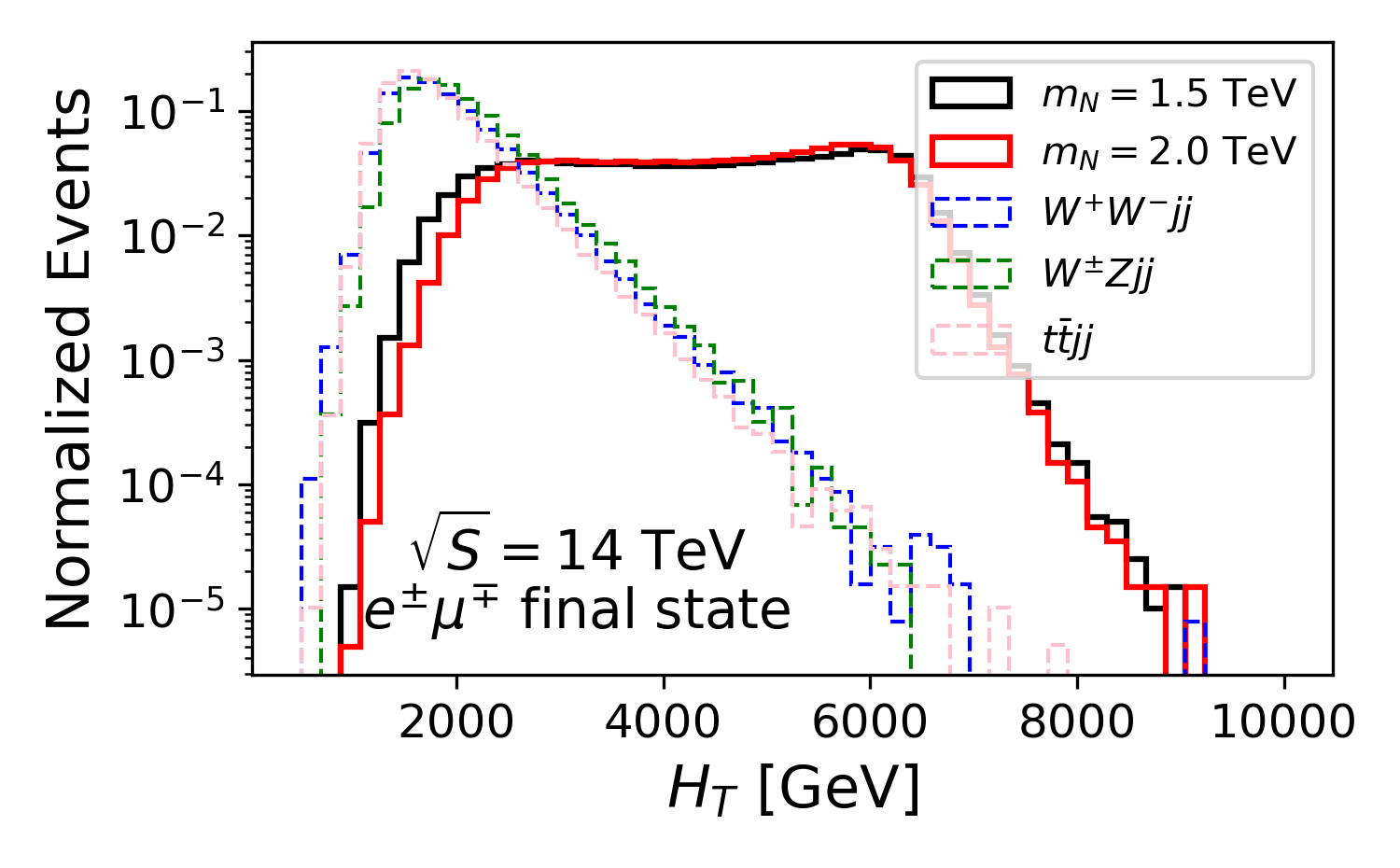}
	\includegraphics[width=0.48\linewidth]{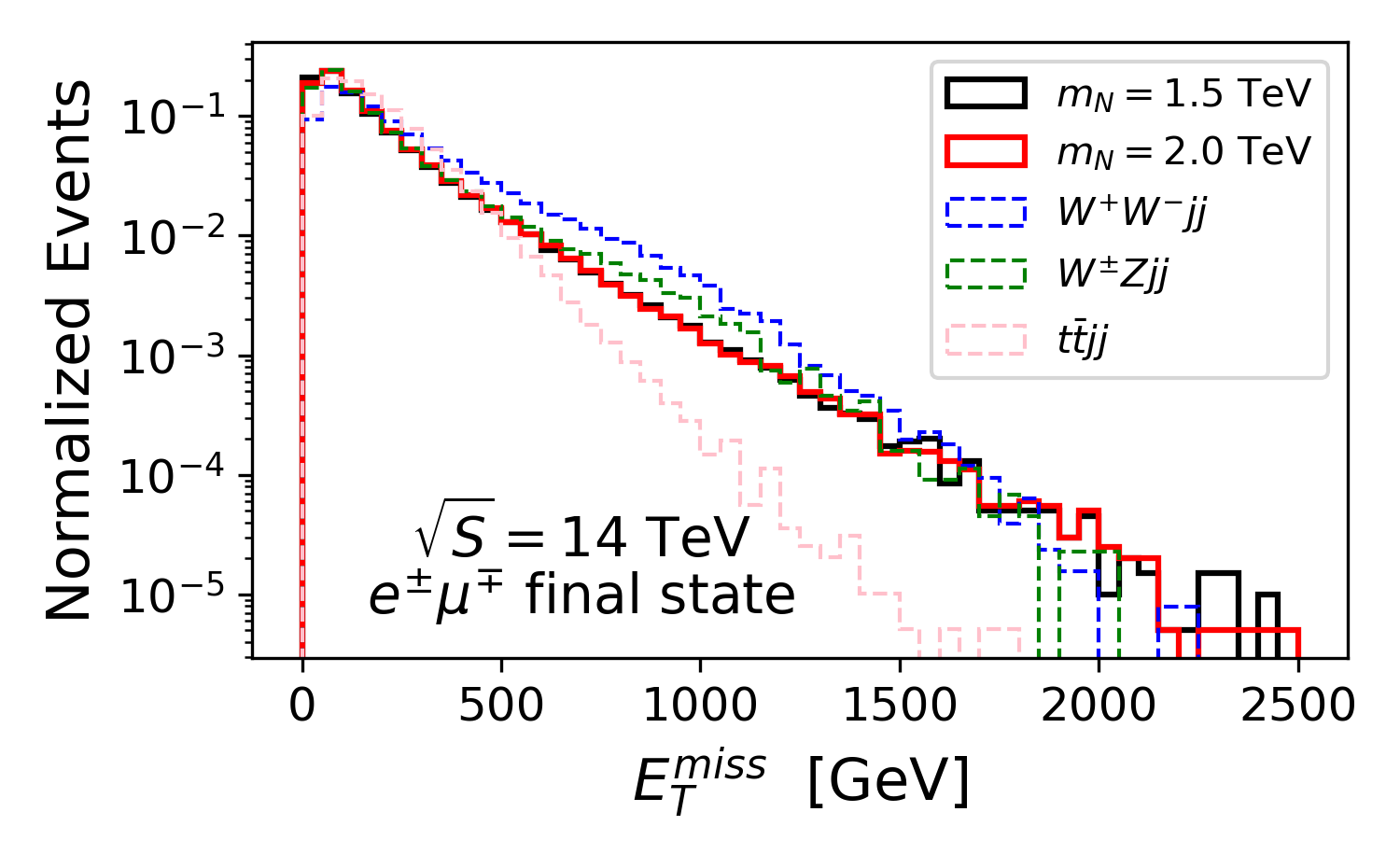}
	\includegraphics[width=0.48\linewidth]{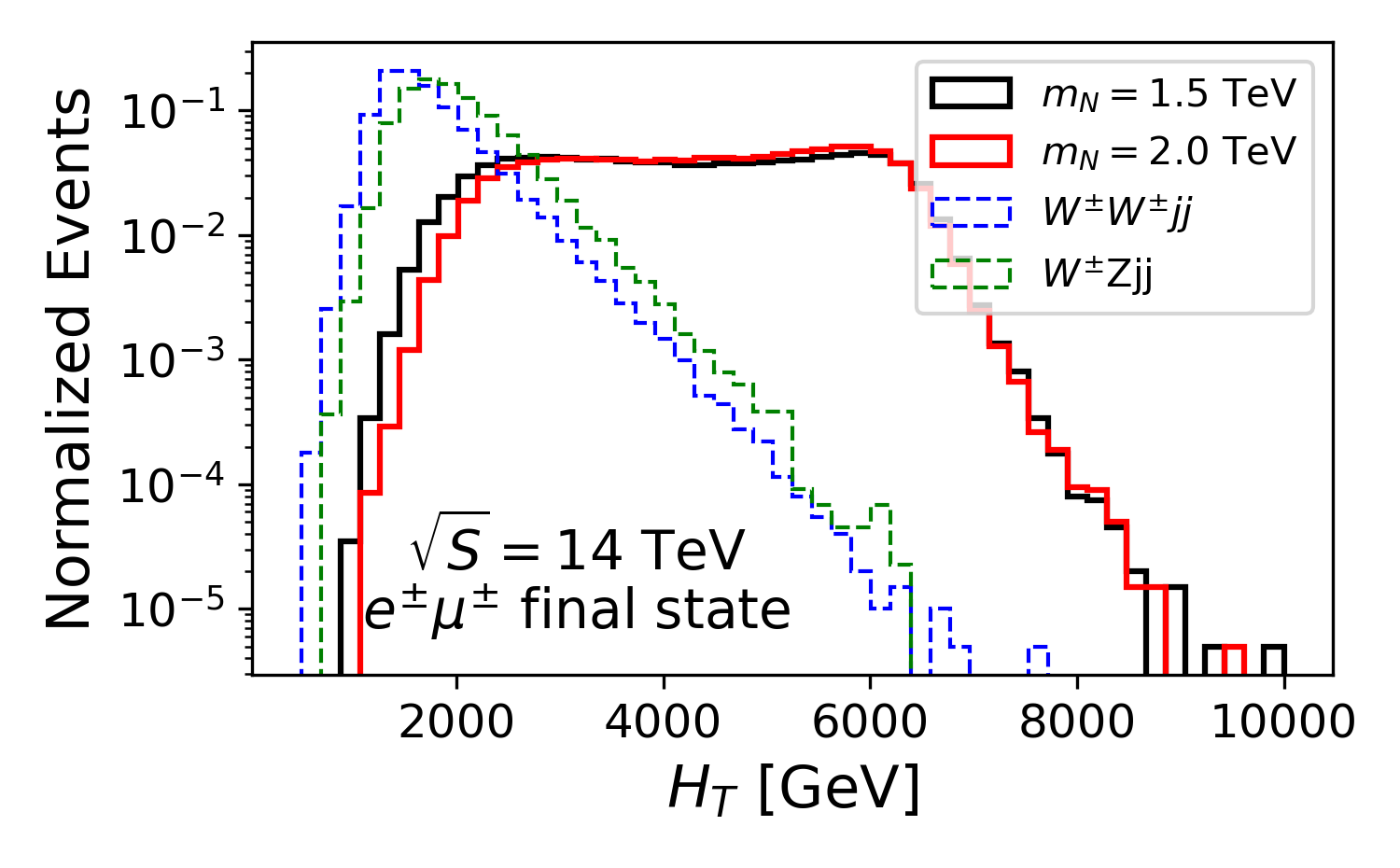}
	\includegraphics[width=0.48\linewidth]{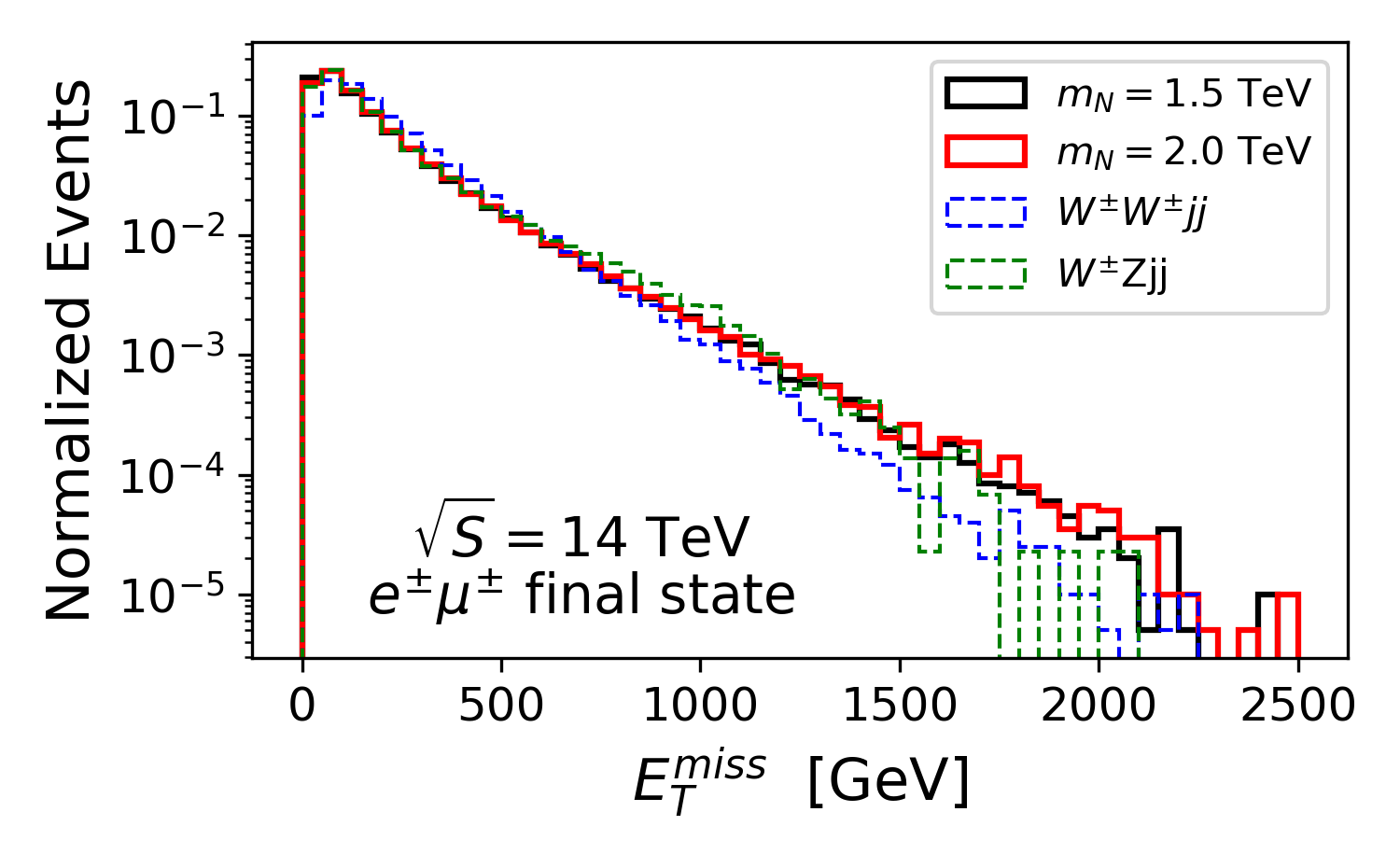}
\caption{The same as Fig.~\ref{fig:HTandMET}, but for the $e\mu$ final state. 
}
	\label{fig:14os_emu_kin_1}
\end{figure}

\begin{figure}[t!]
	\centering
	\includegraphics[width=0.48\linewidth]{"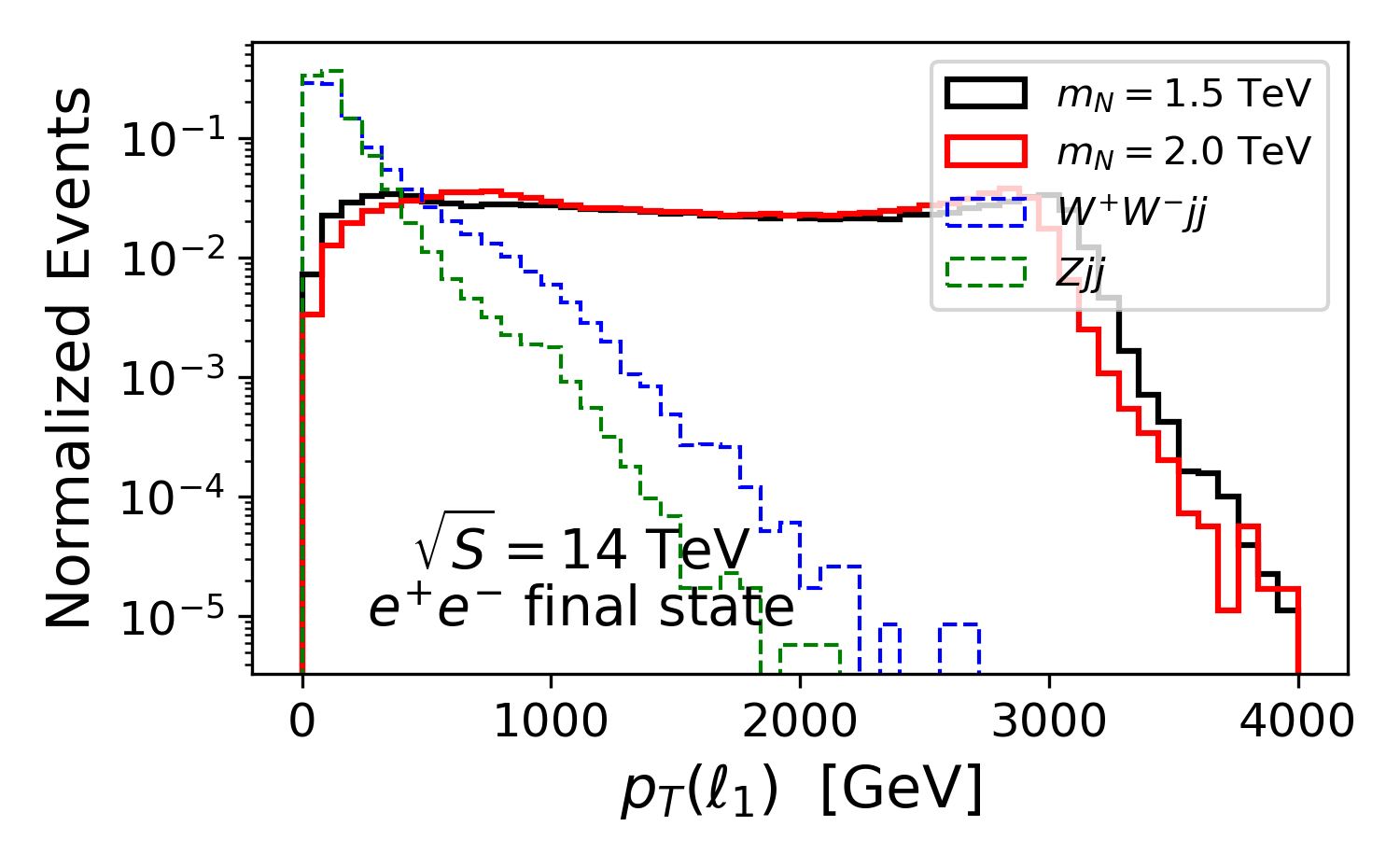"}
	\includegraphics[width=0.48\linewidth]{"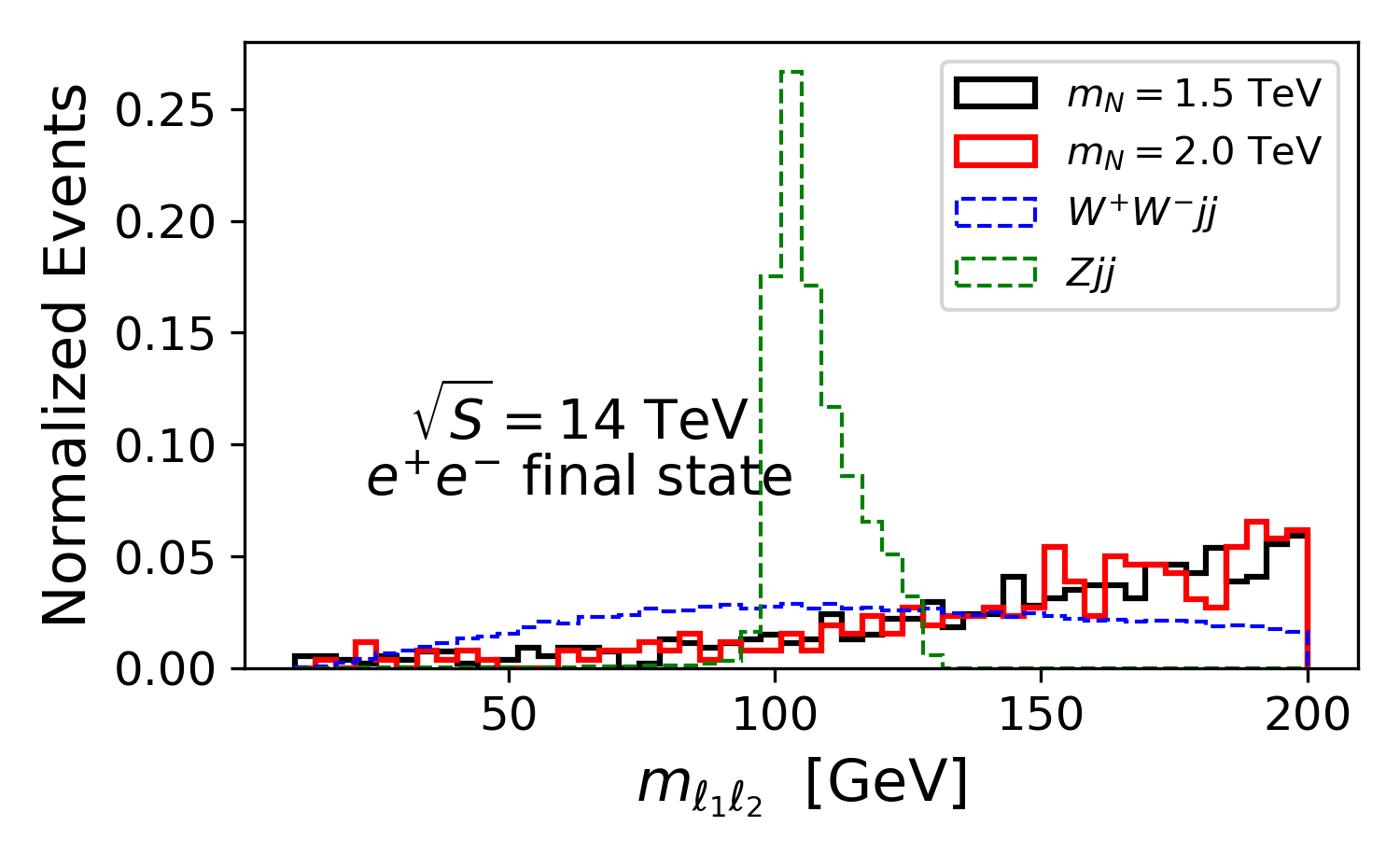"}\\
        \includegraphics[width=0.48\linewidth]{"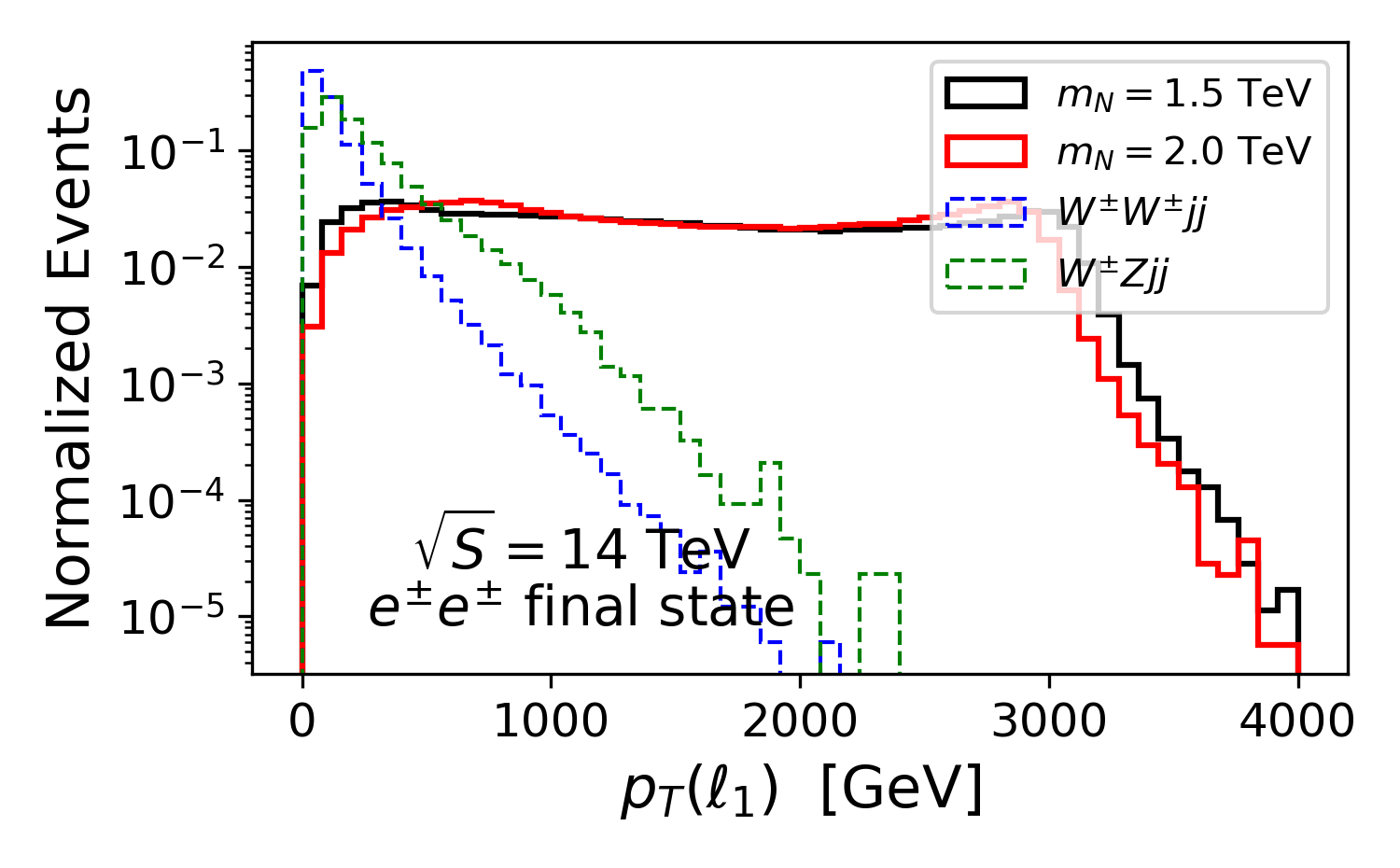"}
	\includegraphics[width=0.48\linewidth]{"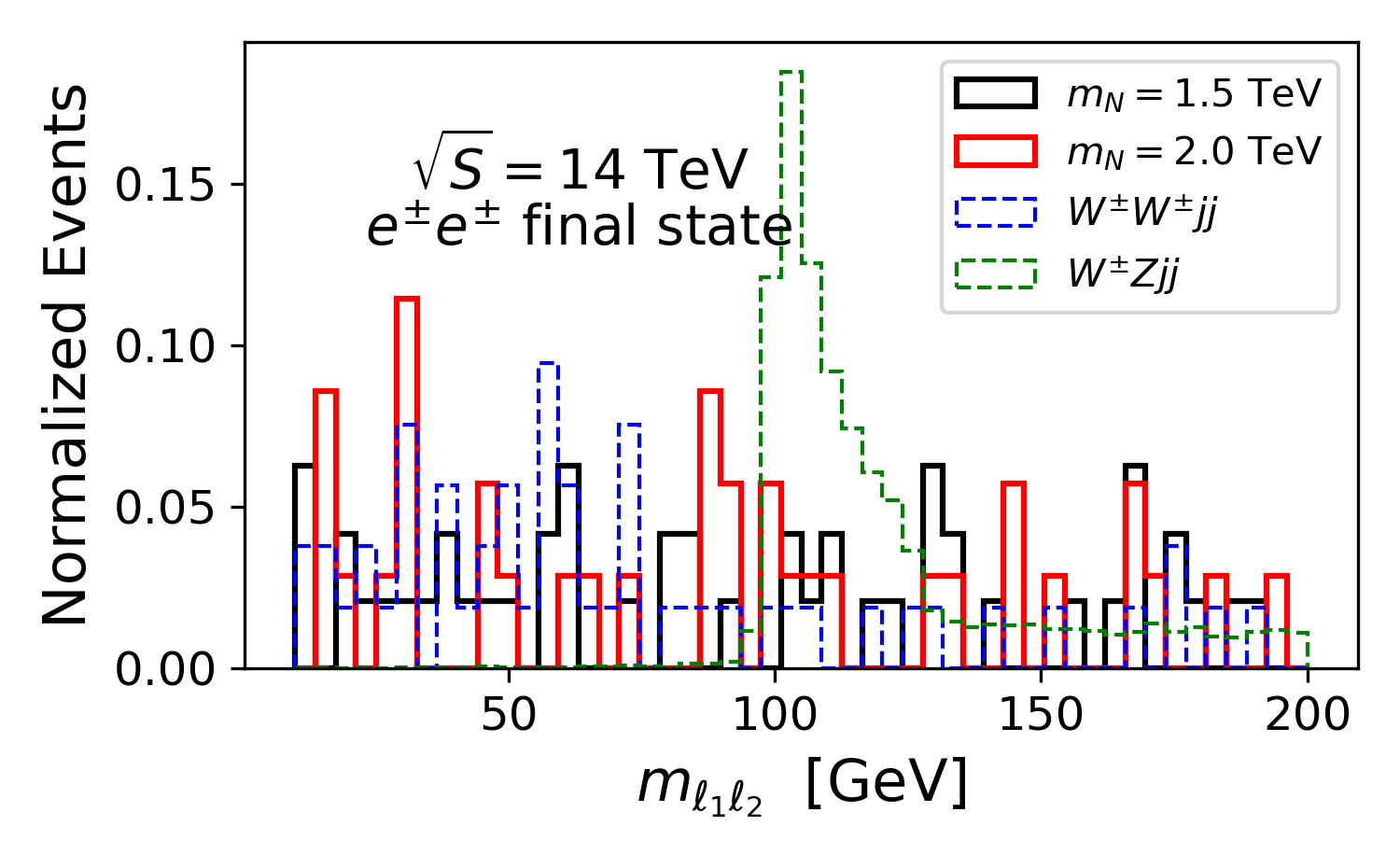"}
	\caption{The same as Fig.~\ref{fig:HTandMET}, but for the transverse momentum $p_T (\ell_1)$ of the leading lepton (left panels) and the invariant mass $m_{\ell_1 \ell_2}$ of the dilepton (right panels), for the $ee$ final state. 
 }
\label{fig:pte:mee}
\end{figure}

\begin{figure}[h!]
	\centering
	\includegraphics[width=0.48\linewidth]{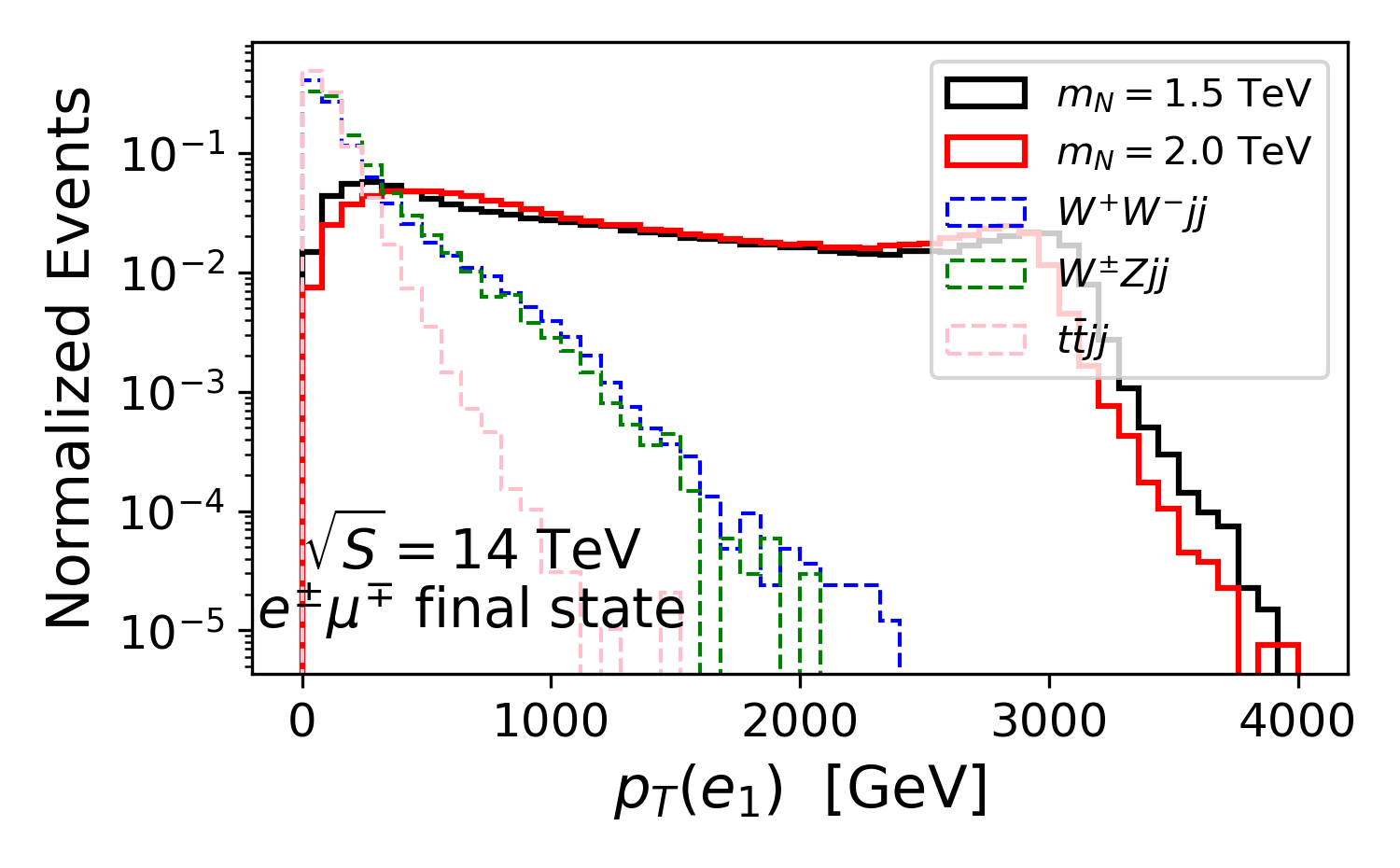}
	\includegraphics[width=0.48\linewidth]{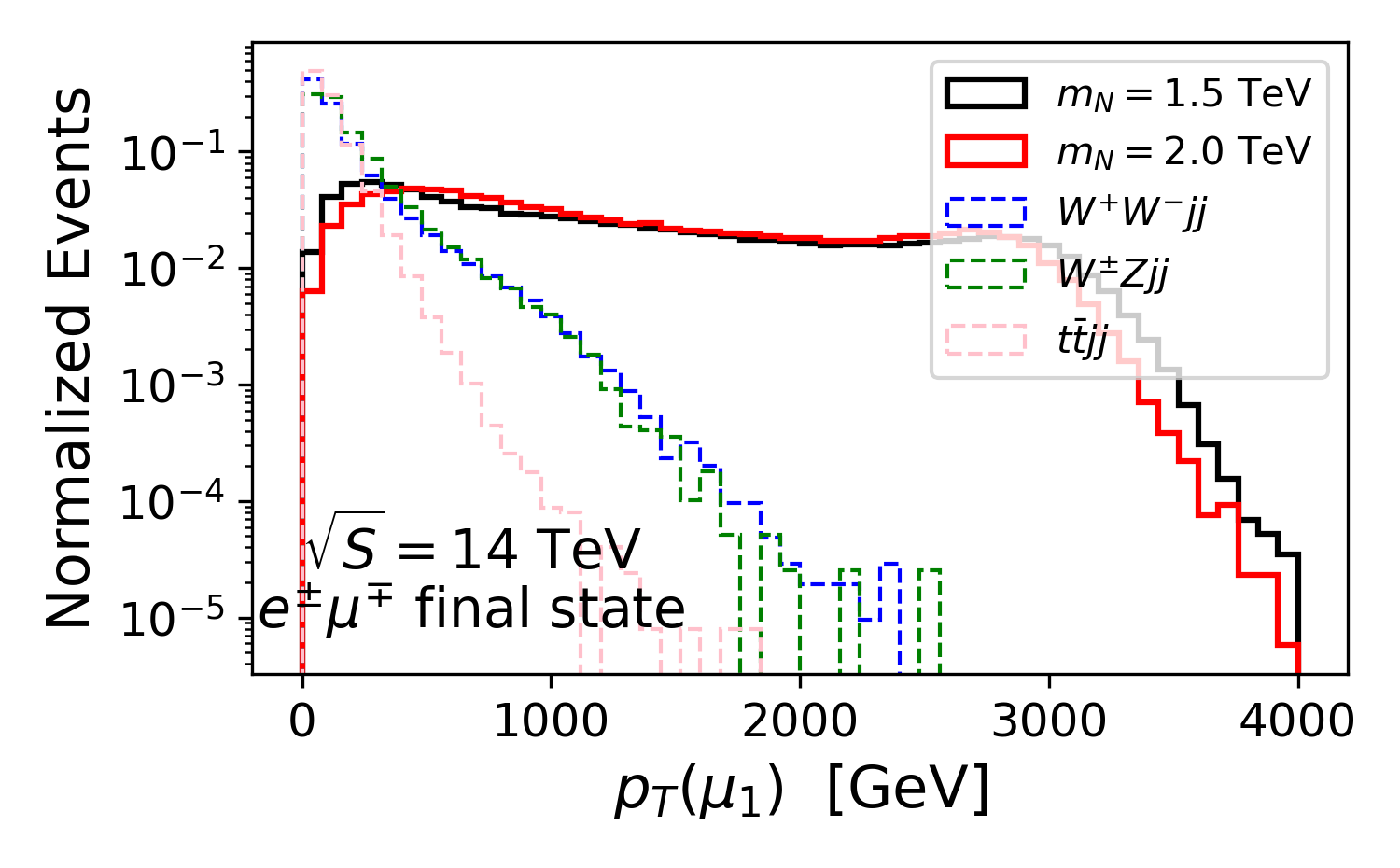}
	\includegraphics[width=0.48\linewidth]{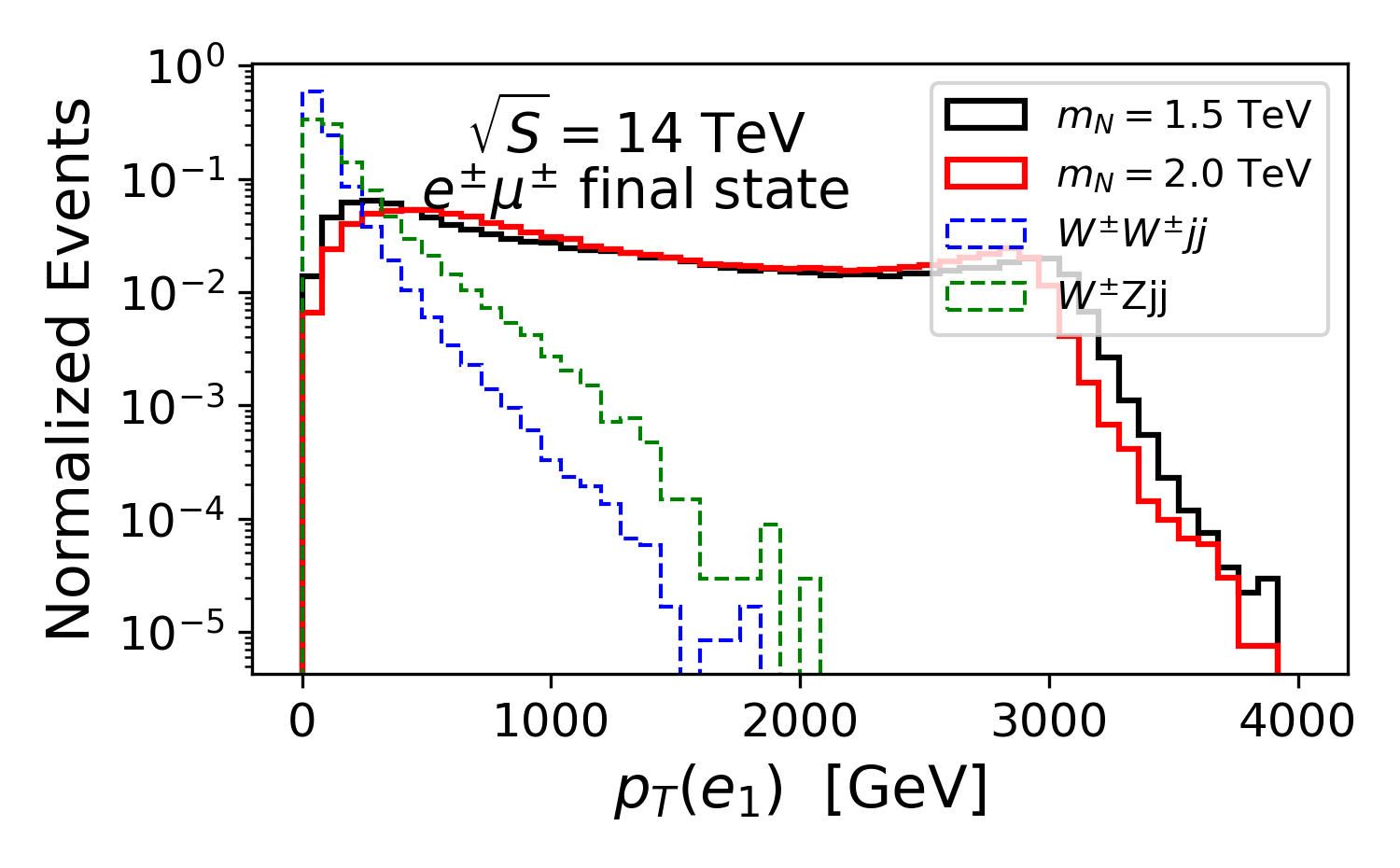}
	\includegraphics[width=0.48\linewidth]{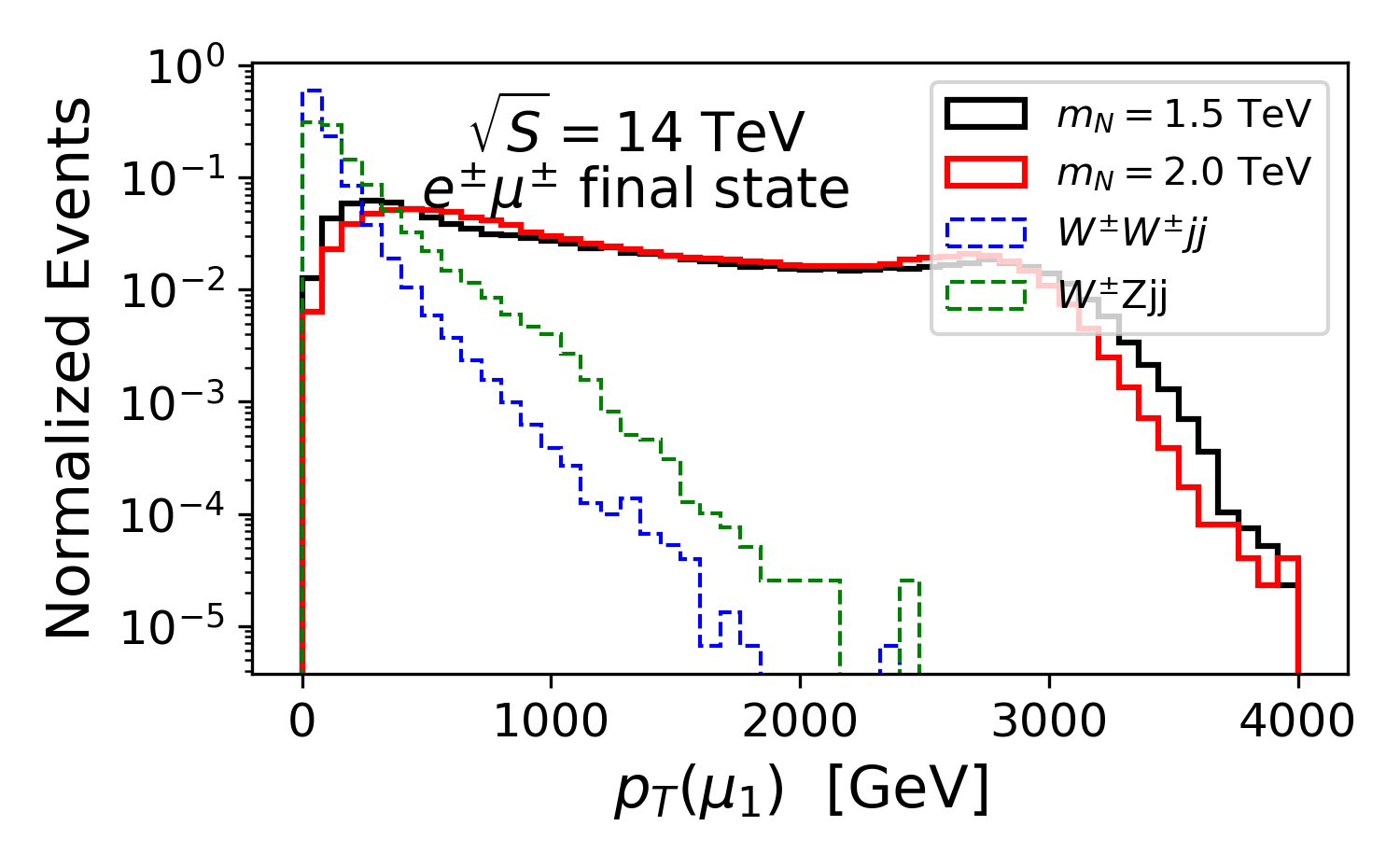}
	\caption{The same as Fig.~\ref{fig:HTandMET}, but for the transverse momenta of the leading electron $p_T (e_1)$ (left panels) and the leading muon $p_T (\mu_1)$ (right panels), for the $e\mu$ final state. 
    }
\label{fig:14os_emu_kin_2}
\end{figure}

\begin{figure}[t!]
	\centering
	\includegraphics[width=0.48\linewidth]{"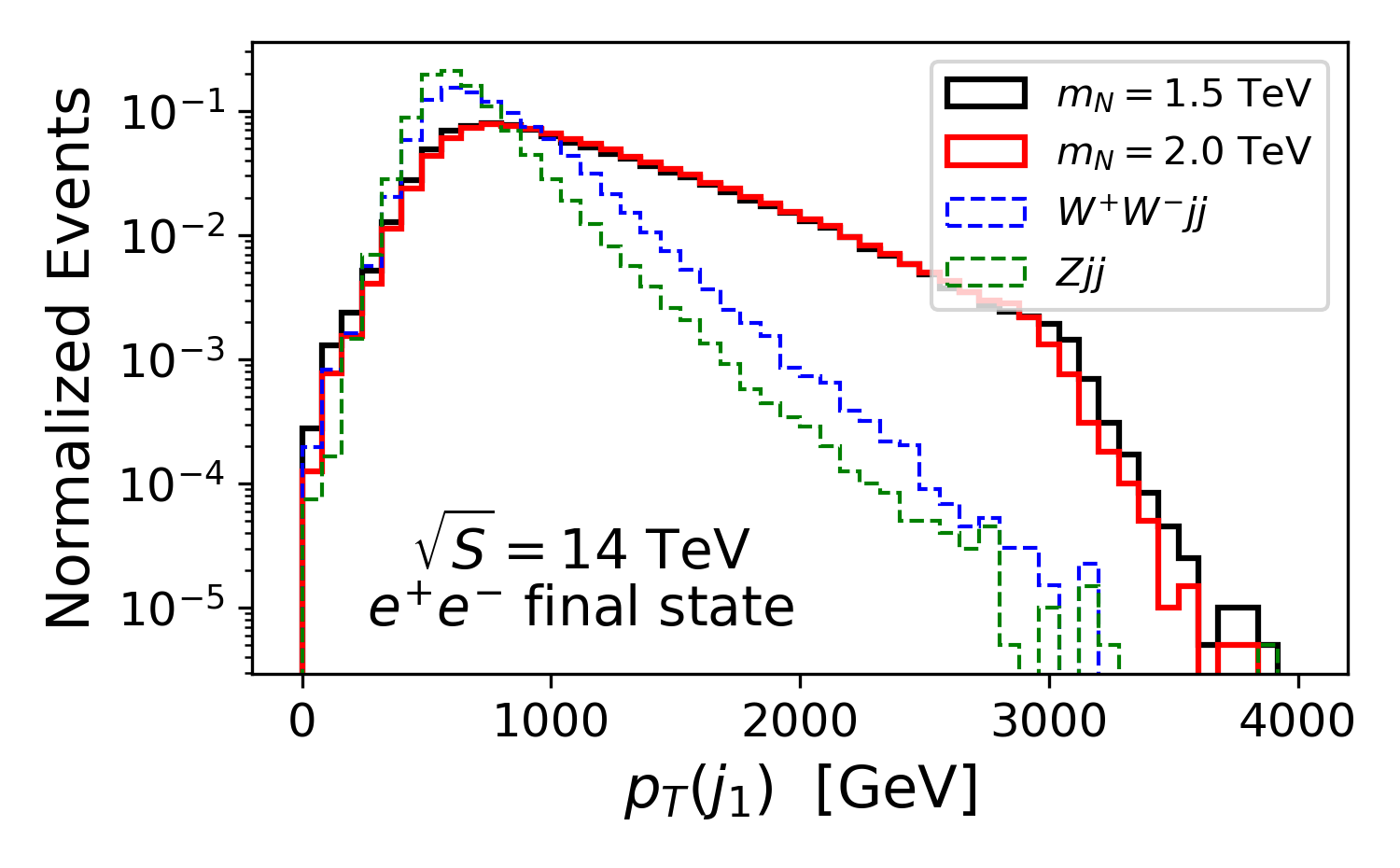"}
	\includegraphics[width=0.48\linewidth]{"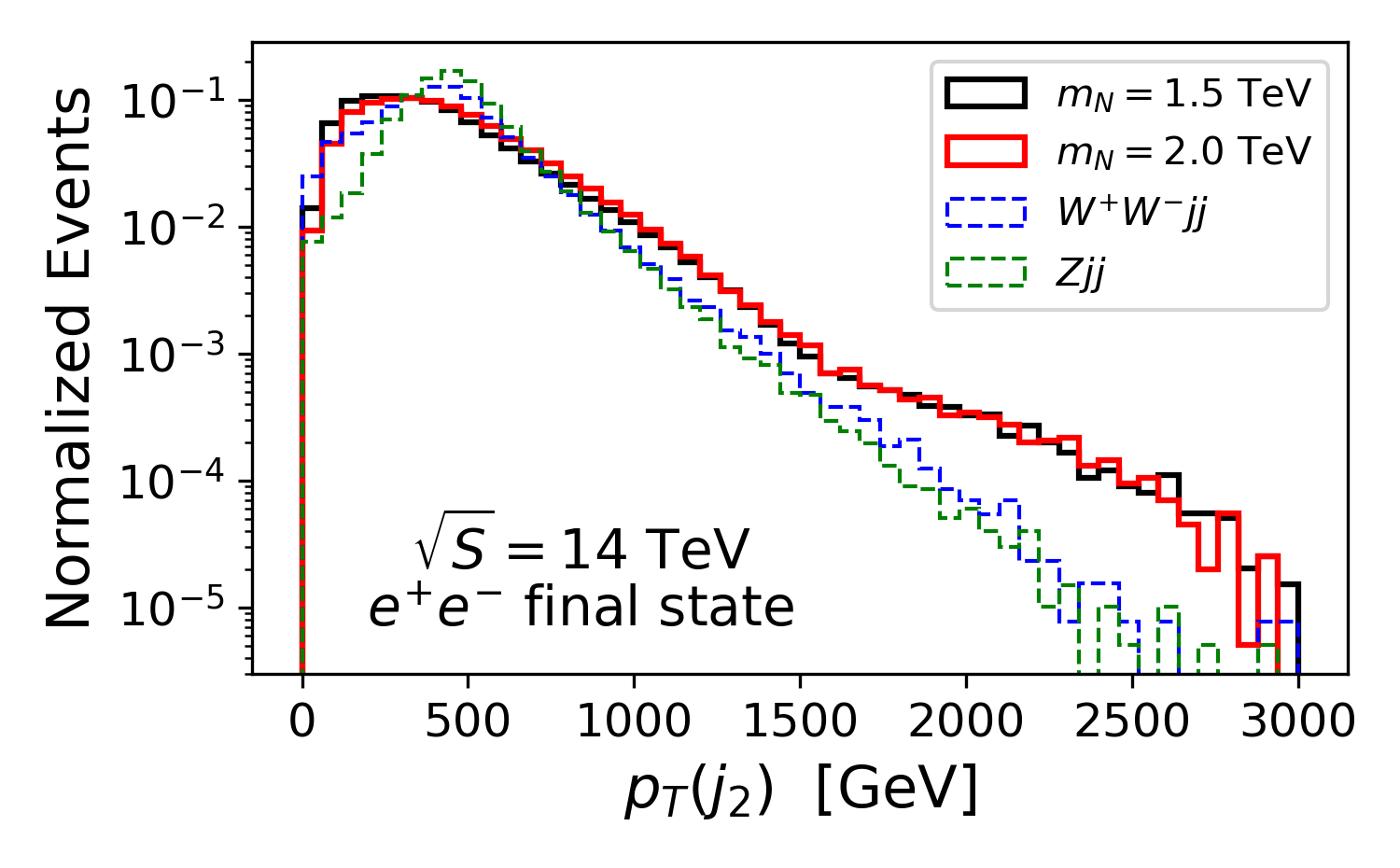"}\\
 \includegraphics[width=0.48\linewidth]{"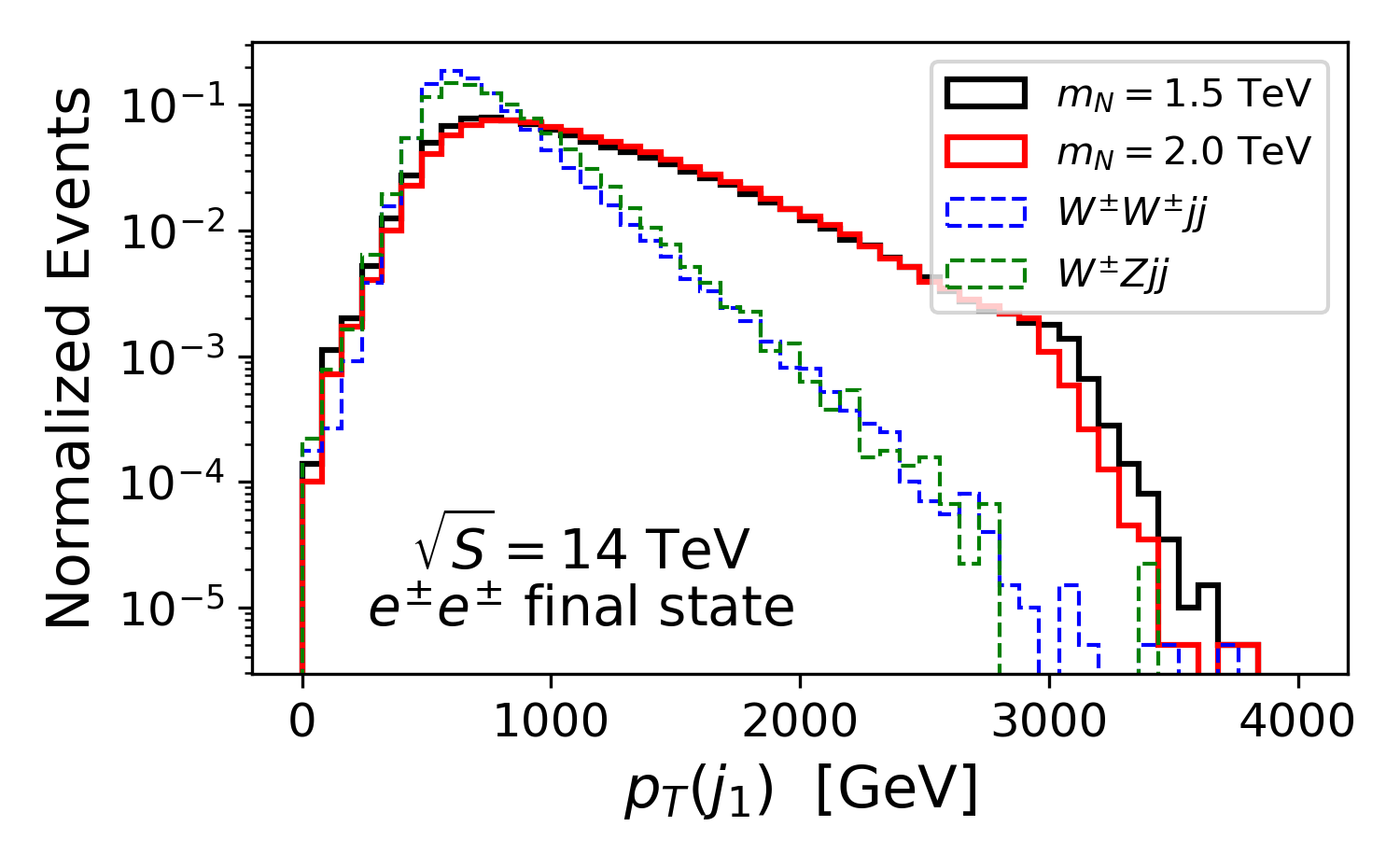"}
	\includegraphics[width=0.48\linewidth]{"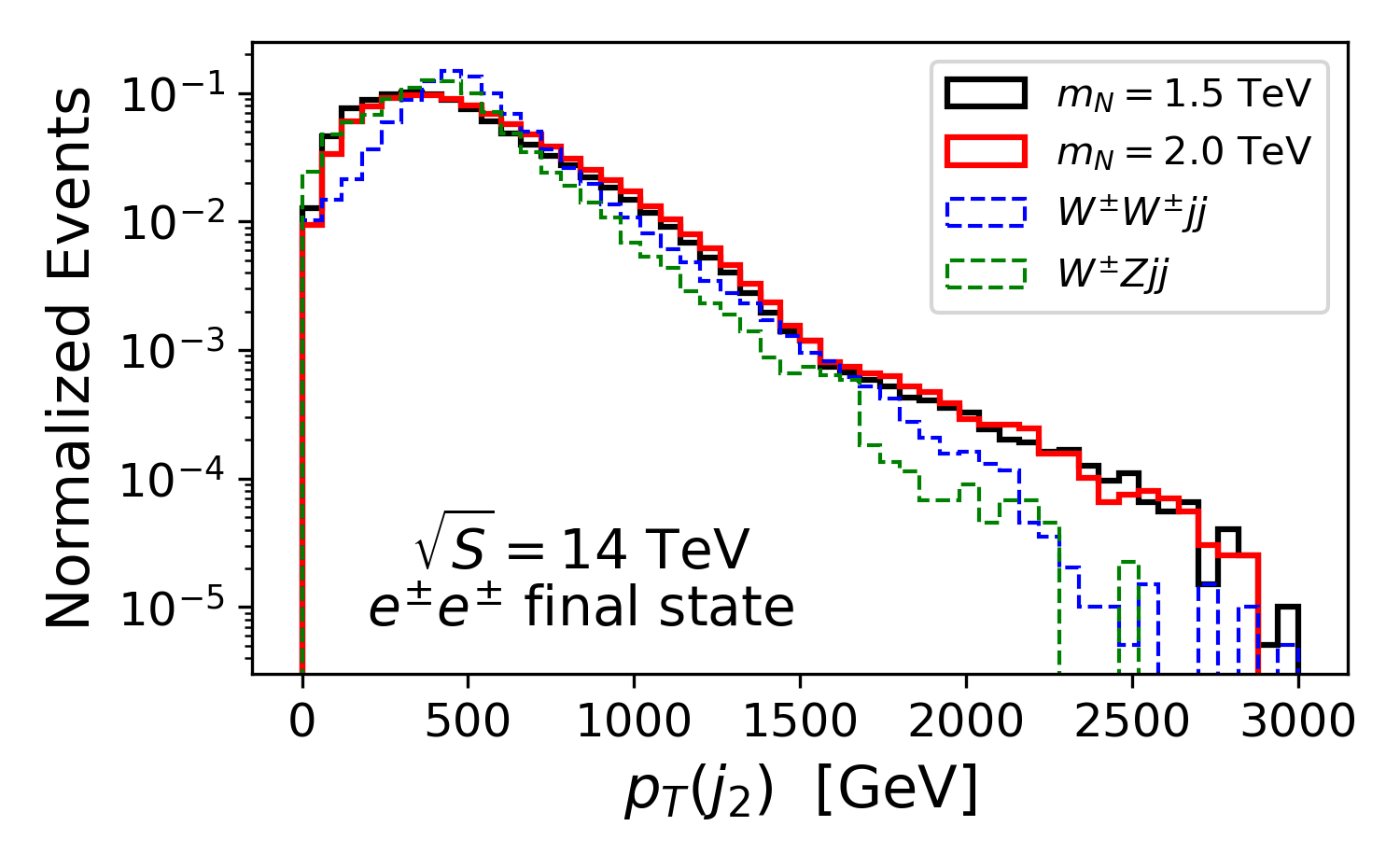"}
	\caption{The same as Fig.~\ref{fig:HTandMET}, but for the transverse momenta of the leading jet $p_T (j_1)$ (left panels) and the next-to-leading jet $p_T (j_2)$ (right panels), for the $ee$ final state.
 }
\label{fig:ptjet}
\end{figure}

\begin{figure}[h!]
	\centering
	\includegraphics[width=0.48\linewidth]{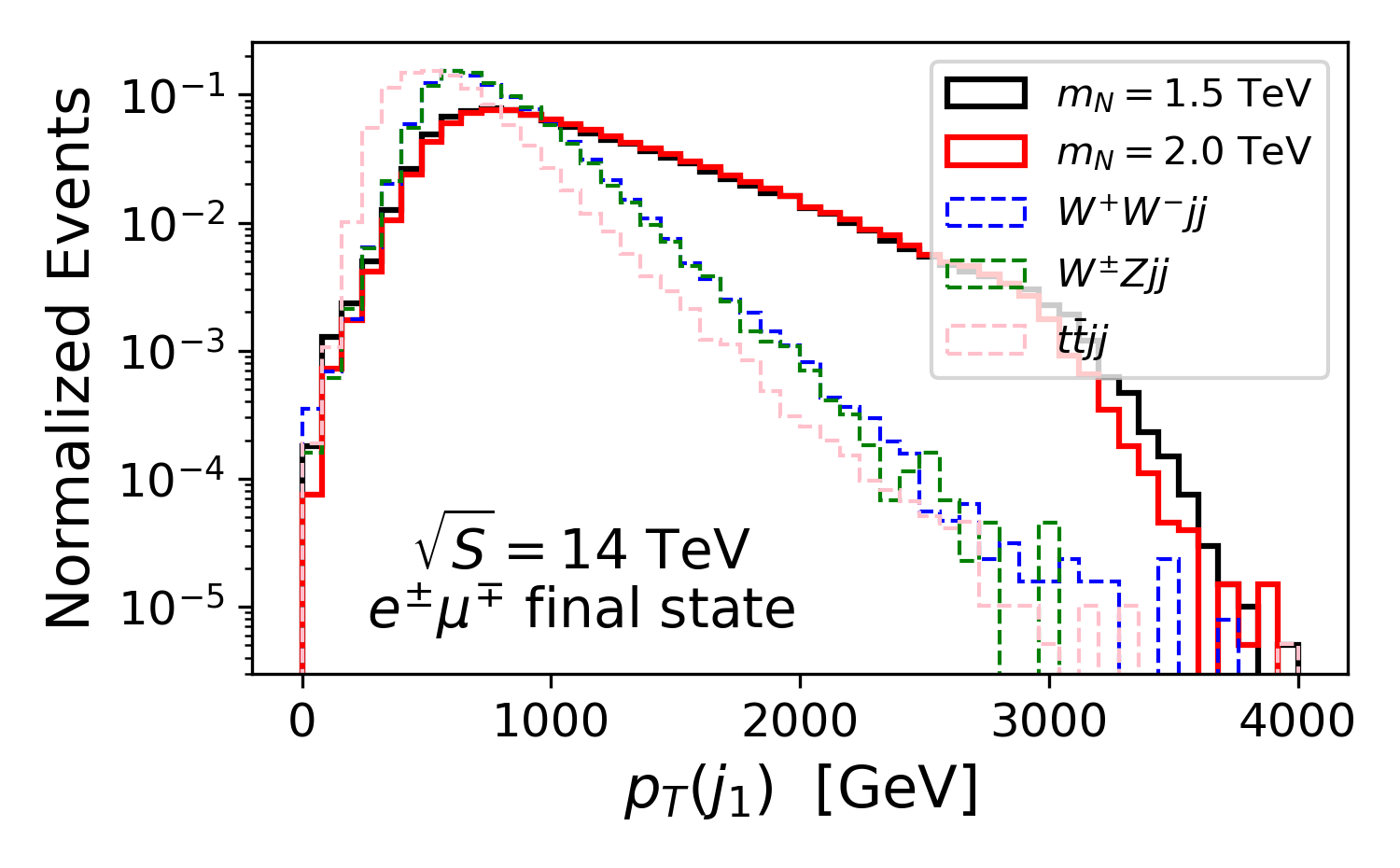}
	\includegraphics[width=0.48\linewidth]{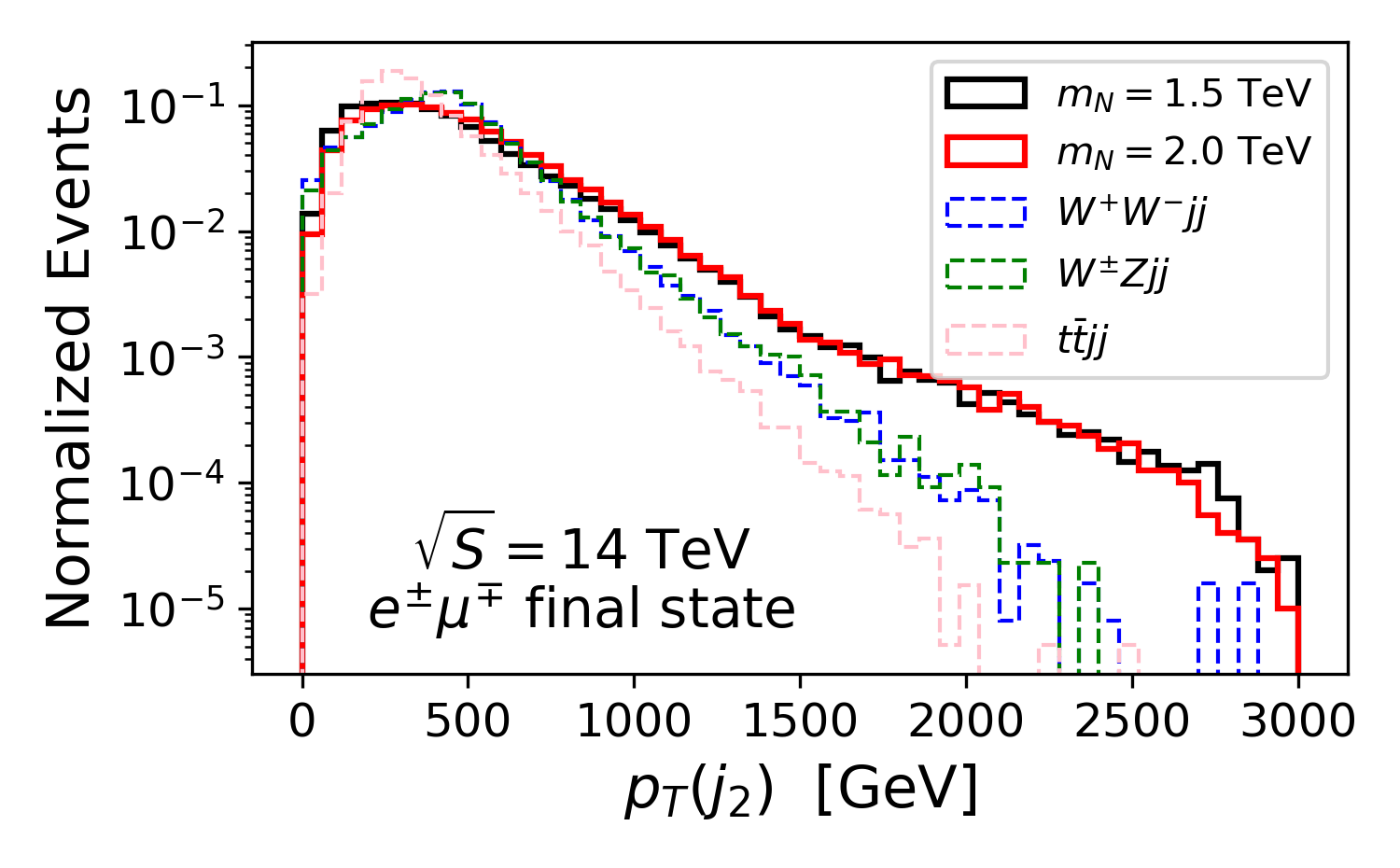}
	\includegraphics[width=0.48\linewidth]{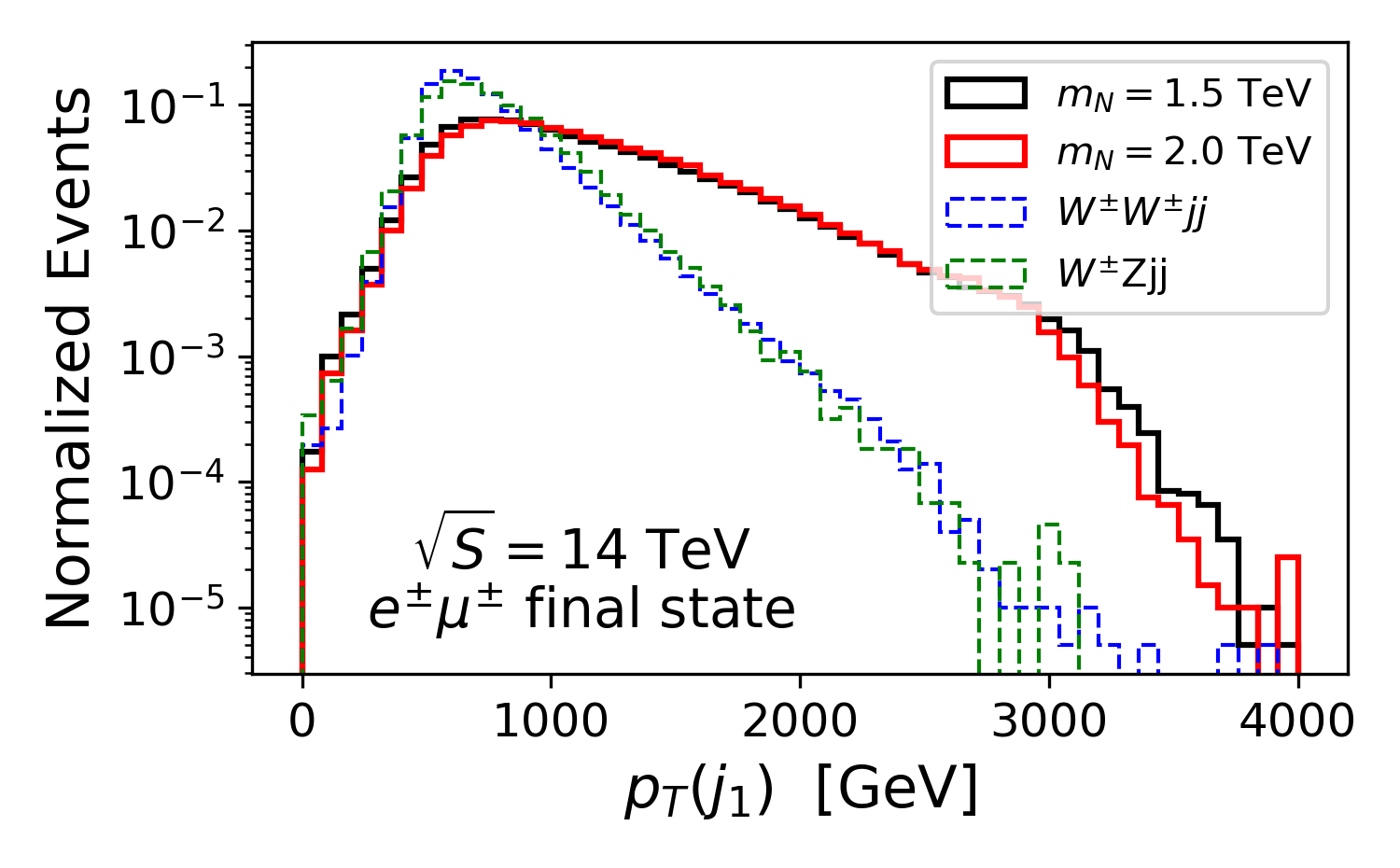}
	\includegraphics[width=0.48\linewidth]{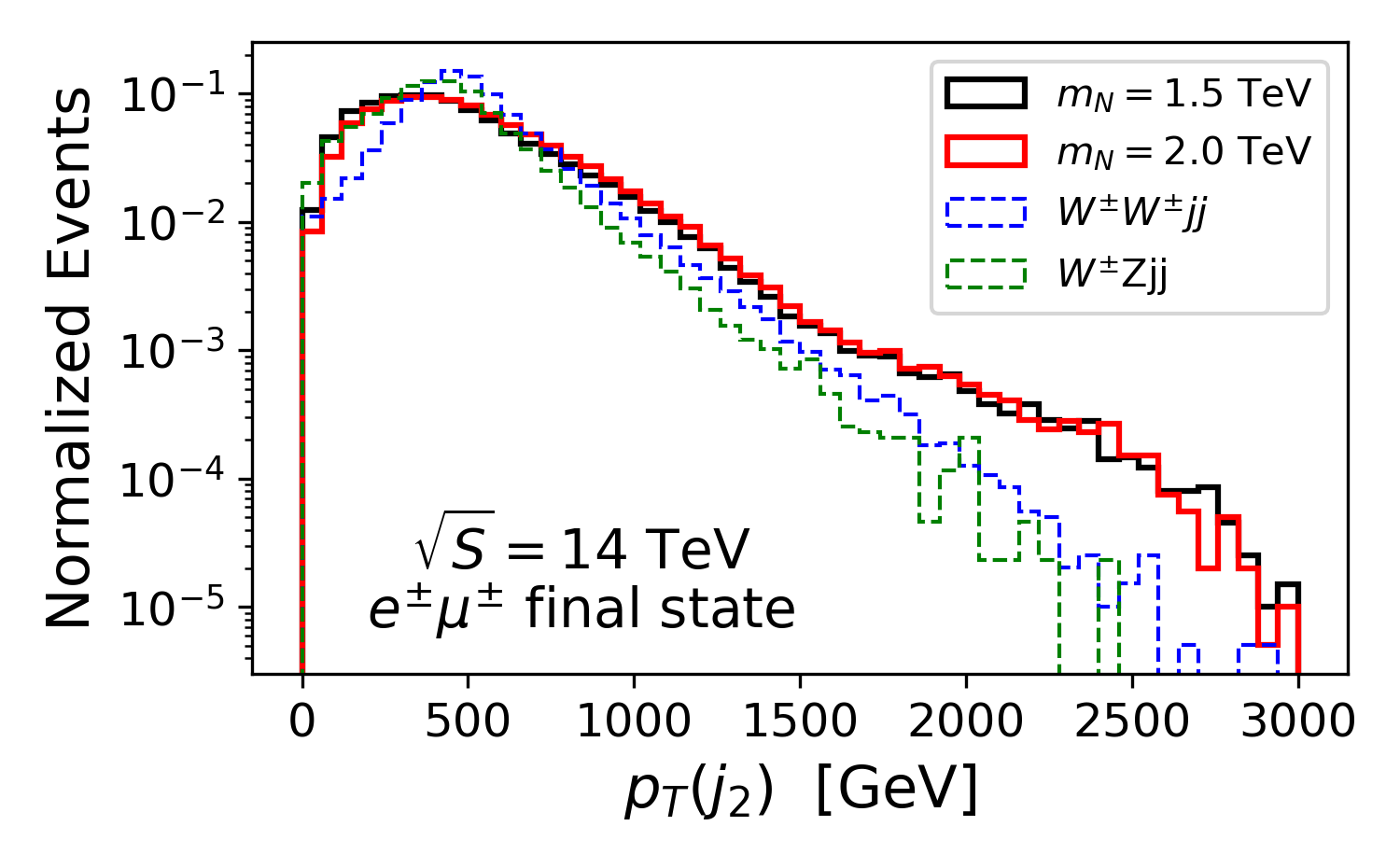}
	\caption{
    The same as Fig.~\ref{fig:HTandMET}, but for the transverse momenta of the leading jet $p_T (j_1)$ (left panels) and the next-to-leading jet $p_T (j_2)$ (right panels), for the $e\mu$ final state.
}
	\label{fig:14os_emu_kin_3}
\end{figure}

\begin{figure}[t!]
	\centering
	\includegraphics[width=0.48\linewidth]{"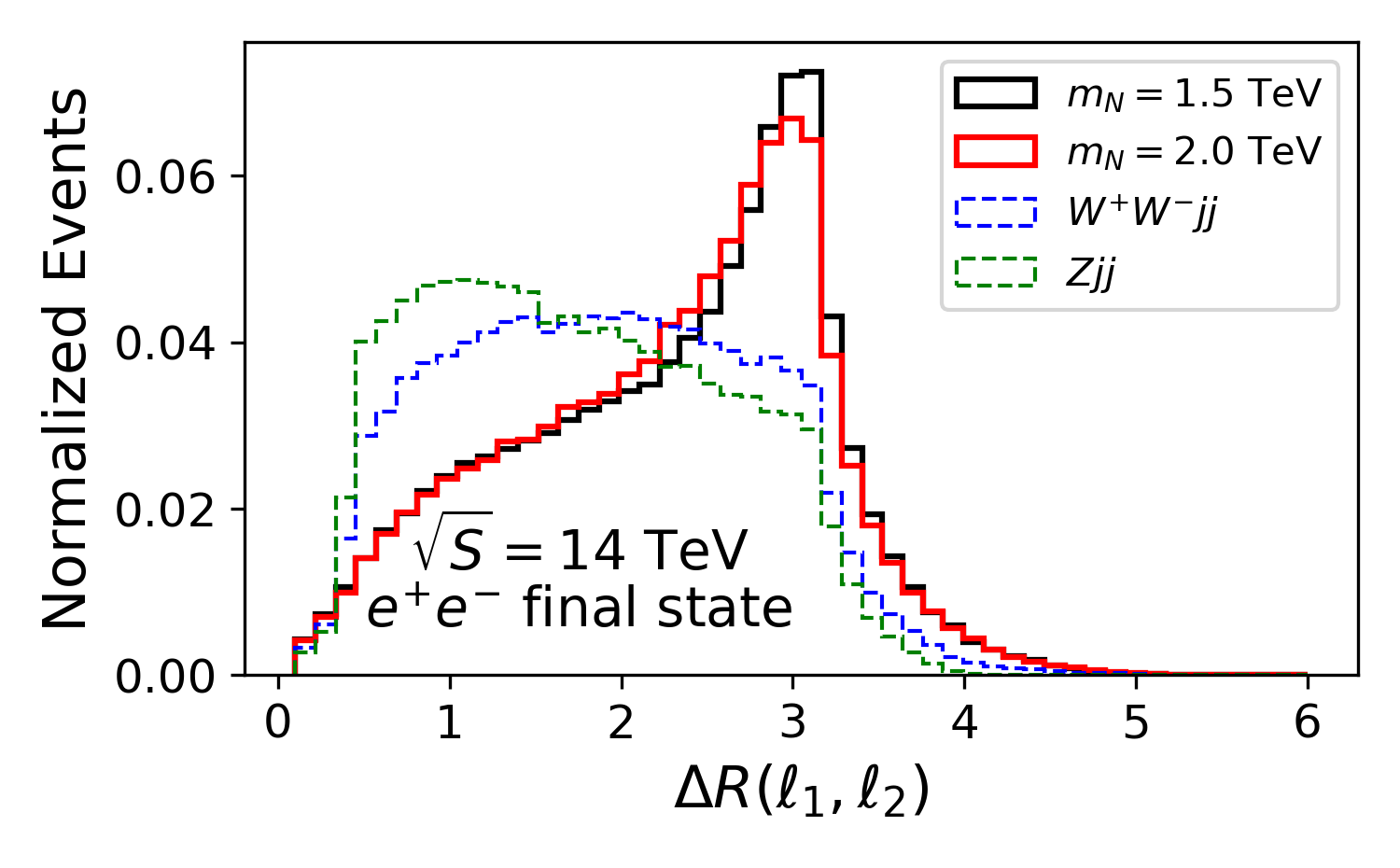"}
	\includegraphics[width=0.48\linewidth]{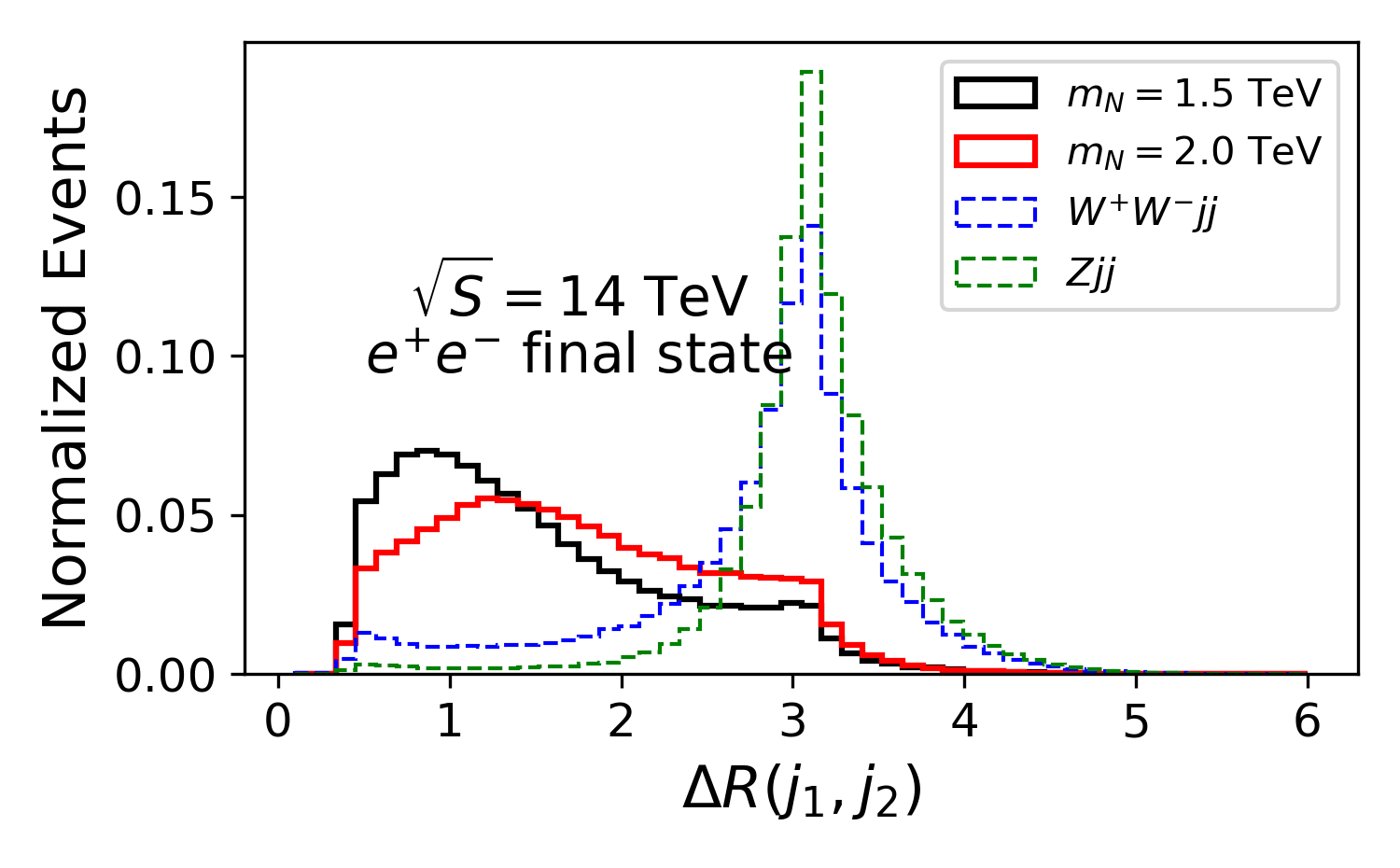}\\
 \includegraphics[width=0.48\linewidth]{"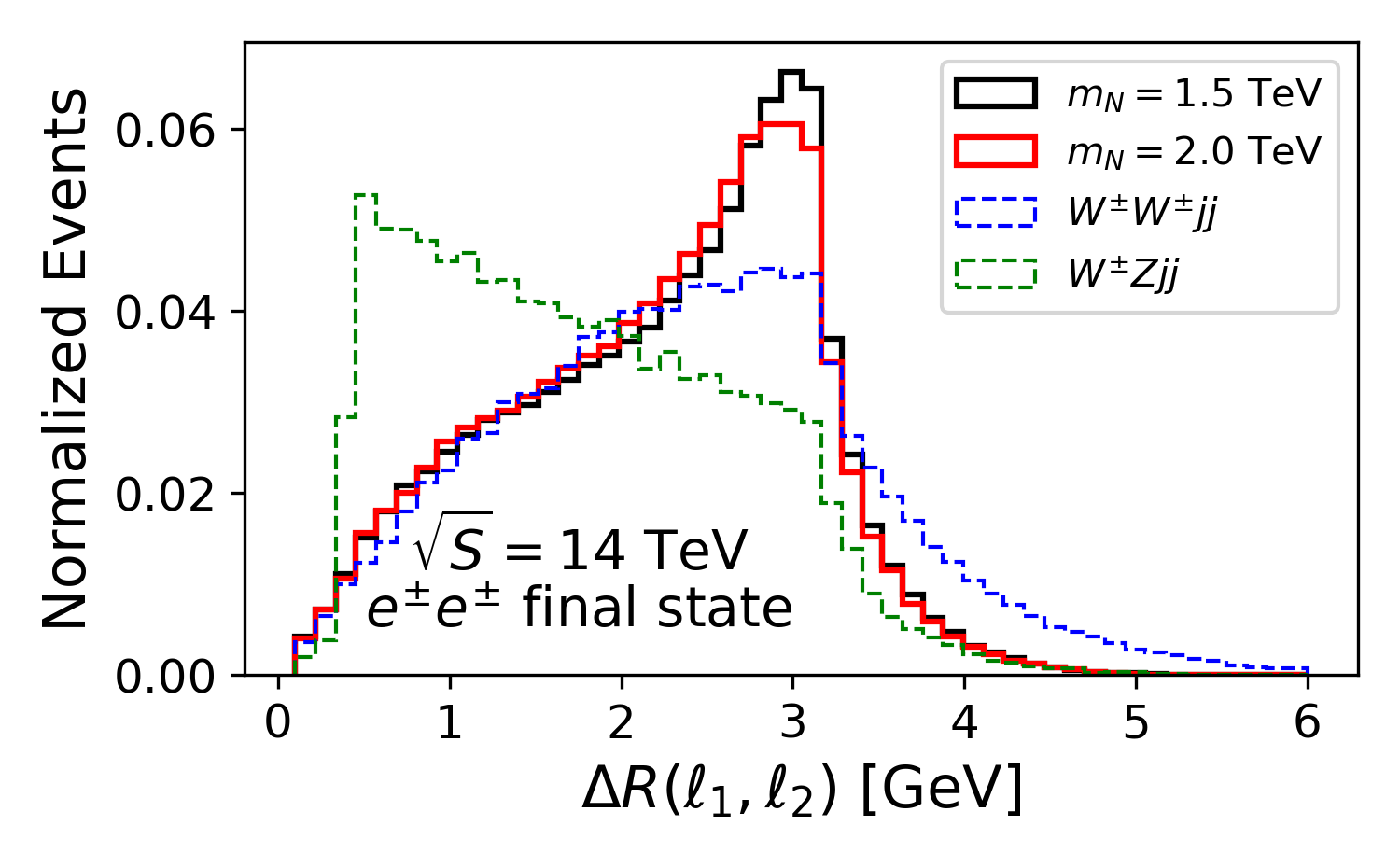"}
	\includegraphics[width=0.48\linewidth]{"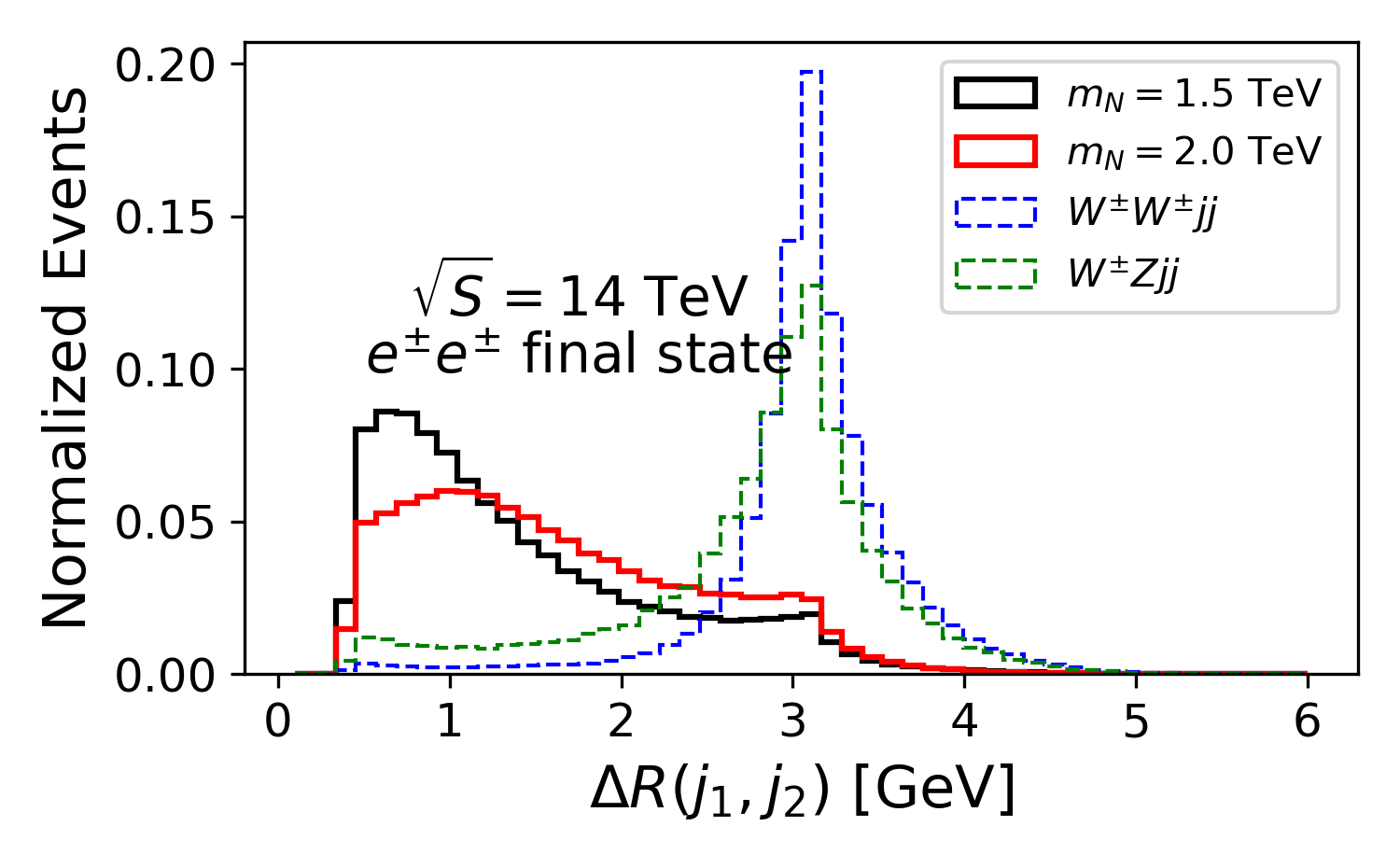"}
	\caption{The same as Fig.~\ref{fig:HTandMET}, but for the angles between the two leptons $\Delta R (\ell_1,\,\ell_2)$ (left panels) and the two jets $\Delta R (j_1,\,j_2)$ (right panels), for the $ee$ final state.
 }
\label{fig:delta}
\end{figure}

\begin{figure}[t!]
	\centering
	\includegraphics[width=0.48\linewidth]{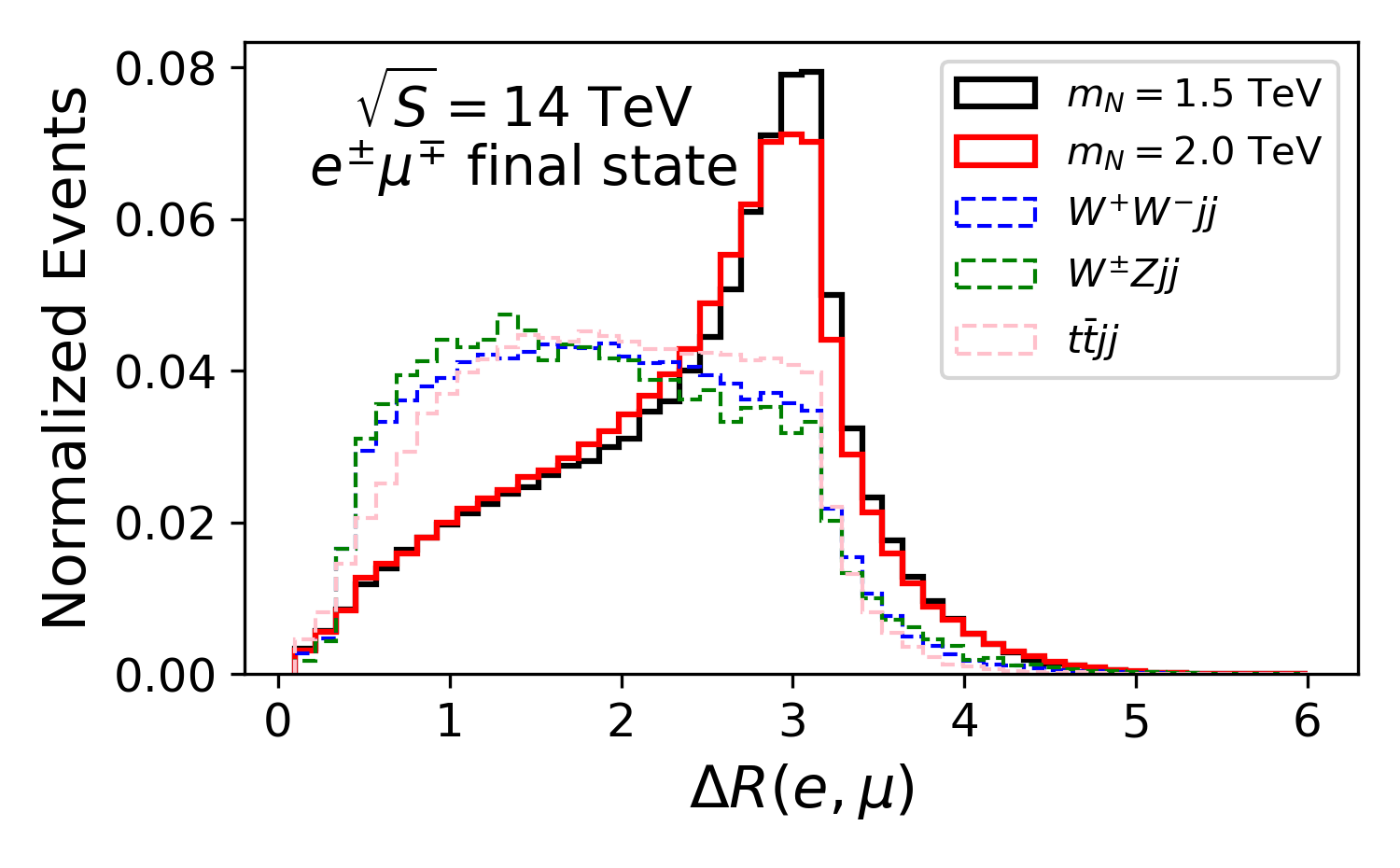}
	\includegraphics[width=0.48\linewidth]{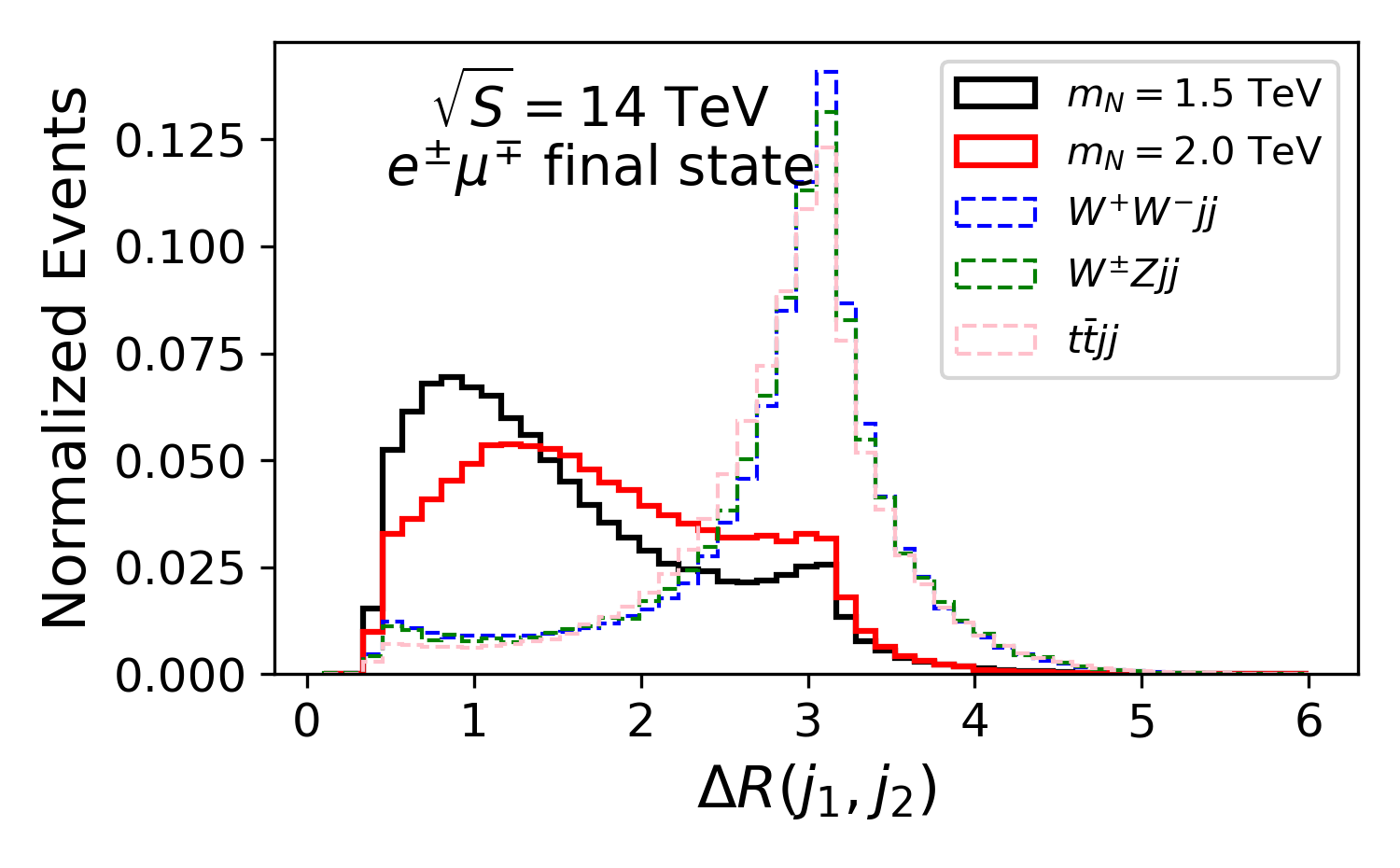}\\
	\includegraphics[width=0.48\linewidth]{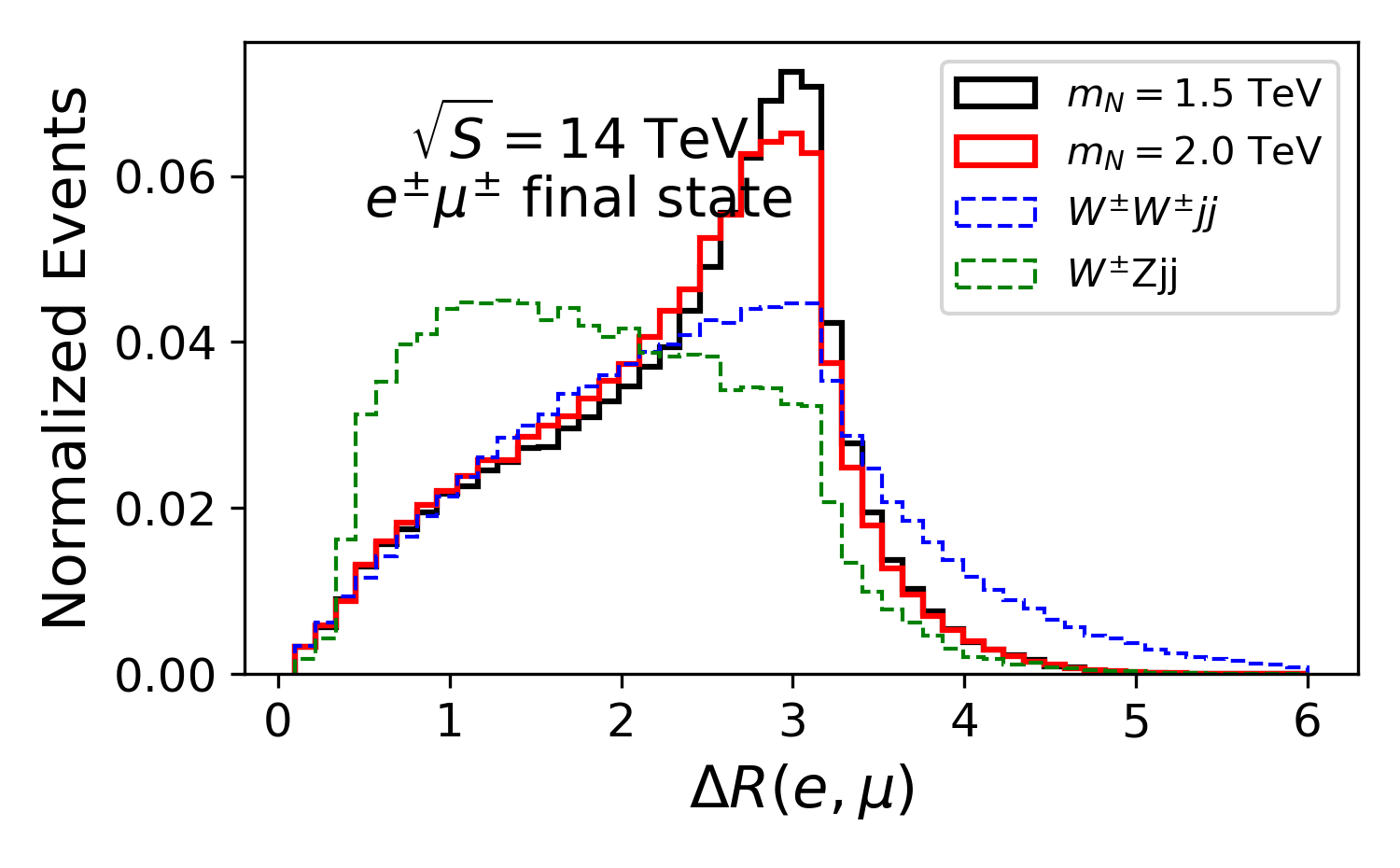}
	\includegraphics[width=0.48\linewidth]{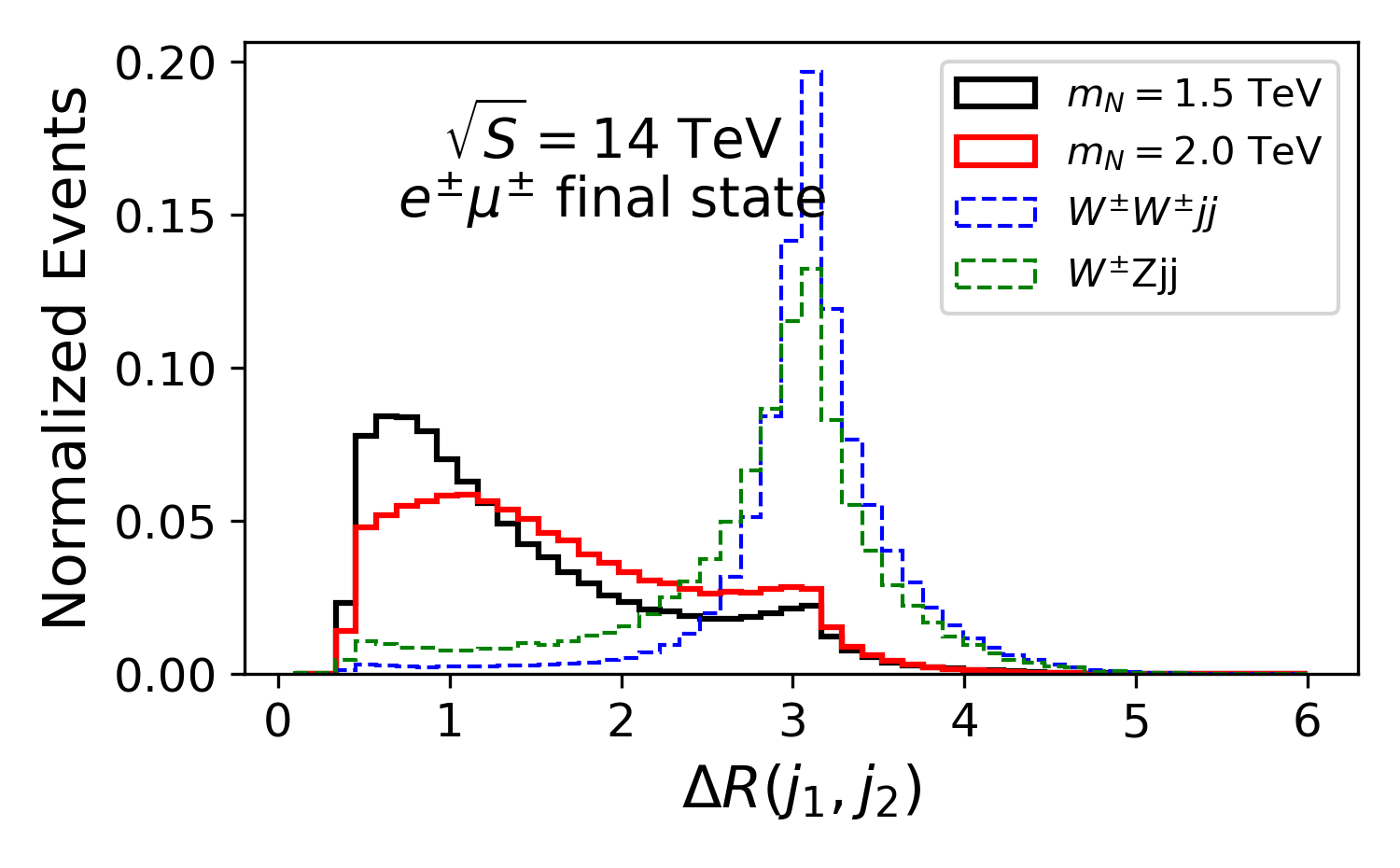}
	\caption{
    The same as Fig.~\ref{fig:HTandMET}, but for the angles between the two leptons $\Delta R (\ell_1,\,\ell_2)$ (left panels) and the two jets $\Delta R (j_1,\,j_2)$ (right panels), for the $e\mu$ final state.
    }
\label{fig:14os_emu_kin_4}
\end{figure}

Based on these pre-selection cuts, we are now ready to compare the distributions of some important observables for the signals and the corresponding SM backgrounds at the high-energy hadron colliders. 
\begin{itemize}
    \item The scalar sum $H_{T}$ of the transverse momenta of jets and charged leptons is a crucial variable for our analysis, which is closely linked to the $W_R$ mass $m_{W_{R}}$ for the signal events. 
    In our simulations $H_{T}$ typically peaks around $m_{W_R}$ for signals, a range uncommon in the SM events, thus making it an effective discriminator between signals and backgrounds. The distributions of $H_{T}$ for the OS $e^+ e^-$ and SS $e^\pm e^\pm$ signals and the corresponding backgrounds at $\sqrt{s}=14$ TeV are shown in the top left and bottom left panels of Fig.~\ref{fig:HTandMET}, respectively. The signal and backgrounds are depicted as the solid and dashed lines, respectively. For the signals, we take have taken the heavy neutrino mass $m_{N} = 1.5$ TeV (black) or 2 TeV (red), and  the $W_R$ mass $m_{W_R} = 6.5$ TeV. For concreteness the heavy neutrino mass splitting is set to be $10^{-7}$ GeV. But this splitting is too small to have any effects on the differential cross sections here. As a comparison, the distributions of the missing transverse momentum $E_{T}^{\rm miss}$ are illustrated in the right panel of Fig.~\ref{fig:HTandMET}, with the top and bottom panels for the OS and SS signals, respectively. Apparently, the spectra of $E_T^{\rm miss}$ of SM backgrounds and signals are rather similar, which thus has much less discriminating power than $H_T$. The distributions of $H_T$ and $E_T^{\rm miss}$ for the $\mu\mu$ signals are very similar to those for $ee$ in Fig.~\ref{fig:HTandMET}. The corresponding distributions for the $e\mu$ final state are also very similar, as shown in Fig.~\ref{fig:14os_emu_kin_1}.
    
    \item The distributions of the transverse momentum $p_T (\ell_1 = e_1)$ of the leading charged lepton for the case of $ee$ signals are shown in the left panels of Fig.~\ref{fig:pte:mee}. 
    No matter the leading charged lepton originates directly from the $W_R$ boson or from the heavy neutrino $N$ (cf. Fig.~\ref{fig:diagram}), its transverse momentum is typically very large, 
    which is very different from the SM background processes. The distributions of the invariant mass $m_{\ell_1 \ell_2}$ of the leading dileptons are presented in the right panels of Fig.~\ref{fig:pte:mee}. Unlike the backgrounds, which display distinct peaks at a relatively low energy, the variable $m_{\ell_1 \ell_2}$ extends to a much higher energy for the signal processes. This feature of $m_{\ell_1 \ell_2}$ is particularly useful for suppressing the SM backgrounds, especially those involving the $Z$ boson. For the $e\mu$ signals, the distributions of transverse momenta of the leading electrons and muons are very similar to the $ee$ case, as seen in Fig.~\ref{fig:14os_emu_kin_2}.

    \item The distributions of the transverse momenta of the leading jet $p_T (j_1)$ and the next-to-leading jet $p_T (j_2)$ for the $ee$ signals and the corresponding SM backgrounds are shown in the left and right panels of Fig.~\ref{fig:ptjet}, respectively. It is clear that the jets in the signal processes tend to have a larger $p_T$ with respect to the  backgrounds, in particular for the leading jet $j_1$. The corresponding distributions of jets $j_{1,\,2}$ for the $e\mu$ signals are quite similar, as depicted in Fig.~\ref{fig:14os_emu_kin_3}.

    \item The distributions of the angular distances between the two charged leptons $\Delta R(\ell_{1},\ell_{2})$ and the two jets $\Delta R(j_{1},j_{2})$ are shown in the left and right panels of Fig.~\ref{fig:delta}, respectively. In the signal processes, the dilepton pair is from the heavy $W_R$ boson decay, and tends to have a back-to-back configuration in the limit of $m_{N} \ll m_{W_R}$, whereas the dijet is more aligned, showing a parallel distribution. These features of signal events are noticeably absent for most of background events, which could help to suppress efficiently the SM backgrounds. The distributions of $\Delta R$ for the $e\mu$ final state is also similar to that above, as shown explicitly in Fig.~\ref{fig:14os_emu_kin_4}.
\end{itemize}



\begin{figure}[h!]
	\centering
	\includegraphics[width=0.48\linewidth]{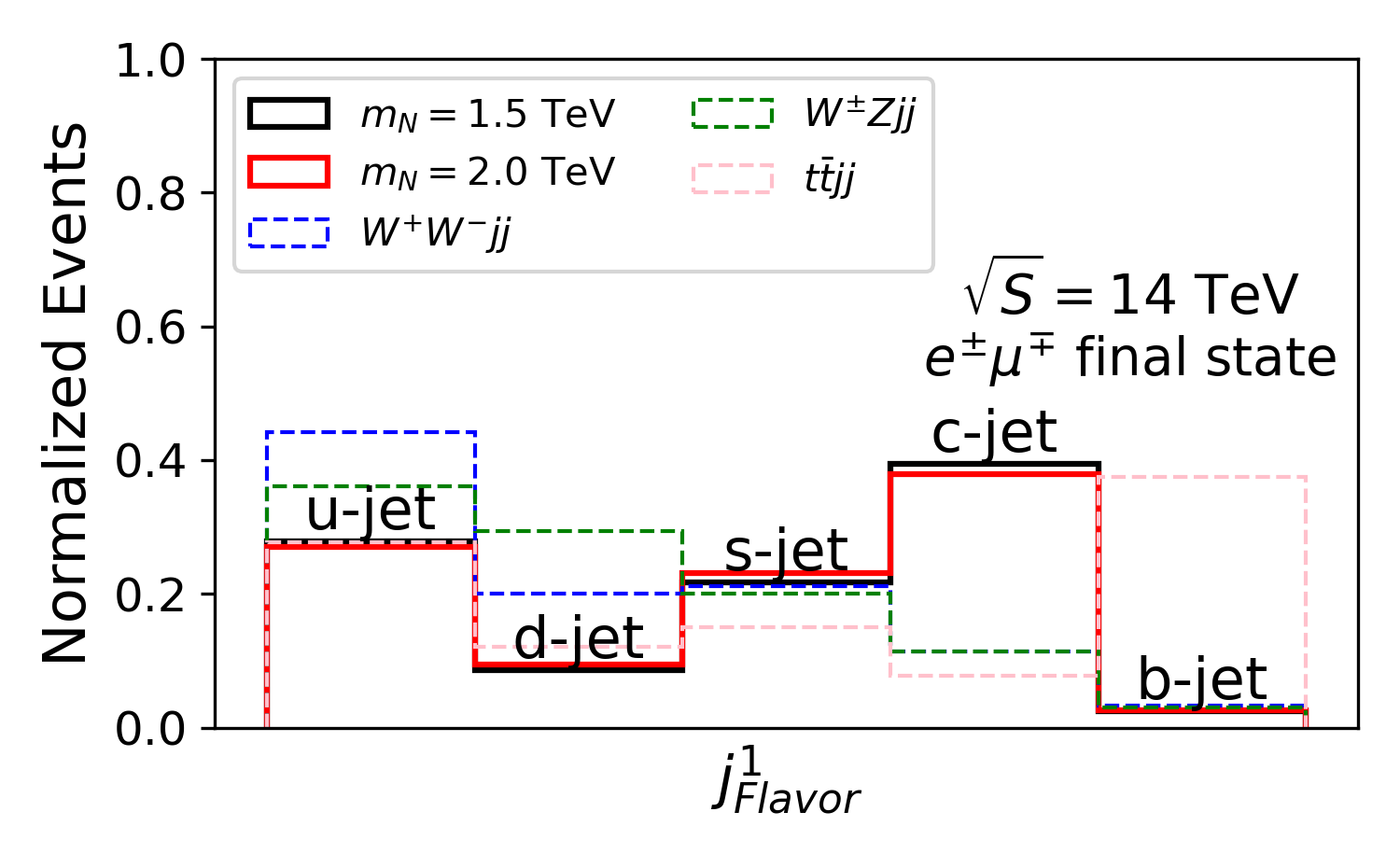}
	\includegraphics[width=0.48\linewidth]{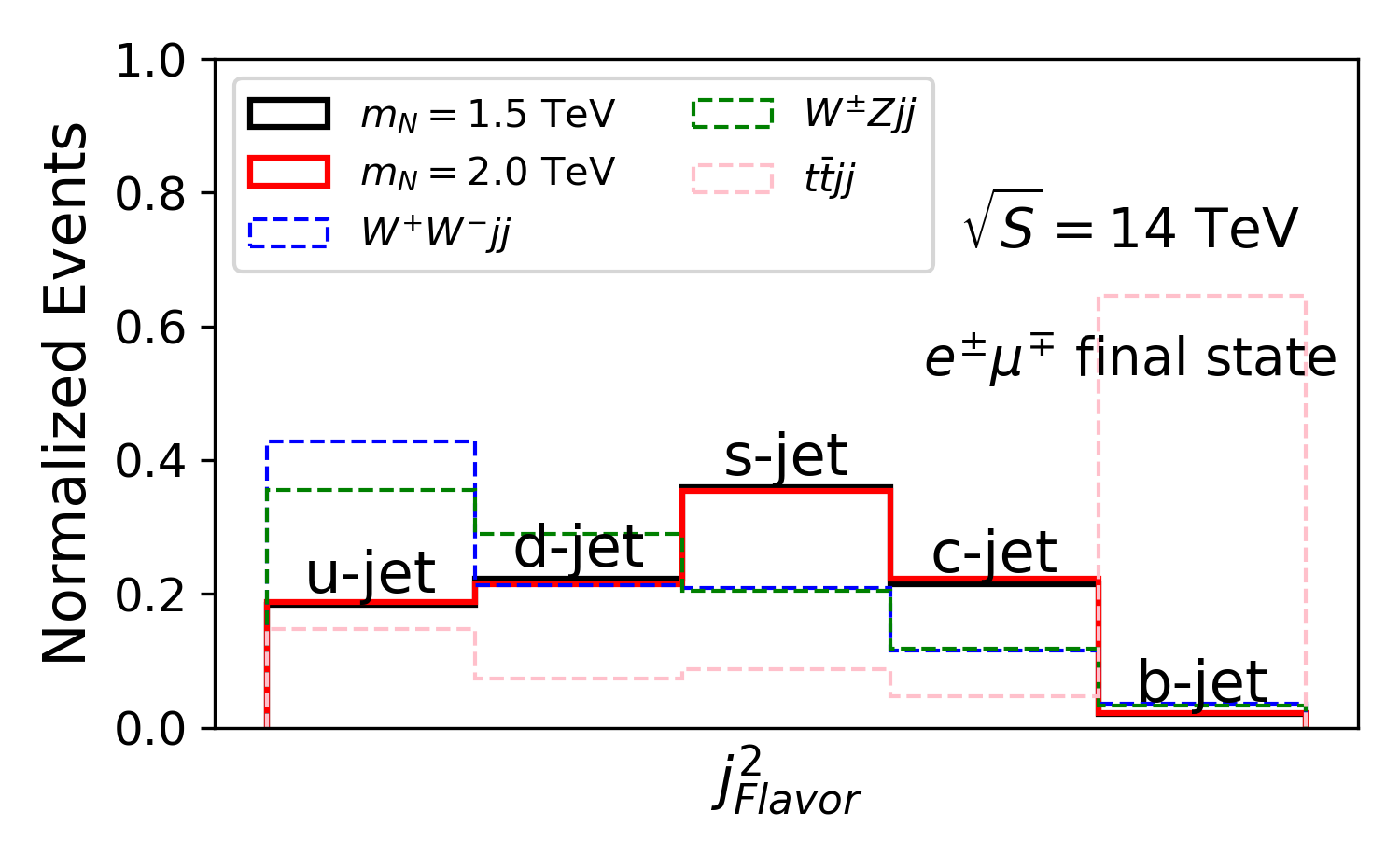}
\caption{The flavor distributions of the two leading jets $j_1$ and $j_2$ in the signal and background processes in the $e^\pm \mu^\mp$ final state. The notations are the same as in Fig.~\ref{fig:HTandMET}.
    }
	\label{fig:14os_emu_flavor}
\end{figure}

The key distinction between the $t\bar{t}$ background and the signal processes is the presence of two $b$-jets in the $t\bar{t}$ process. In the machine learning analysis, we will incorporate the jet flavor as an additional input during the training process. The normalized distributions of jet flavors in the $e^\pm \mu^\mp$ final state at $\sqrt{s}=14$ TeV are shown in Fig.~\ref{fig:14os_emu_flavor}. The left and right panels are for the leading jet $j_1$ and the next-to-leading jet $j_2$, respectively.

\begin{figure}[t!]
    \centering
    \includegraphics[width=0.48\linewidth]{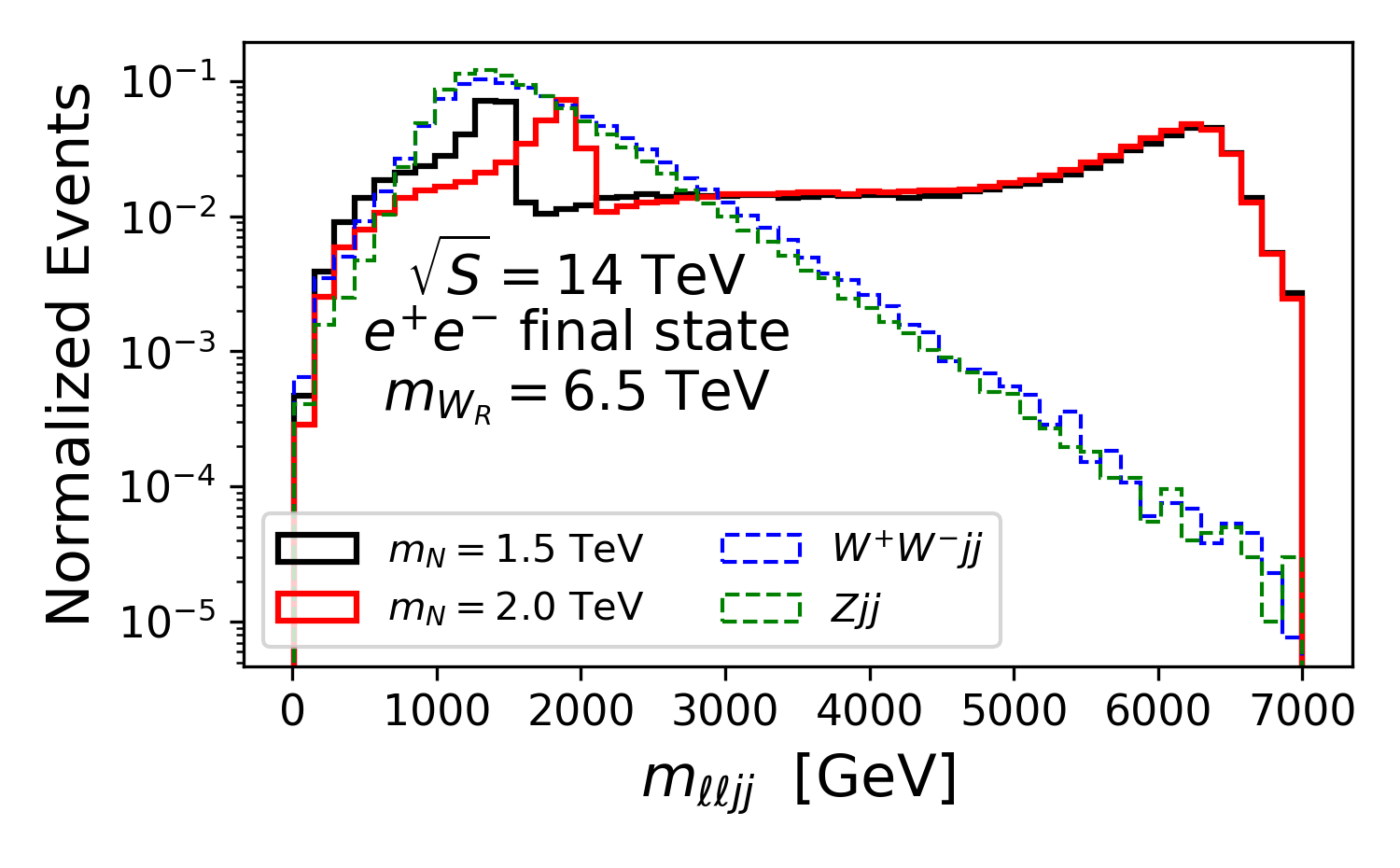}
    \caption{
    The same as the top left panel of Fig.~\ref{fig:HTandMET}, but for the invariant mass $m_{\ell\ell jj}$ of the dilepton plus two jets.
}
    \label{fig:mWR}
\end{figure}

The primary distinction between the signal processes and the SM backgrounds lies in the presence of a \(W_R\) boson in the signal, characterized by a mass significantly larger than those of the SM particles. Consequently, the \(W_R\) boson mass serves as a critical feature for distinguishing signals from backgrounds. It is expected that the invariant mass $m_{\ell \ell jj}$ of the dilepton pair and the two jets should be close to the $W_R$ mass for the signals, i.e. $m_{\ell \ell jj} \sim m_{W_R}$. The distributions of $m_{\ell \ell jj}$ for the OS $e^+ e^-$ signals and the corresponding backgrounds at $\sqrt{s} = 14$ TeV are shown in Fig.~\ref{fig:mWR}. We have taken again $m_{N} = 1.5$ TeV (black) or 2 TeV (red) with $m_{W_R} = 6.5$ TeV. It is clear that both the signal distributions have two peaks: the one at 6.5 TeV corresponds to the $W_R$ mass, while the other one is at the heavy neutrino mass of 1.5 TeV or 2 TeV.   
This dual-peak structure arises from the loss of kinematic information for certain electrons (positrons) in some events, preventing complete reconstruction of the \(W_R\) mass and leaving only the heavy neutrino mass reconstructable. Since the $m_{\ell \ell jj}$ and $H_T$ variables encapsulate nearly identical kinematic information, we have trained machine learning models separately, using each of these variables in combination with other kinematic distributions. As seen in Fig.~\ref{fig:mWR}, the signal peaks around the heavy neutrino mass overlap partially with the SM backgrounds. This makes the model trained with $m_{\ell \ell jj}$ perform approximately 10\% worse than the model with $H_T$. Therefore, the machine learning study in this paper will be based on the $H_T$ variable but not on $m_{\ell \ell jj}$, although the latter one is also very physically informative.

\section{ Machine learning analysis}
\label{ML}






Machine learning methods, particularly those capable of handling large-scale data such as Extreme Gradient Boosting (XGBoost), can significantly enhance the ability of distinguishing signals from backgrounds~\cite{Chen:2016btl}. In this work, we use XGBoost to perform the machine learning analysis, which is a robust machine learning algorithm that extends the Gradient Boosted Decision Trees (GBDT)~\cite{Friedman:2001wbq}. It combines gradient boosting with regularization techniques to improve both the predictive power and generalizability of the model, by optimizing the loss function through a second-order Taylor expansion that includes a regularization term. It uses information gain for making decisions on node splits and controls the learning rate, which contribute to its popularity as one of the most effective machine learning algorithms in active use today.

\subsection{Model training}

\begin{figure}[t!]
	\centering
	\includegraphics[width=0.49\linewidth]{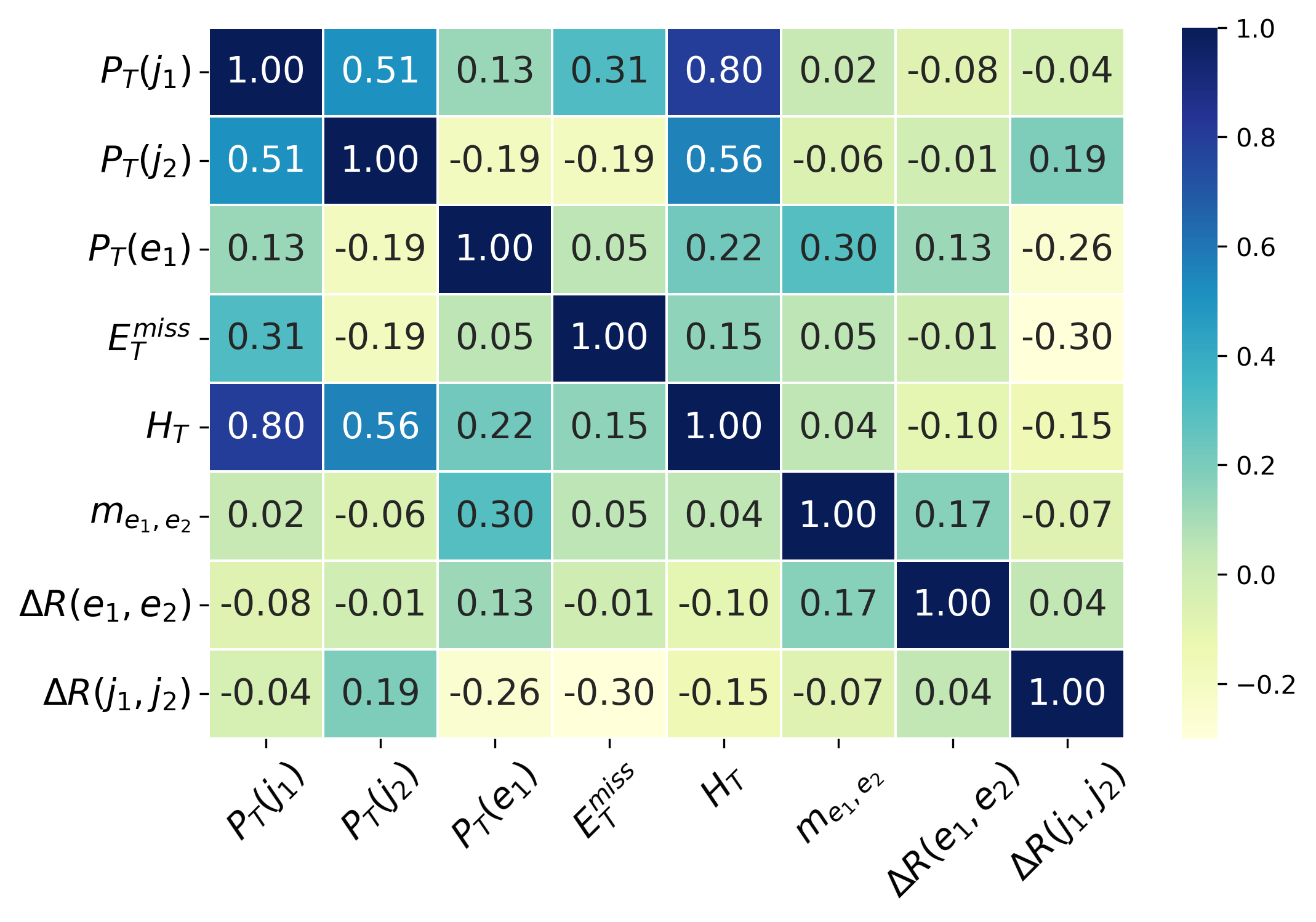}
	\includegraphics[width=0.49\linewidth]{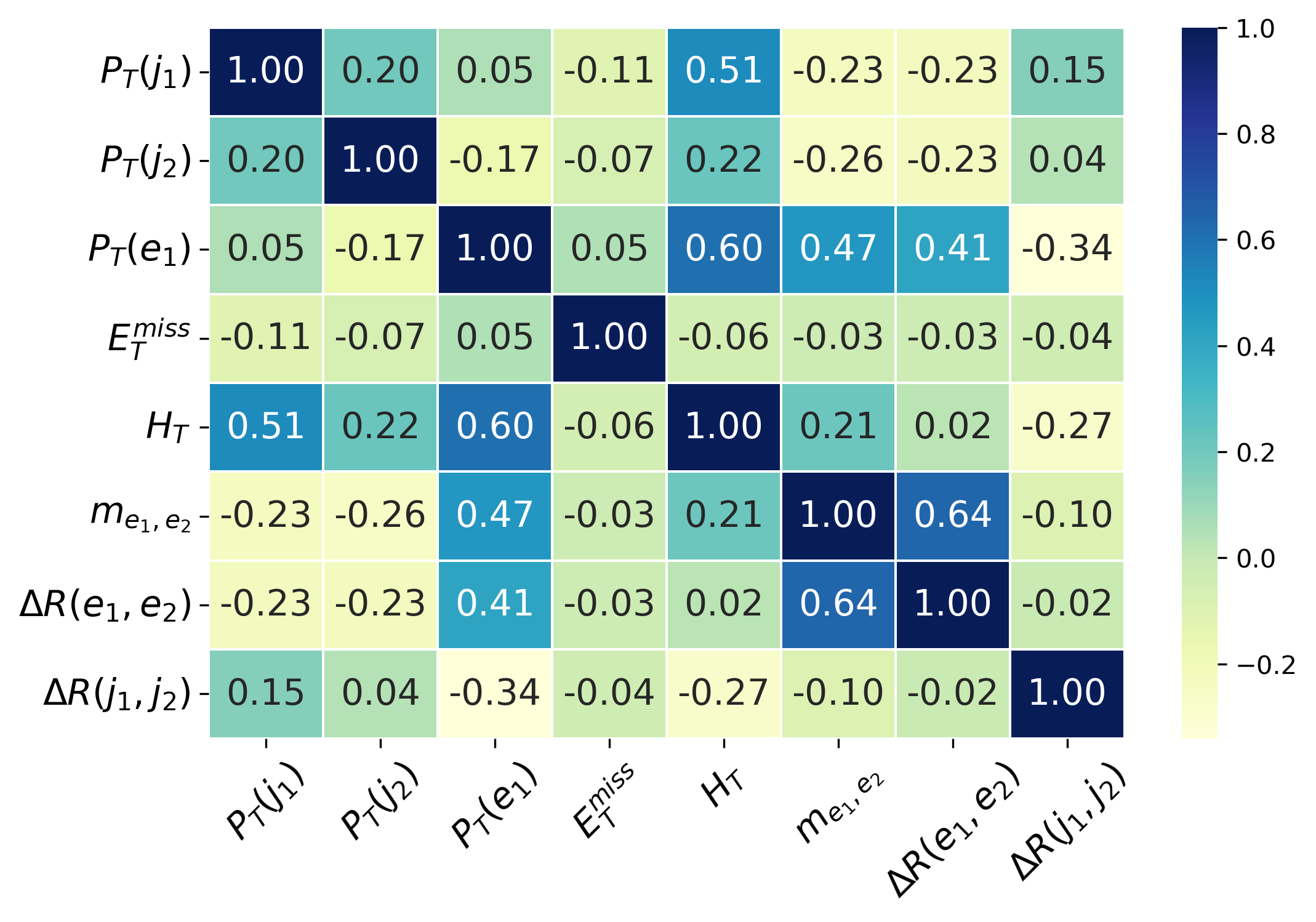} \\
 	\includegraphics[width=0.49\linewidth]{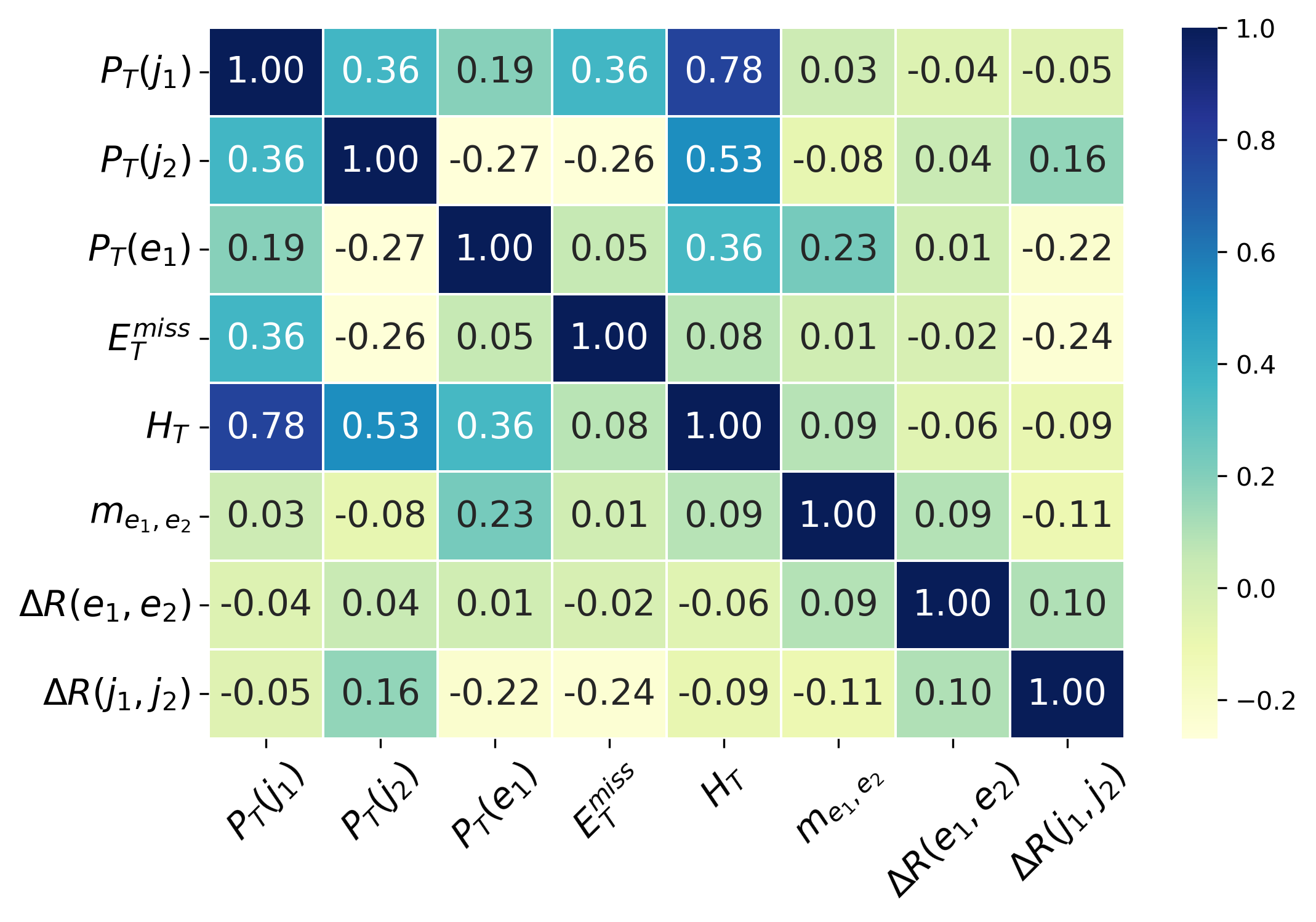}
	\includegraphics[width=0.49\linewidth]{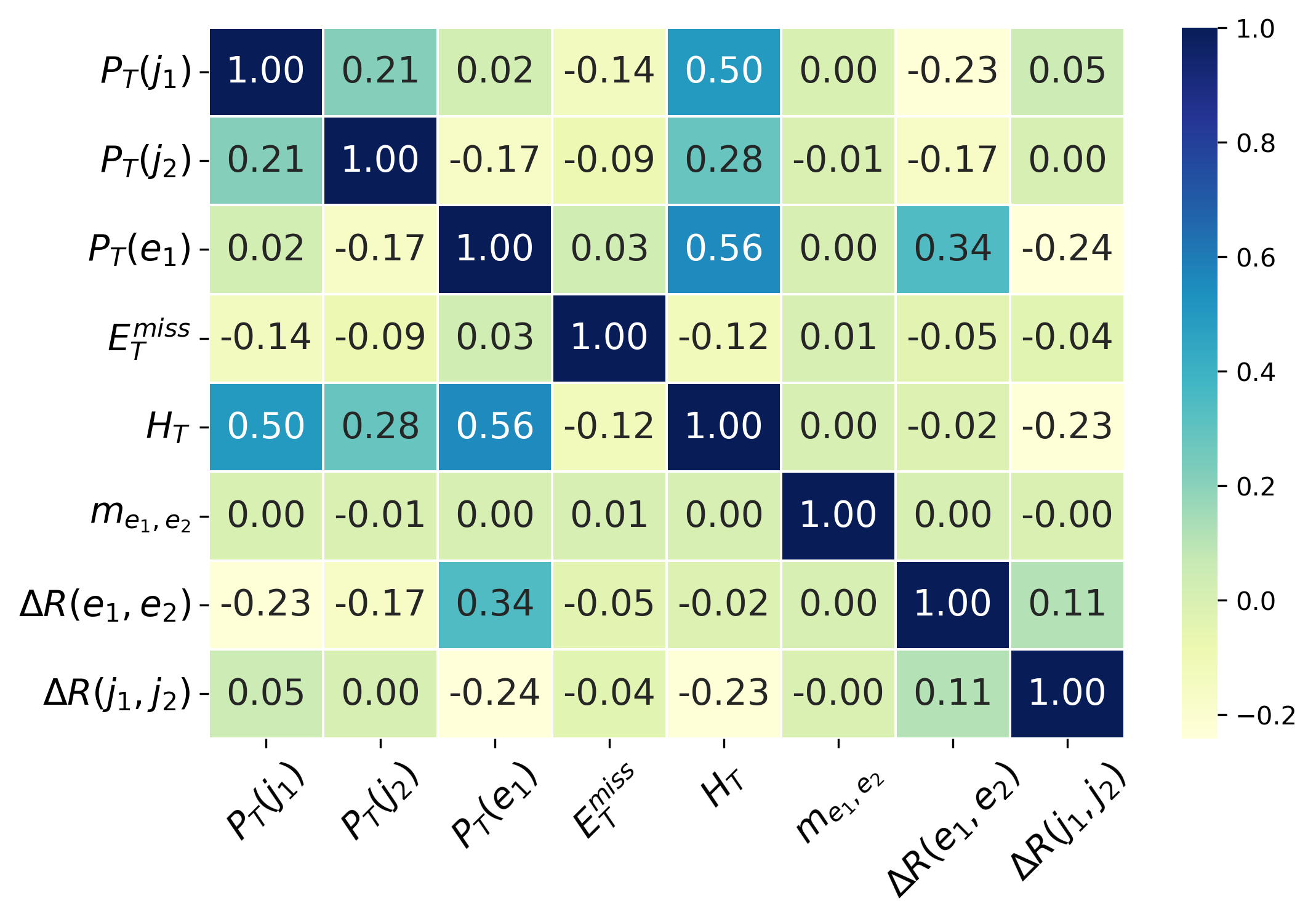}
	\caption{Correlation matrix of the observables in Eq.~(\ref{eqn:observables}) for the OS (upper panels) and SS (lower panels) processes in the $ee$ final state. The left and right panels are for the backgrounds and singals, respectively.
    }
	\label{cos}
\end{figure}

Feature selection is a crucial aspect of model building in machine learning. Its primary objective is to identify the most important features while eliminating irrelevant, redundant, and noisy features. Before training the model, we first evaluate the correlation among the observables:
\begin{equation}
\label{eqn:observables}
    p_T (j_1),\; p_T (j_2),\; 
    p_T (\ell_1),\; E_T^{\rm miss},\;
    H_T (j),\; m_{\ell_1 \ell_2},\;
    \Delta R (\ell_1, \ell_2),\; \Delta R (j_1,j_2) \,.
\end{equation}
The distributions of these observables for the OS and SS signals and the corresponding backgrounds are presented in Figs.~\ref{fig:HTandMET} to \ref{fig:14os_emu_kin_4}. As a concrete example, the correlation matrix charts for these observables in the $ee$ channel are shown in Fig.~\ref{cos}, with the upper and lower panels for the OS and SS processes, respectively. The correlations for backgrounds and signals are shown in the left and right panels, respectively. As can be seen in this figure, there are noticeable differences in the correlations between the signal and background variables, and most of the variables have relatively weak correlations.


\begin{figure}[t!]
	\centering
	\includegraphics[width=0.48\linewidth]{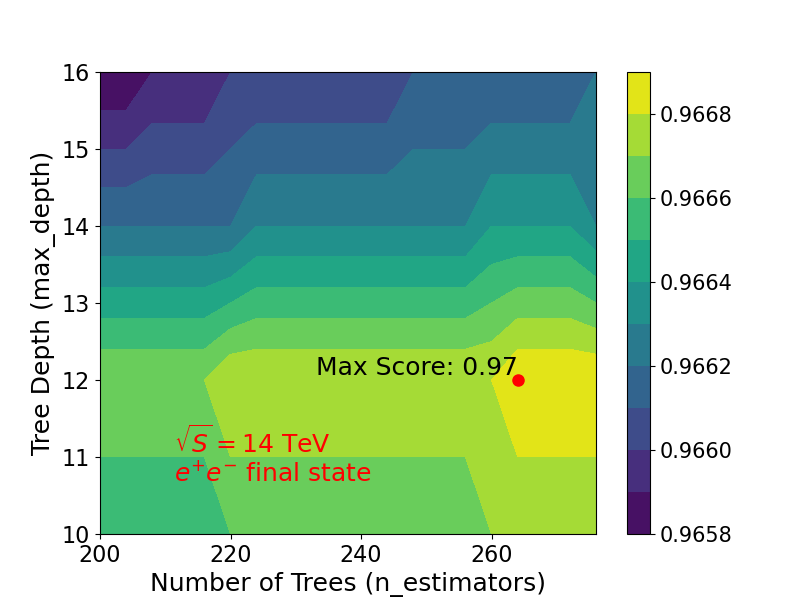}
	\includegraphics[width=0.48\linewidth]{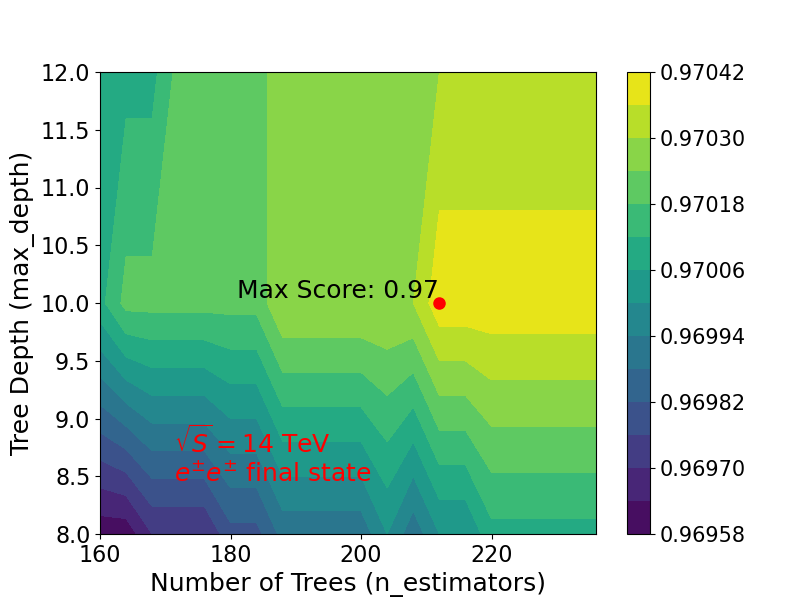}
	\includegraphics[width=0.48\linewidth]{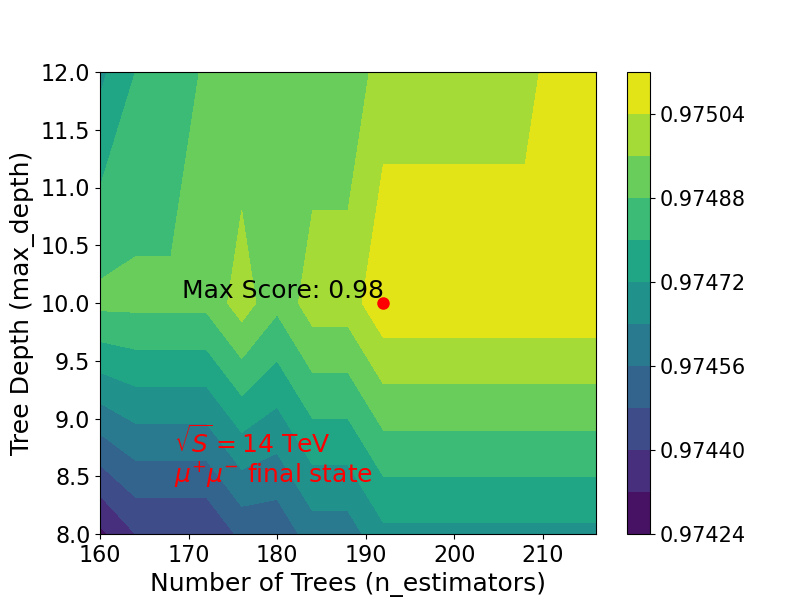}
	\includegraphics[width=0.48\linewidth]{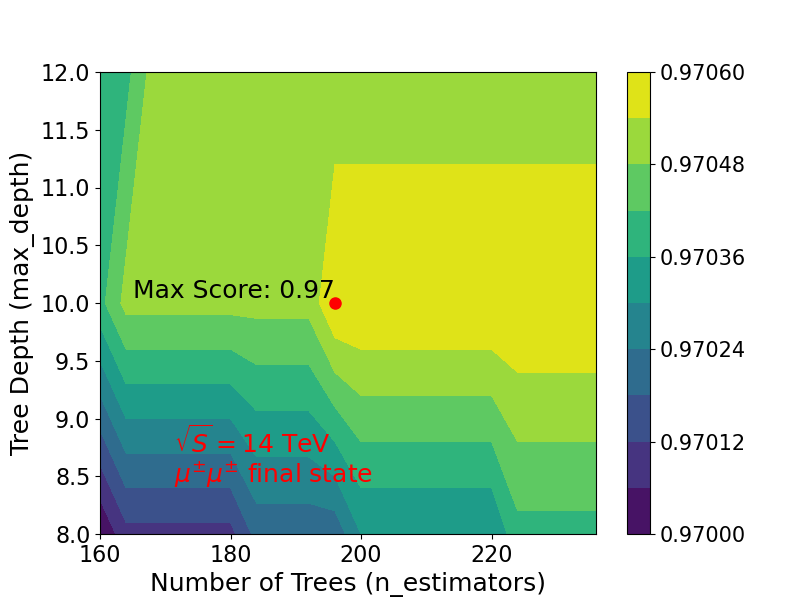}
	\includegraphics[width=0.48\linewidth]{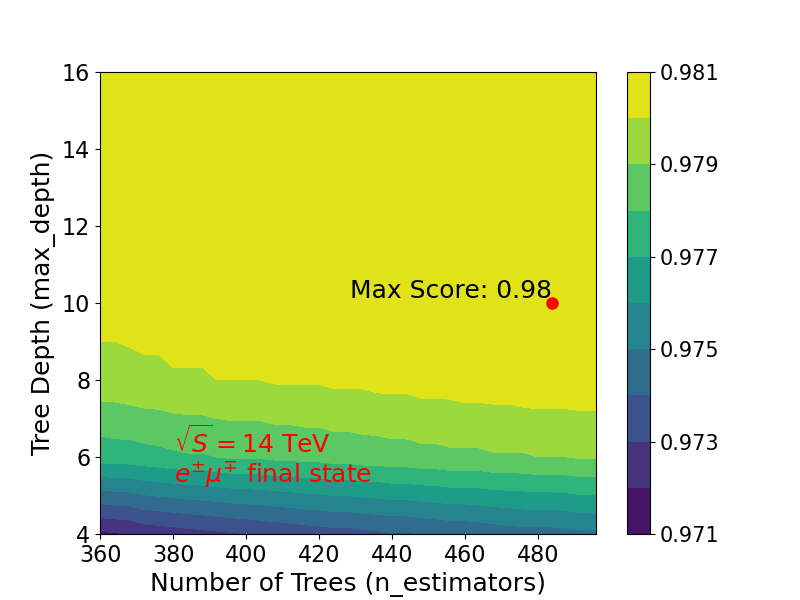}
	\includegraphics[width=0.48\linewidth]{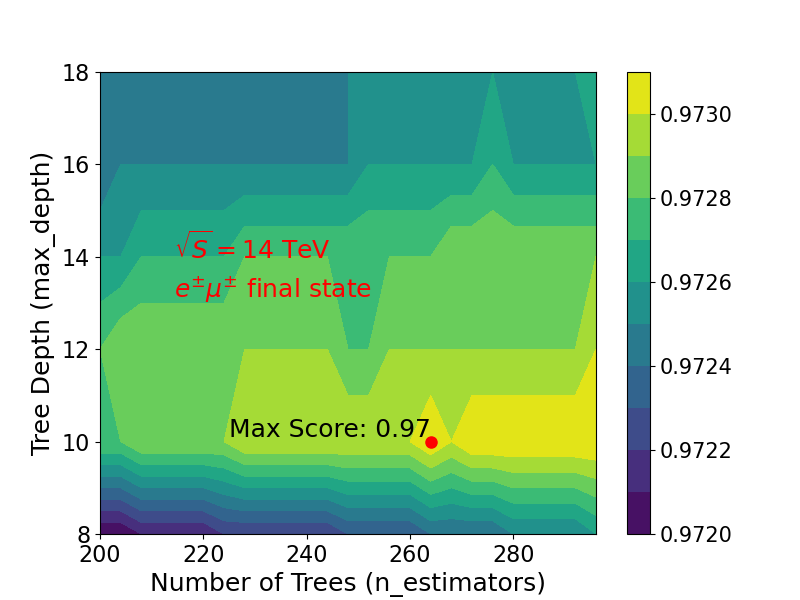}
	\caption{Models' scores on the test set, as functions of the number of trees and tree depth, for the OS (left panels) and SS (right panels) dilepton processes at $\sqrt{s}=14$ TeV. The top, middle and bottom panels are for the $ee$, $\mu\mu$ and $e\mu$ flavors, respectively. The red points in the plots indicate the positions of the highest scores. 
    }
	\label{fig:e_score}
\end{figure}

\begin{figure}[h!]
	\centering
	\includegraphics[width=0.48\linewidth]{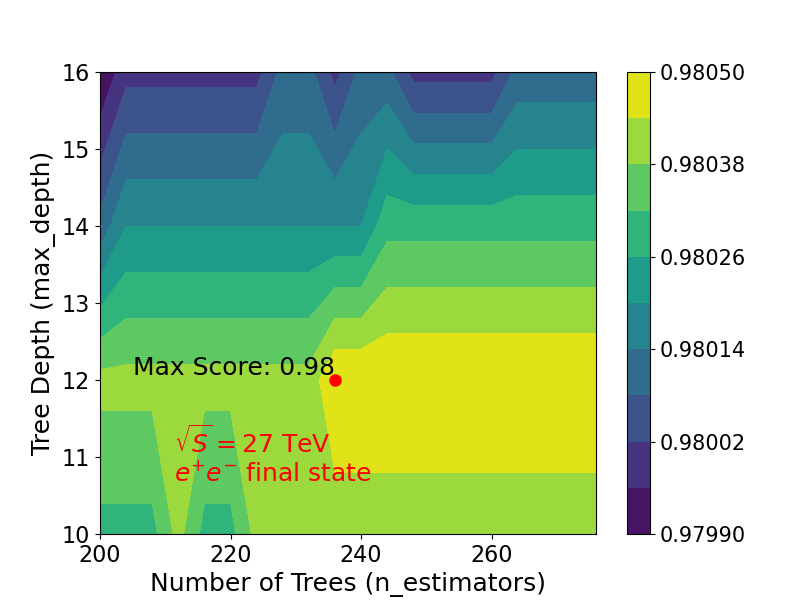}
	\includegraphics[width=0.48\linewidth]{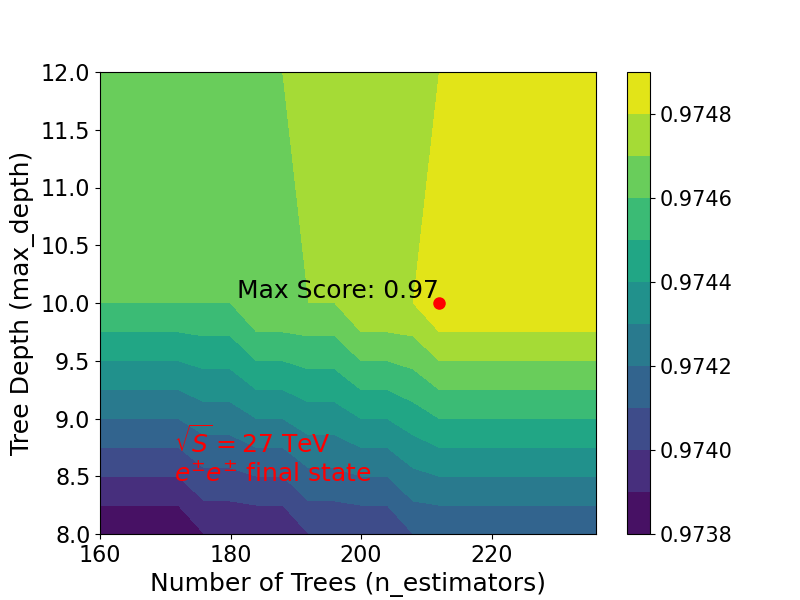}
    \includegraphics[width=0.48\linewidth]{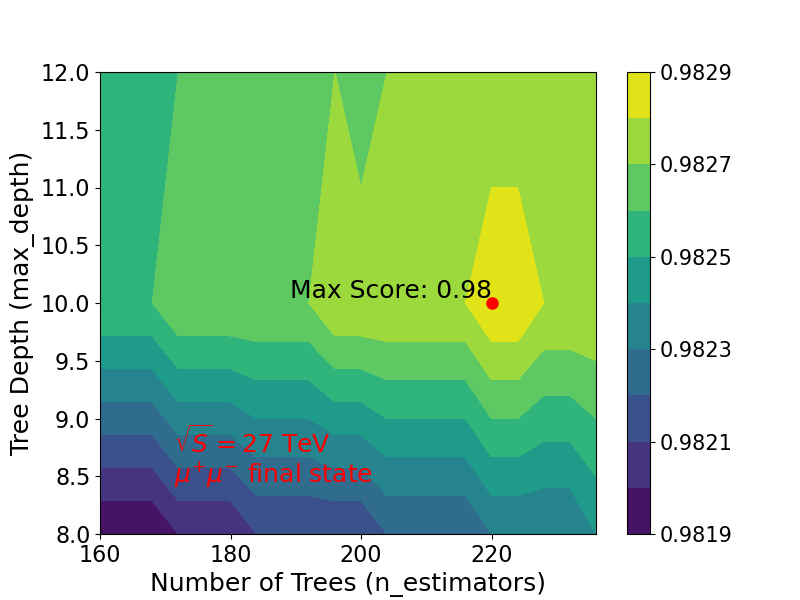}
	\includegraphics[width=0.48\linewidth]{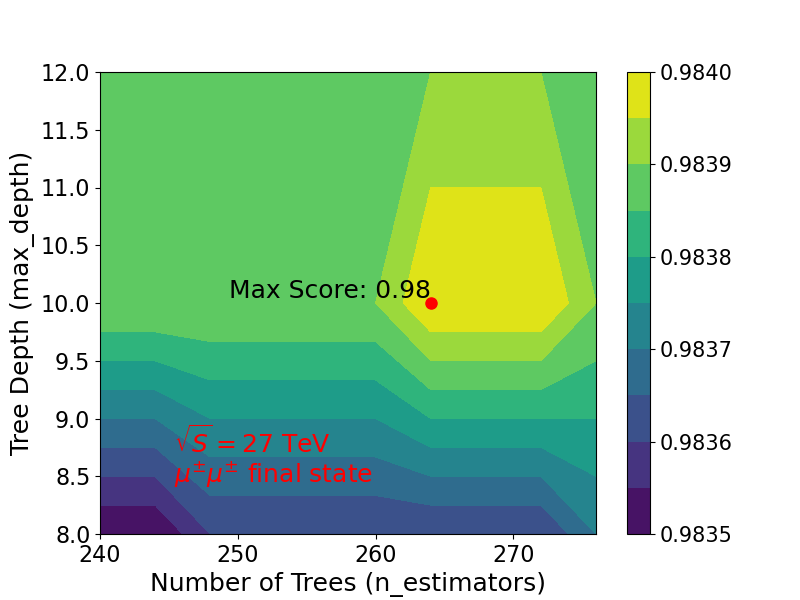}
    \includegraphics[width=0.48\linewidth]{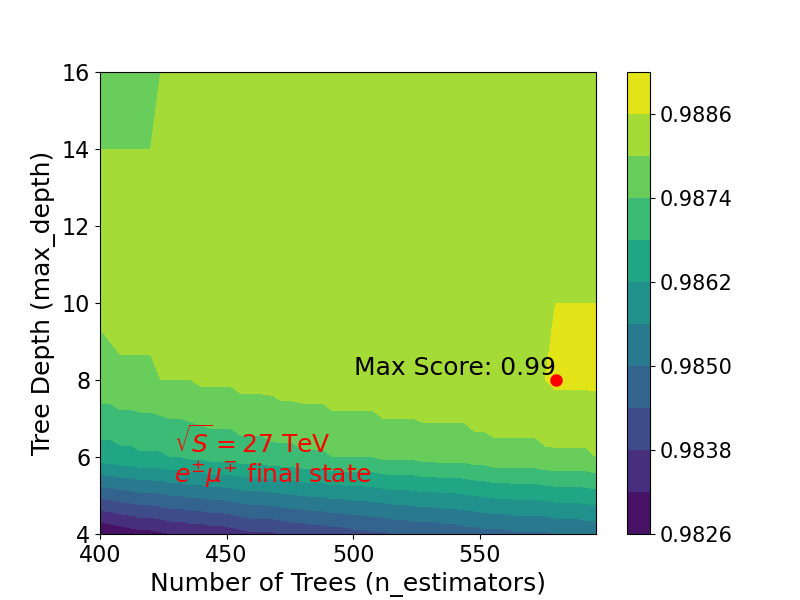}
	\includegraphics[width=0.48\linewidth]{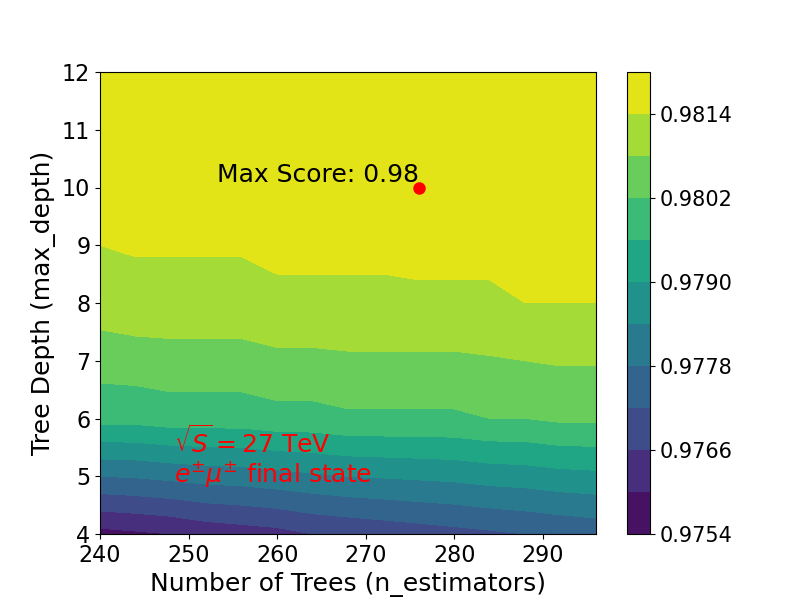}
	\caption{The same as Fig.~\ref{fig:e_score}, but for the center-of-mass energy of $\sqrt{s}=27$ TeV. 
    }
	\label{fig:mu_score}
\end{figure}

\begin{figure}[h!]
	\centering
	\includegraphics[width=0.48\linewidth]{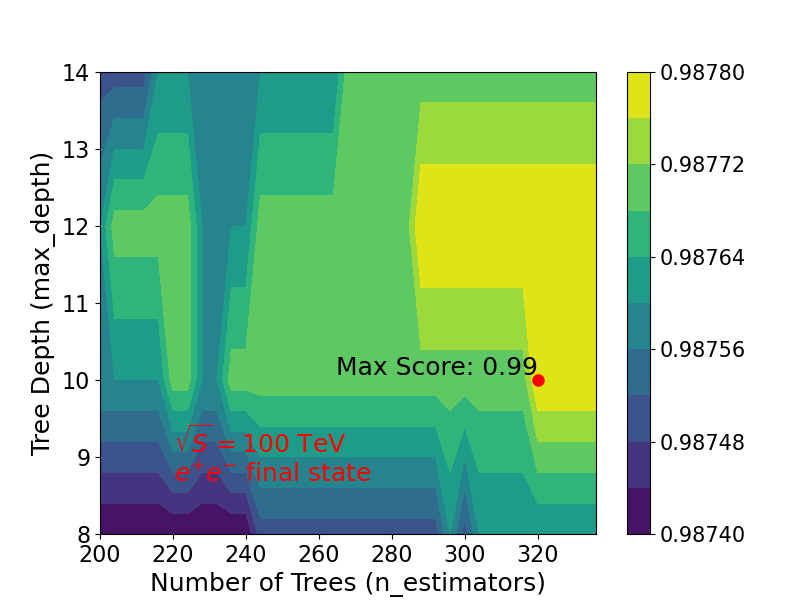}
	\includegraphics[width=0.48\linewidth]{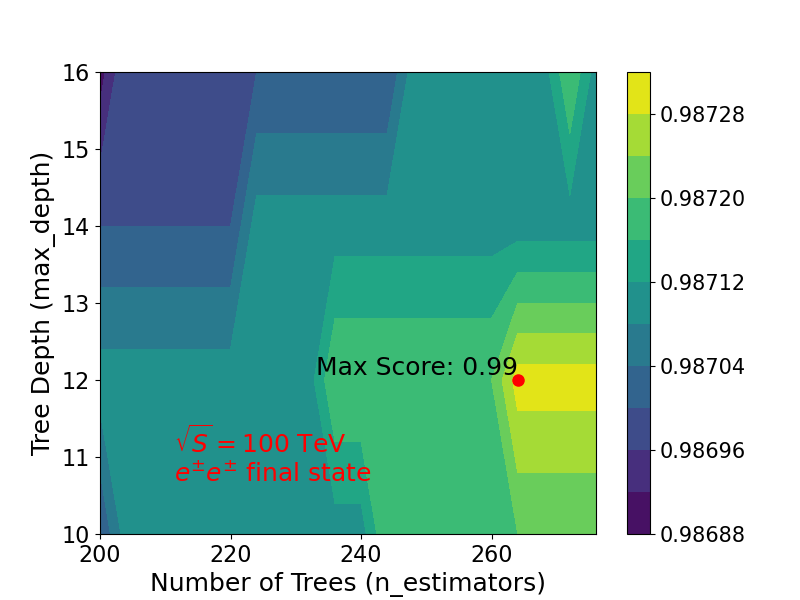}
	\includegraphics[width=0.48\linewidth]{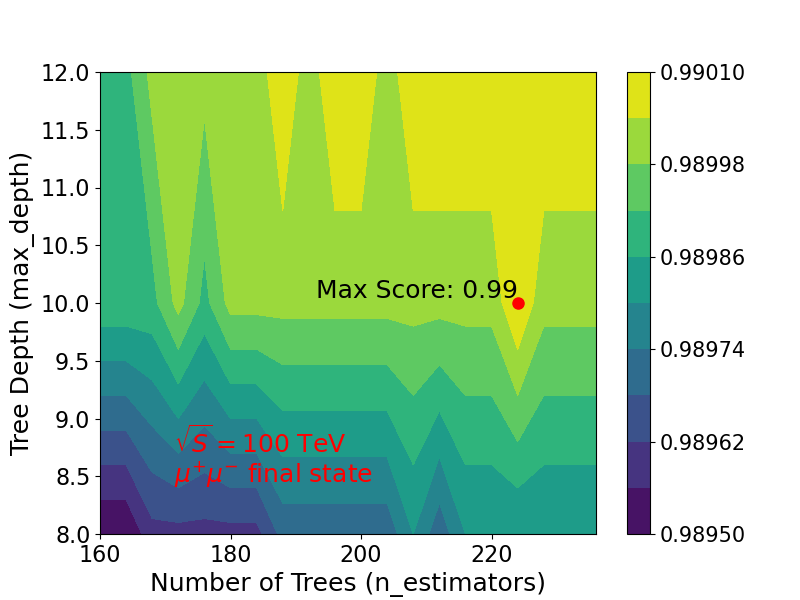}
	\includegraphics[width=0.48\linewidth]{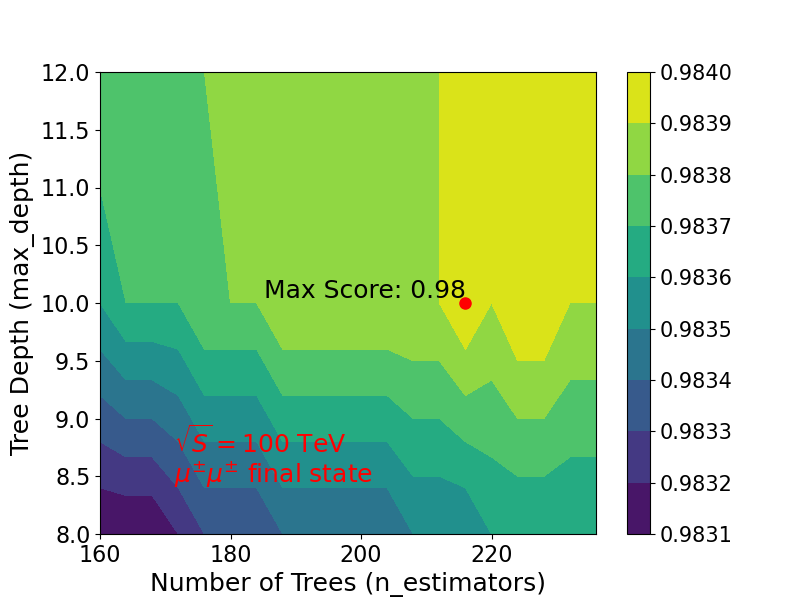}
	\includegraphics[width=0.48\linewidth]{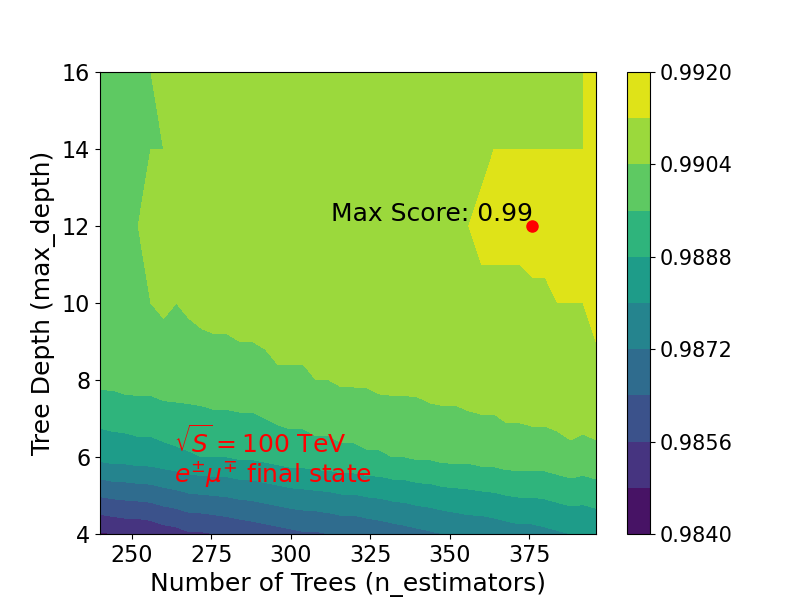}
	\includegraphics[width=0.48\linewidth]{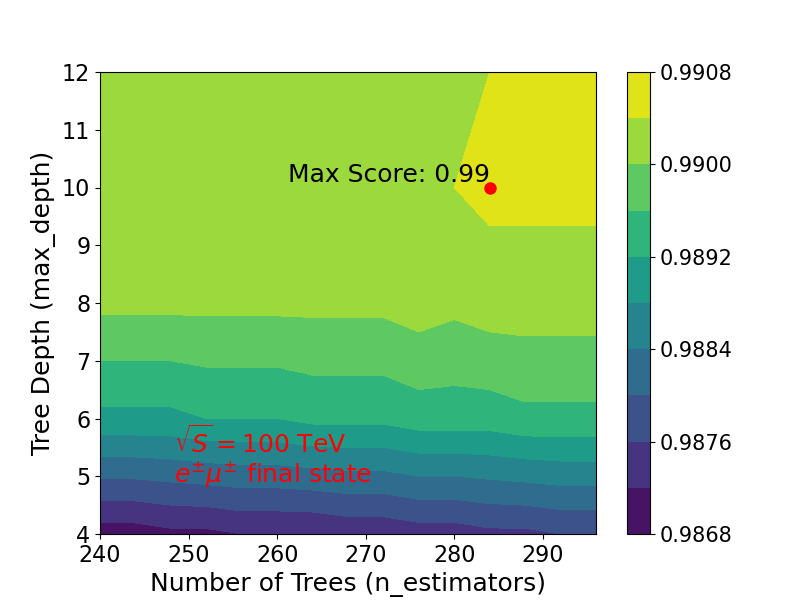}
	\caption{The same as Fig.~\ref{fig:e_score}, but for the center-of-mass energy of $\sqrt{s}=100$ TeV. 
    }
	\label{fig:emu_score}
\end{figure}

Machine learning methods can not only utilize all the information from the distributions in Figs.~\ref{fig:HTandMET} through to \ref{fig:14os_emu_kin_4}, but also take into account the correlations among the variables in Fig.~\ref{cos}. We use the observables in Eq.~(\ref{eqn:observables})  as inputs for the machine learning analysis. In the training, there are some key parameters of XGBoost that are crucial for the analysis:
\begin{enumerate}
	\item Learning rate: controlling the step size for each iteration. A smaller learning rate makes the model more stable, but may require more iterations. In this study, we set the learning rate of all machine learning models to be 0.02.
	\item Tree depth: limiting the maximum depth of each tree to prevent overfitting.
	\item Regularization parameters: used to control model complexity and prevent overfitting. The regularization parameter is set to the default value of the XGBoost model.
	\item Number of trees: specifying the number of trees in the iteration. Too many trees can lead to overfitting.
\end{enumerate}
During model training, the grid search strategy has been performed to determine the optimal count of trees and depth. We use metrics such as the logarithmic loss, area under the ROC curve (AUC), and error rate for model evaluations. 
During the training, the optimal combination of parameters is determined by evaluating the models' performance on the test set across various parameter settings. 
The performance scores of machine learning training for the heavy neutrino signals at $\sqrt{s}=14$ TeV, 27 TeV and 100 TeV are shown in Figs.~\ref{fig:e_score}, \ref{fig:mu_score} and \ref{fig:emu_score}, respectively. In these figures, the left and right panels are for the OS and SS signals, and the top, middle and bottom panels correspond to the $ee$, $\mu\mu$ and $e\mu$ flavors in the final state, respectively. The bright yellow regions with a higher score correspond to the parameter settings where the model can achieve a better performance.
The positions with the highest scores are indicated by the red pionts in the figures.

\begin{figure}[t!]
	\centering
	\includegraphics[width=0.48\linewidth]{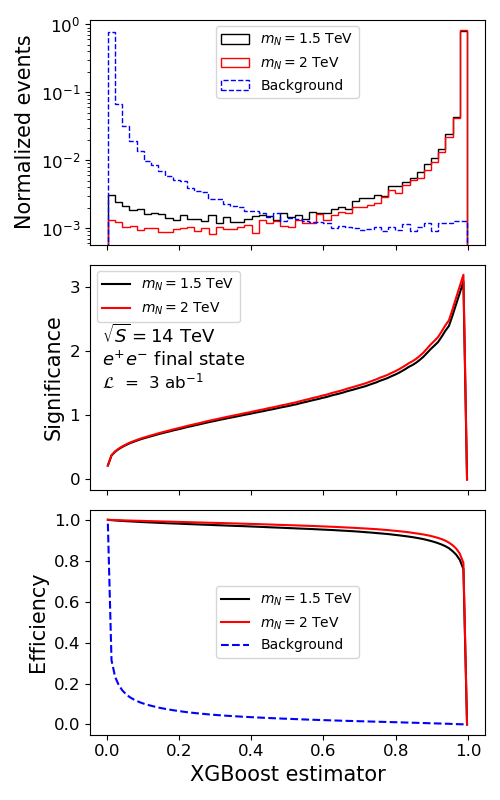}
	\includegraphics[width=0.48\linewidth]{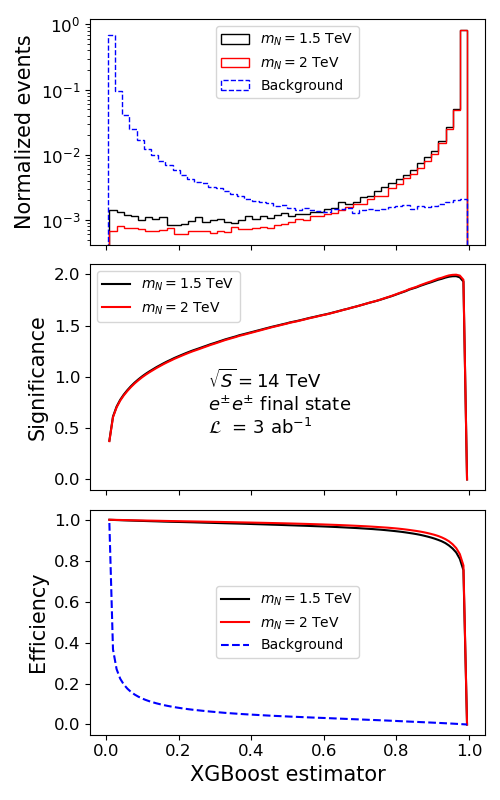}
\caption{Machine learning results for the OS (left) and SS (right) signals in the $ee$ final state at $\sqrt{s}=14$ TeV. The normalized event distributions for the signal $m_N = 1.5$ TeV (solid black) or 2 TeV (solid red) and backgrounds (dashed blue) are shown in the top panels. 
The signal significances are presented in the middle panels, with an integrated luminosity of \( \mathcal{L} = 3 \, \text{ab}^{-1} \), and the efficiency curves for signals and backgrounds are shown in the bottom panels. 
}
\label{fig:14Events}
\end{figure}

\begin{figure}[h!]
	\centering
	\includegraphics[width=0.48\linewidth]{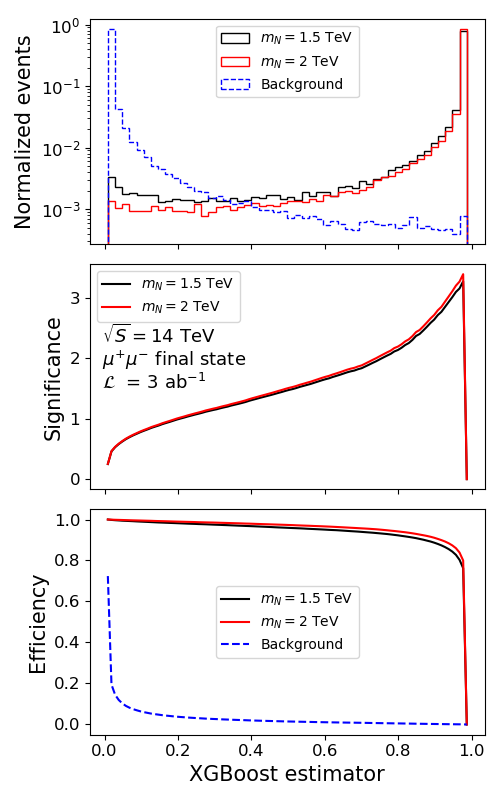}
	\includegraphics[width=0.48\linewidth]{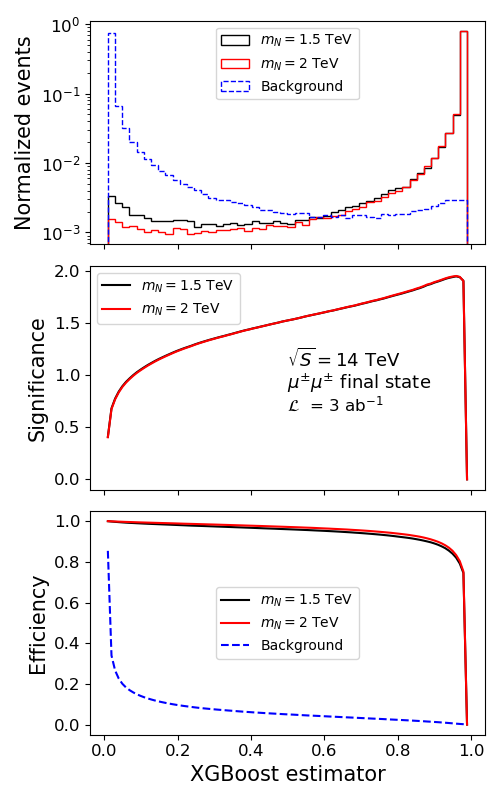}
	\caption{The same as Fig~\ref{fig:14Events} but for the $\mu\mu$ final state  at $\sqrt{s}=14$ TeV.
    }
	\label{fig:14os_mu}
\end{figure}

\begin{figure}[h!]
	\centering
	\includegraphics[width=0.48\linewidth]{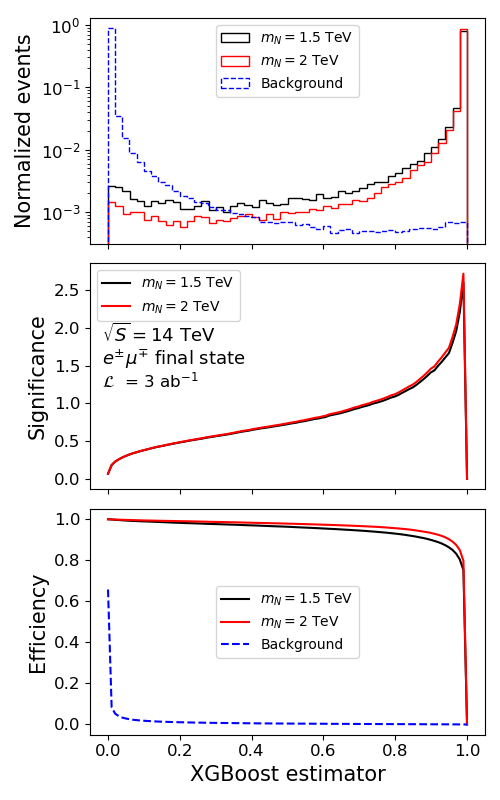}
	\includegraphics[width=0.48\linewidth]{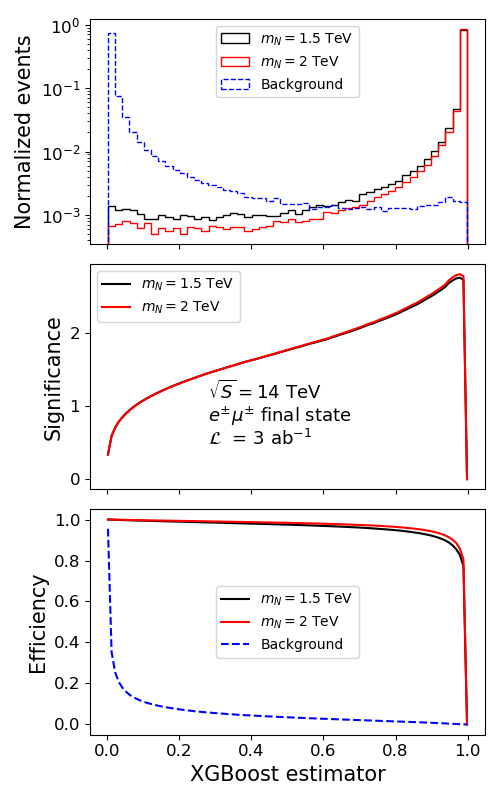}
	\caption{The same as Fig~\ref{fig:14Events} but for the $e\mu$ final state at $\sqrt{s}=14$ TeV. 
    }
	\label{fig:14os_emu}
\end{figure}

\begin{figure}[t!]
	\centering
	\includegraphics[width=0.48\linewidth]{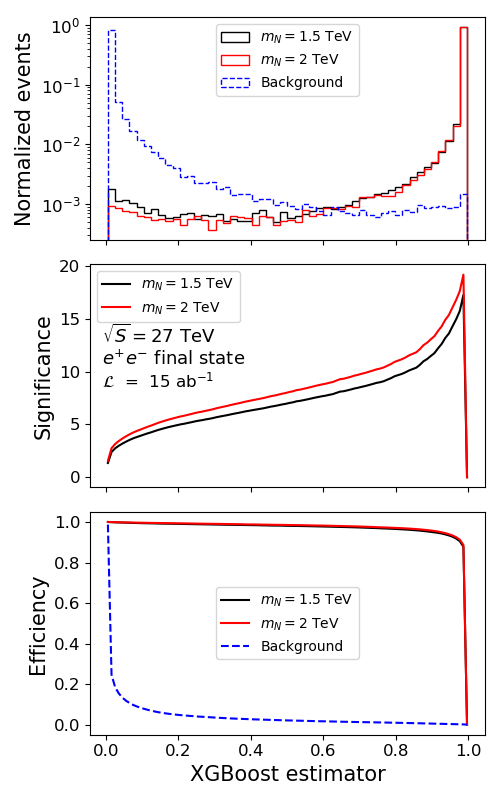}
	\includegraphics[width=0.48\linewidth]{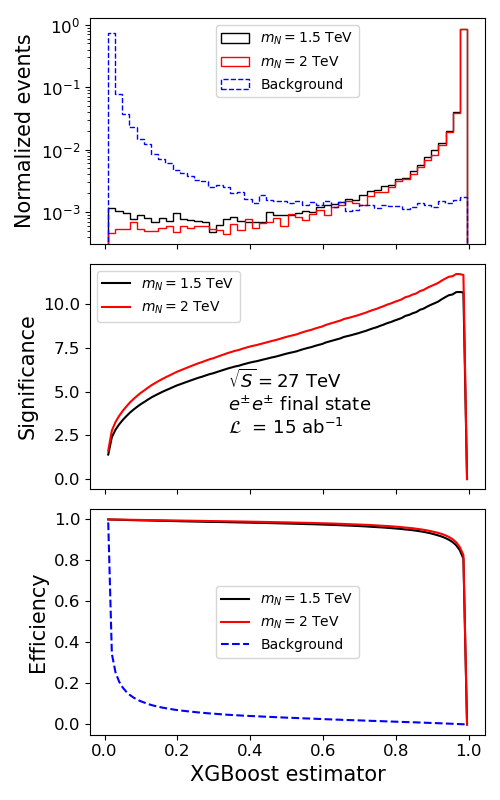}
	\caption{The same as Fig~\ref{fig:14Events}, but for the $ee$ final state at $\sqrt{s}=27$ TeV with an integrated luminosity of 15 ab$^{-1}$.
    }
	\label{fig:27Events}
\end{figure}

\begin{figure}[h!]
	\centering
	\includegraphics[width=0.48\linewidth]{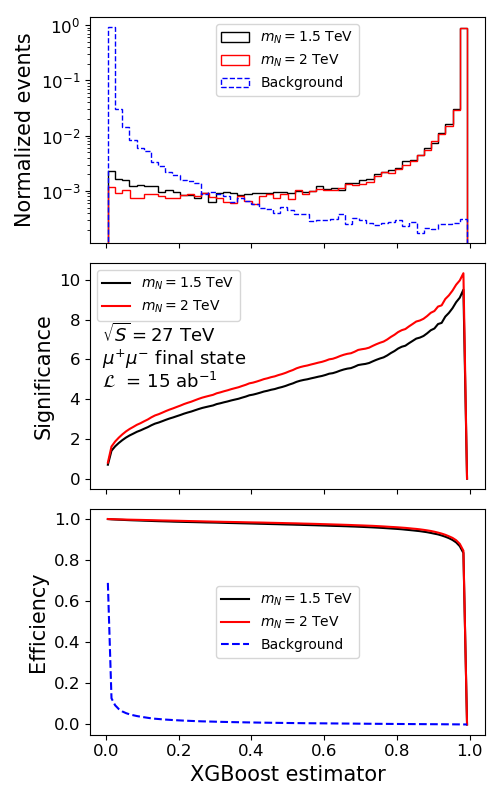}
	\includegraphics[width=0.48\linewidth]{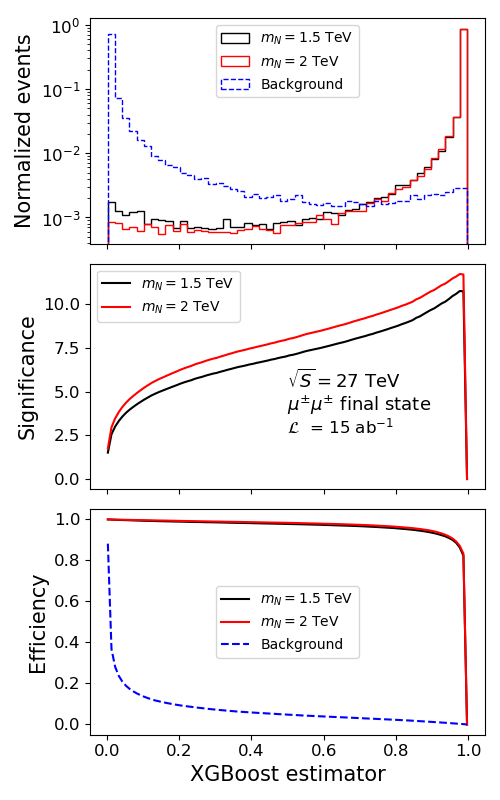}
	\caption{The same as Fig~\ref{fig:14Events}, but for the $\mu\mu$ final state at $\sqrt{s}=27$ TeV with an integrated luminosity of 15 ab$^{-1}$. 
   }
	\label{fig:27os_mu}
\end{figure}

\begin{figure}[h!]
	\centering
	\includegraphics[width=0.48\linewidth]{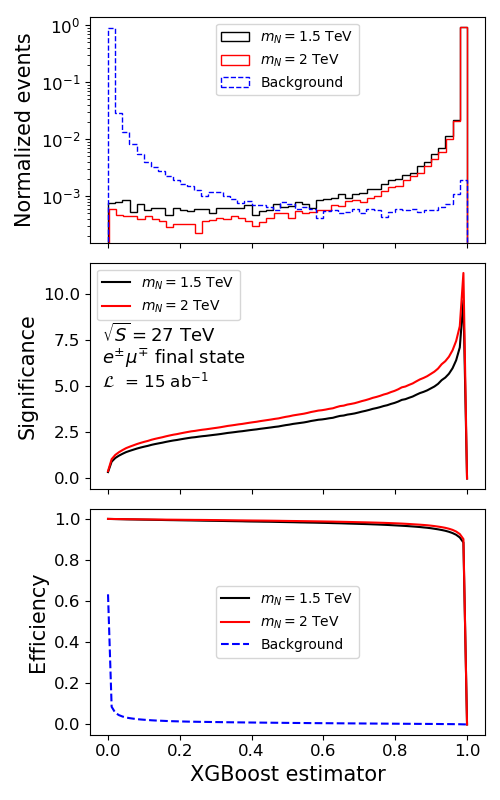}
	\includegraphics[width=0.48\linewidth]{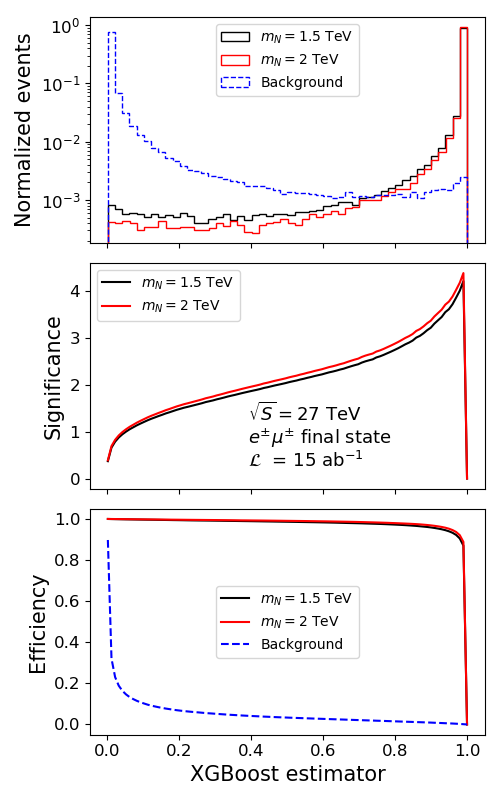}
	\caption{The same as Fig~\ref{fig:14Events} but for the $e\mu$ final state at $\sqrt{s}=27$ TeV with an integrated luminosity of 15 ab$^{-1}$. 
    }
	\label{fig:27os_emu}
\end{figure}

\begin{figure}[t!]
	\centering
	\includegraphics[width=0.48\linewidth]{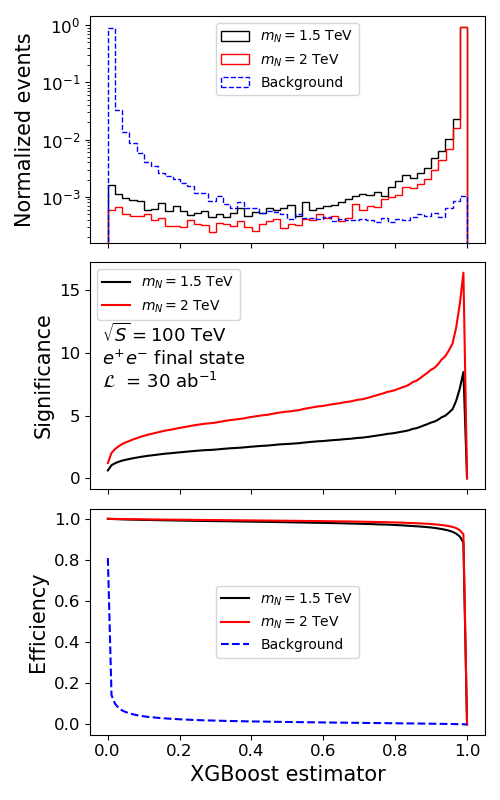}
	\includegraphics[width=0.48\linewidth]{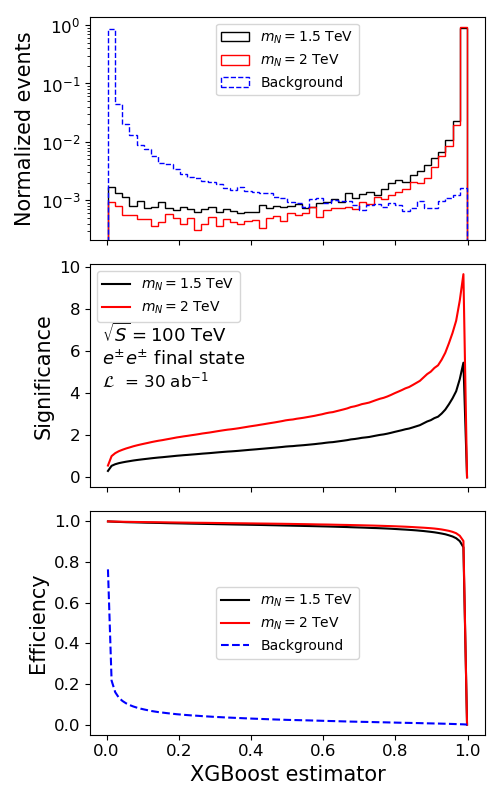}
	\caption{The same as Fig~\ref{fig:14Events} but for  the $ee$ final state at $\sqrt{s}=100$ TeV with an integrated luminosity of 30 ab$^{-1}$.
    }
	\label{fig:100Events}
\end{figure}

\begin{figure}[h!]
	\centering
	\includegraphics[width=0.48\linewidth]{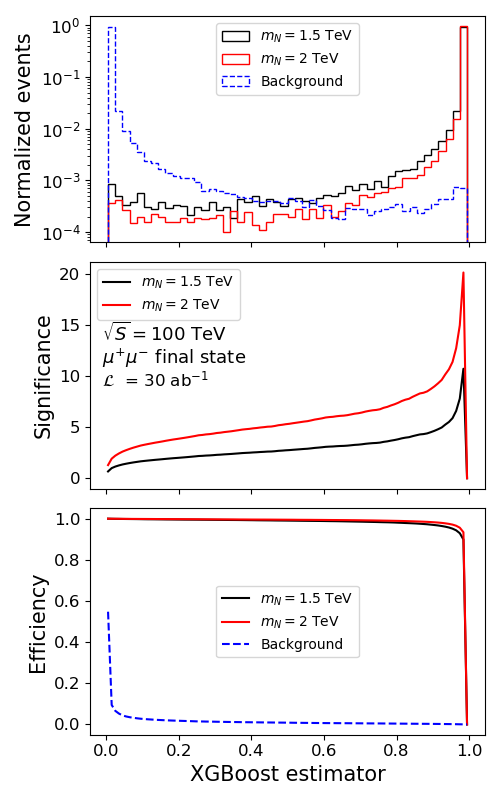}
	\includegraphics[width=0.48\linewidth]{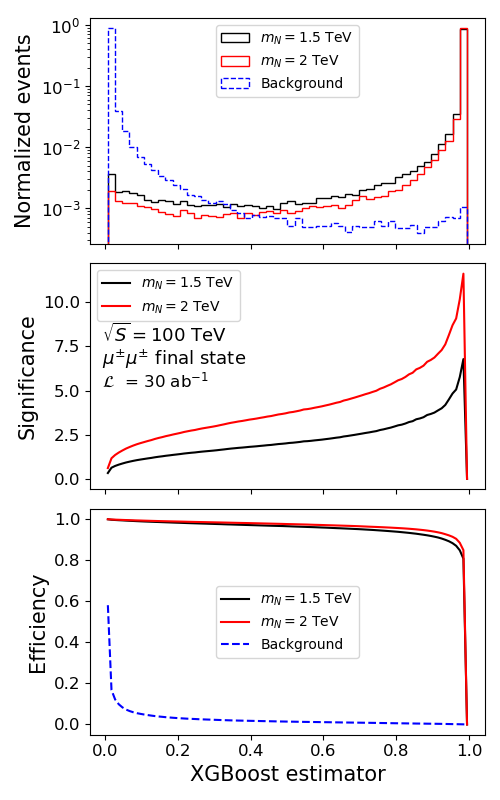}
	\caption{The same as Fig~\ref{fig:14Events}, but for the $\mu\mu$ final state at $\sqrt{s}=100$ TeV with an integrated luminosity of 30 ab$^{-1}$.
     }
	\label{fig:100os_mu}
\end{figure}

\begin{figure}[h!]
	\centering
	\includegraphics[width=0.48\linewidth]{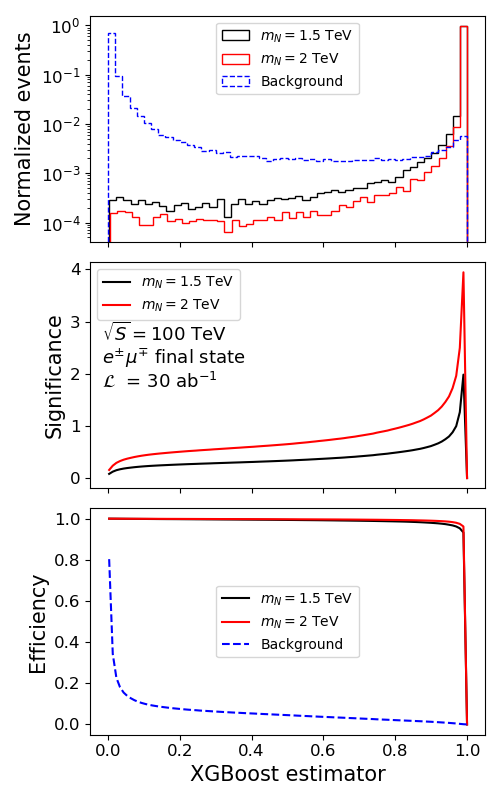}
	\includegraphics[width=0.48\linewidth]{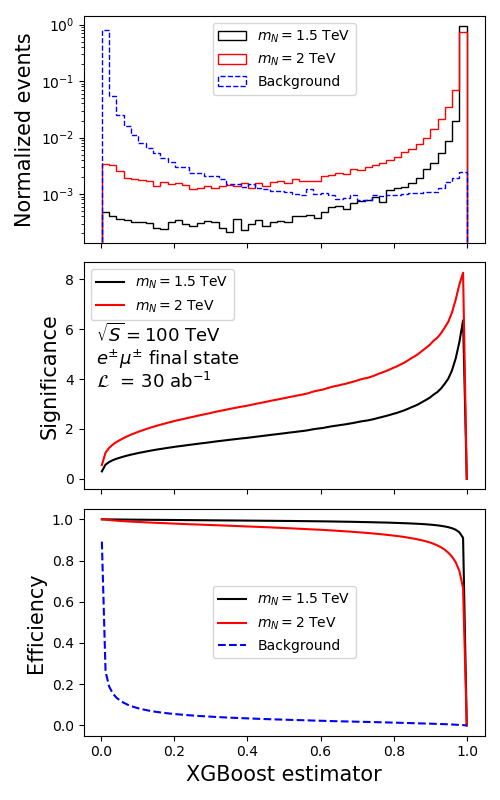}
	\caption{The same as Fig~\ref{fig:14Events}, but for the $e\mu$ final state at $\sqrt{s}=100$ TeV with an integrated luminosity of 30 ab$^{-1}$. 
    }
	\label{fig:100os_emu}
\end{figure}

For the OS and SS processes, although the same observables in Eq.~(\ref{eqn:observables}) are used, the corresponding backgrounds are significantly different. Therefore, the models are trained separately for each flavor combination of $ee$, $\mu\mu$ and $e\mu$ at $\sqrt{s}=14$ TeV, 27 TeV and 100 TeV. The corresponding optimized machine learning parameters are summarized in Table~\ref{tab:parameters summarized}, and the results of the multivariate analysis with machine learning are visualized in Figs.~\ref{fig:14Events} through \ref{fig:100os_emu}.
The classification performance of the XGBoost classifier for the OS and SS $ee$ dilepton signals at $\sqrt{s}=14$ TeV is presented in the left and right panels of Fig.~\ref{fig:14Events}, respectively. The cases with heavy neutrino masses $m_N = 1.5$ TeV and 2.0 TeV are again depicted as the solid black and red lines, respectively, while the backgrounds are indicated by the dashed blue lines. 
The normalized event distributions are shown in the top panels. For both OS and SS processes, background events are predominantly distributed near a classifier output of 0, while signal events dominate the region close to 1, indicating that the classifier distinguishes effectively signals from backgrounds. The significances of signals are presented in the middle panels of Fig.~\ref{fig:14Events}, as functions of the estimator. A similar trend can be observed for both OS and SS signals with an integrated luminosity of \( \mathcal{L} = 3 \, \text{ab}^{-1} \) at 14 TeV, with a slightly higher significance for the signal of $m_N = 2.0$ TeV at high classifier outputs. Furthermore, the OS and SS processes exhibit differences in the significance peaks, reflecting the impact of different background compositions. The efficiency curves for the signals and backgrounds are shown in the bottom panels of Fig.~\ref{fig:14Events}. 
In the OS process, the efficiencies of the $W^{+}W^{-}jj$ and $Zjj$ backgrounds drop rapidly, reaching 0.18\% and 0.039\% at the classifier output of 0.98, respectively. For the SS signal, the efficiencies of the $W^{\pm}W^{\pm}jj$ and $W^{\pm}Zjj$ backgrounds are suppressed to 0.053\% and 0.36\%, respectively. Signal efficiencies decrease gradually in both processes; at a classifier output of 0.98, the efficiencies for the signals of $m_N = 1.5$ TeV and 2.0 TeV are 75.4\% and 78.3\% in the OS process, and 75.4\% and 77.7\% in the SS process, respectively. These results demonstrate that the classifier achieves substantial background suppression while maintaining high signal efficiencies for both the OS and SS signals, providing robust support for distinguishing signals from backgrounds. 
The analysis of the flavors of $\mu\mu$ and $e\mu$ in the final state at $\sqrt{s} = 14$ TeV is presented in Figs.~\ref{fig:14os_mu} and \ref{fig:14os_emu}, respectively. The results for the flavors of $ee$, $\mu\mu$ and $e\mu$ at the center-of-mass energy of $\sqrt{s}=27$ TeV are shown in Figs.~\ref{fig:27Events}, \ref{fig:27os_mu} and \ref{fig:27os_emu}, respectively, while the corresponding results at $\sqrt{s}=100$ TeV are presented in Figs.~\ref{fig:100Events}, \ref{fig:100os_mu} and \ref{fig:100os_emu}, respectively. 
The optimized machine learning thresholds $\chi$ and the corresponding signal and background efficiencies for all these flavor combinations at $\sqrt{s} = 14$ TeV, 27 TeV and 100 TeV are collected in Table~\ref{tab:eff_e_table}.

\begin{table}[t!]
	\centering
    \caption{Optimized machine learning parameters for training the models, for the OS and SS processes with the flavors of $ee$, $\mu\mu$ and $e\mu$ in the final state at $\sqrt{s}=14$ TeV, 27 TeV and 100 TeV.}
    \vspace{5pt}
	\label{tab:parameters summarized}
	\begin{tabular}{|c|c|ccc|ccc|ccc|}
	\hline
	\multicolumn{2}{|c|}{$\sqrt{s}$ [TeV]} & \multicolumn{3}{c|}{14} & \multicolumn{3}{c|}{27} & \multicolumn{3}{c|}{100} \\ \hline
     \multicolumn{2}{|c|}{flavors}	& $ee$ & $\mu\mu$ & $e\mu$ & $ee$ & $\mu\mu$ & $e\mu$ & $ee$ & $\mu\mu$ & $e\mu$ \\ \hline\hline
\multirow{2}{*}{OS} & {\tt max\_depth} & 12 & 10 & 10 & 12 & 10 & 8 & 10 & 10 & 12 \\ \cline{2-11} 
& {\tt n\_estimators} & 264 & 192 & 484 & 236 & 220 & 578 & 320 & 224 & 376 \\ \hline\hline
\multirow{2}{*}{SS} & {\tt max\_depth} & 10 & 10 & 10 & 10 & 10 & 10 & 12 & 10 & 10 \\ \cline{2-11}
& {\tt n\_estimators} & 212 & 196 & 262 & 212 & 264 & 276 & 264 & 216 & 284 \\ \hline
\end{tabular}
\end{table}

\subsection{Sensitivities at high-energy hadron colliders}
\label{sec:sensitivities}

The prospects of the KS process in the LRSM at future high-energy hadron colliders depend on many factors, including the masses $m_{N}$ and $m_{W_{R}}$, heavy neutrino mixing angle $\alpha$, cut efficiency, center-of-mass energy, luminosity, and detector performance. We are now ready to estimate the prospects of heavy neutrinos and the sine of the mixing angle $s_\alpha$ in both the OS and SS processes for the three collider setups: the 14 TeV HL-LHC with an integrated luminosity of 3 ab$^{-1}$, the 27 TeV HE-LHC with the luminosity of 15 ab$^{-1}$\cite{ATLAS:2019mfr}, and the future 100 TeV collider such as the FCC-hh and SPPC, with the luminosity of 30 ab$^{-1}$\cite{FCC:2018vvp,Tang:2015qga}.

\begin{figure}[h!]
	\centering
\includegraphics[width=0.6\linewidth]{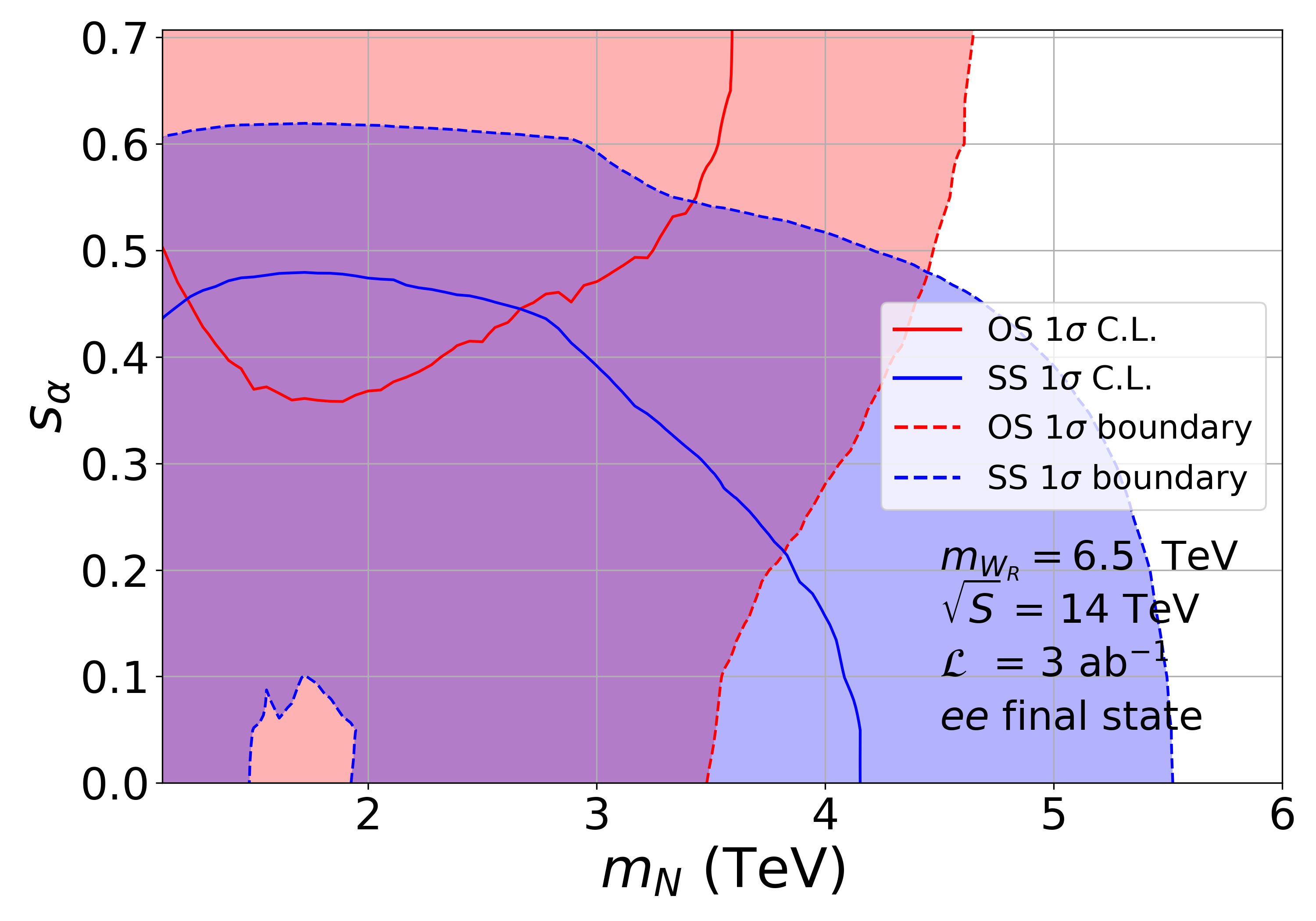}
\includegraphics[width=0.6\linewidth]{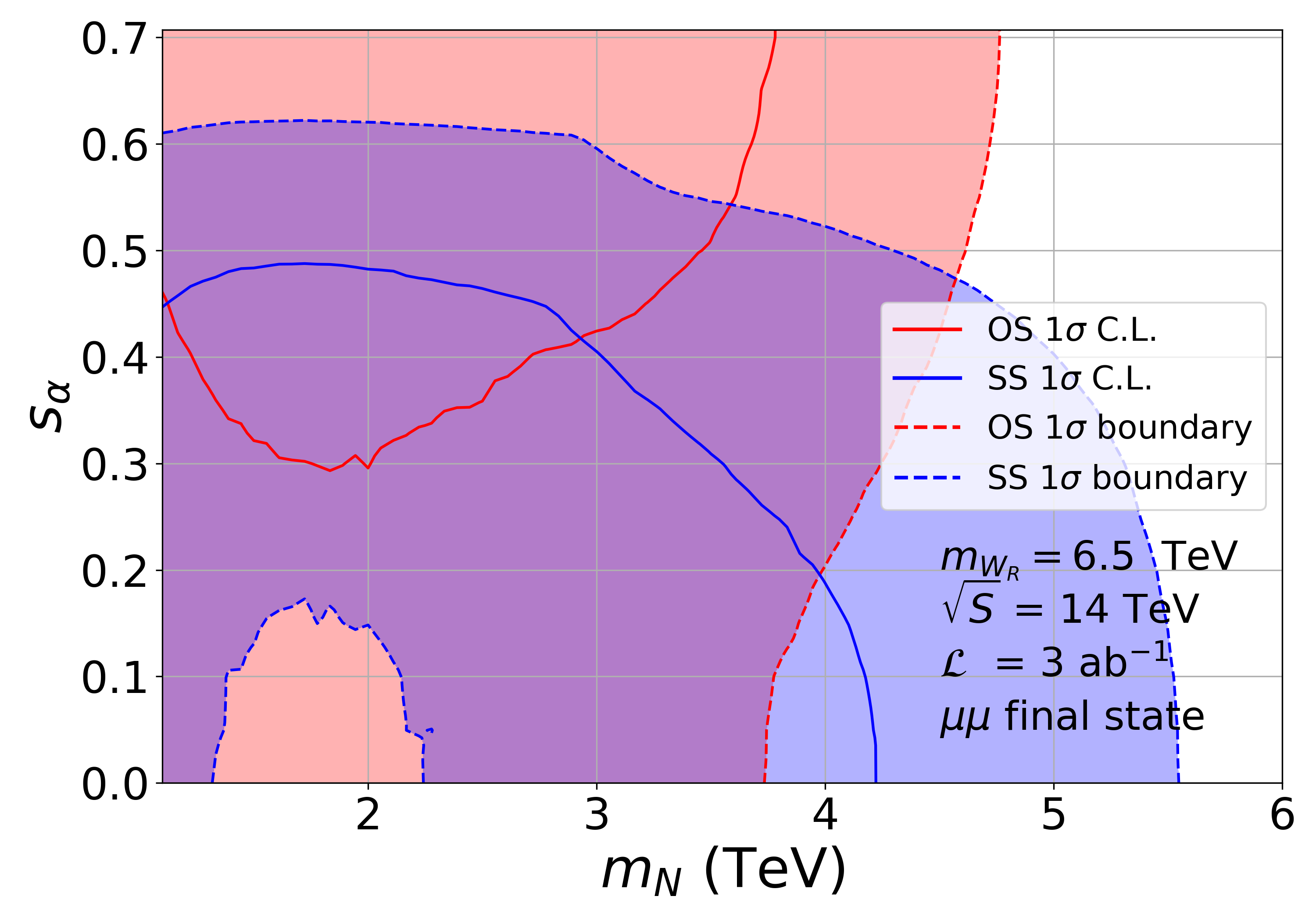}
\includegraphics[width=0.6\linewidth]{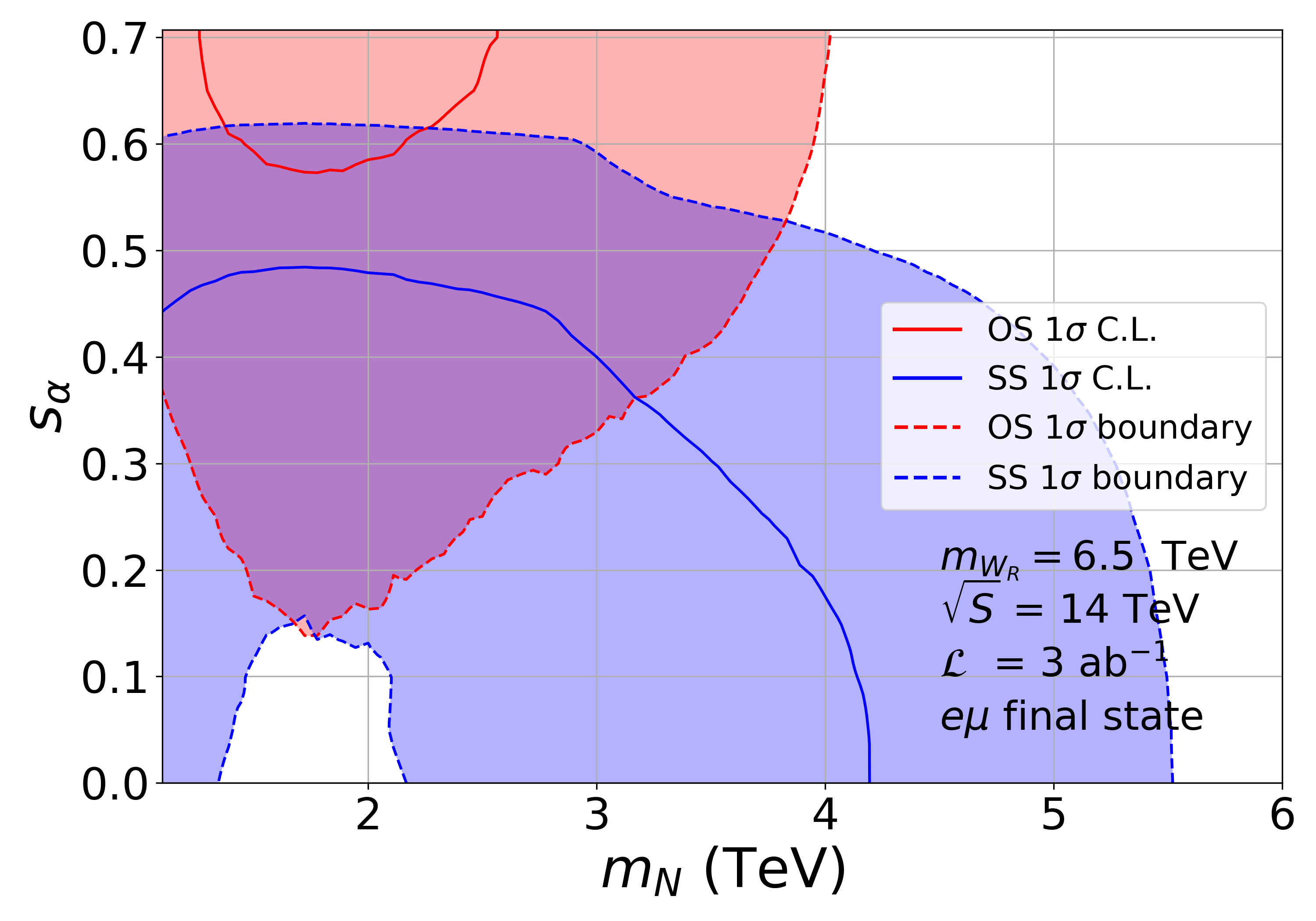}
\caption{Sensitivities of the heavy neutrino mass $m_N$ and mixing angle $s_\alpha$ in the final state of $ee$ (top), $\mu\mu$ (middle) and $e\mu$ (bottom) at $\sqrt{s}=14$ TeV with an integrated luminosity of 3 ab$^{-1}$. The solid red and blue lines represent the sensitivities at the $1\sigma$ C.L. for the OS and SS processes, respectively, while the color bands are the corresponding \(1\sigma\) uncertainties. 
}
\label{fig:sensitivity:14}
\end{figure}

The machine learning cuts are set at the optimized thresholds in Table~\ref{tab:eff_e_table} that yield the maximum significances to distinguish the signals from backgrounds. The significance ${\cal Z}$ is calculated using the formula of 
\begin{equation}
{\cal Z} = \frac{{\cal N}_{\rm S}}{\sqrt{{\cal N}_{\rm S}+{\cal N}_{\rm B}}} \,,
\end{equation}
where \({\cal N}_{\rm S}\) and \({\cal N}_{\rm B}\) are the numbers of events for the signals and backgrounds after applying the machine learning cuts, respectively. 
A parameter space scan has been performed using {\tt EasyScan$\_$HEP}~\cite{Shang:2023gfy}, and the sensitivity regions are determined at ${\cal Z}=2$, which correspond to the sensitivities at the 95\% confidence level (C.L.). The heavy neutrino mass range considered in our analysis here spans from 1 TeV to the $W_R$ mass, with the sine of the mixing angle $s_\alpha \in [0,\, \sqrt2/2]$ (cf. Table~\ref{tab:parameter}). The sensitivities of the heavy neutrino mass $m_{N_{}}$ and the sine of the mixing angle $s_{\alpha}$ at the center-of-mass energy of $\sqrt{s}=14$ TeV are presented in Fig.~\ref{fig:sensitivity:14}. The top, middle and bottom panels are for the final states of $ee$, $\mu\mu$ and $e\mu$, respectively. The prospects of the OS and SS processes are represented by solid red and blue lines, respectively. The color bands indicate the corresponding $1\sigma$ statistical uncertainties. For simplicity we have neglected the systematic uncertainties. It is clear in this figure that the statistical uncertainties are rather large for all the three signals at 14 TeV. In substantial regions of the plane of $m_N$ and $s_\alpha$ the significance can not reach the 95\% C.L. This is mainly due to the large $W_R$ mass of 6.5 TeV, with respect to the center-of-mass energy of 14 TeV.

\begin{table}[h!]
\centering
\caption{Optimized machine learning thresholds $\chi$ and the corresponding signal and background efficiencies.}
\vspace{5pt}
\label{tab:eff_e_table}
\resizebox{\textwidth}{!}{%
\begin{tabular}{|c|c|c|c|c|c|c|c|c|c|c|c|c|}
\hline
\multicolumn{2}{|c|}{$\sqrt{s}$ [TeV]} & \multicolumn{3}{c|}{14} & \multicolumn{3}{c|}{27} & \multicolumn{3}{c|}{100} \\ \hline
\multicolumn{2}{|c|}{flavors} & $ee$ & $\mu\mu$ & $e\mu$ & $ee$ & $\mu\mu$ & $e\mu$ & $ee$ & $\mu\mu$ & $e\mu$ \\ \hline\hline
\multirow{9}{*}{OS} & $W^+ W^-jj$ & 0.18\% & 0.18\% & 0.08\% & 0.49\% & 0.07\% & 0.5\% & 0.59\% &  0.3\% & 0.77\% \\ \cline{2-11}
& $Zjj$ & 0.039\% & 0.01\% & $-$ & 0.058\% & 0.006\% & $-$ &  0.026\% & 0.006\% & $-$ \\ \cline{2-11}
& $W^{\pm}Zjj$ & $-$ & $-$ & 0.09\% & $-$ & $-$ & 0.4\% & $-$ & $-$ & 0.6\% \\ \cline{2-11}
& $t\bar{t}jj$ & $-$ & $-$ & 0.009\% & $-$ & $-$ & 0.04\% & $-$ & $-$ & 0.1\% \\ \cline{2-11}
& \makecell[c]{$m_{N}$=\\  1.5 TeV} & 75.4\% & 76.3\% & 75.0\% &  87.8\% & 83.4\% & 88.0\% & 89.5\% & 90.3\% & 93.2\%  \\ \cline{2-11}
& \makecell[c]{$m_{N}$=\\  2.0 TeV} & 78.3\% & 79.9\%  & 80.0\% & 88.6\% & 84.5\% & 90.0\% & 89.6\% & 93.6\% & 96.2\%  \\ \cline{2-11}
& $\chi$ &  98.7\% & 97.7\% & 98.0\% & 98.6\% & 98.0\% & 99.0\% & 98.3\% & 98.4\% & 98.9\% \\ \hline\hline
\multirow{7}{*}{SS} & $W^\pm W^\pm jj$ & 0.053\%  & 0.05\% & 0.037\% & 0.063\% & 0.06\% & 0.07\%  & 0.07\% & 0.05\% & 0.07\%  \\ \cline{2-11}
& $W^\pm Zjj$ & 0.36\% & 0.7\% & 0.29\% & 0.23\% & 0.6\% & 0.41\% & 0.23\% & 0.09\% & 0.8\%   \\ \cline{2-11}
& \makecell[c]{$m_{N}$=\\  1.5 TeV} & 75.4\% & 74.5\% & 77.0\% & 80.9\% & 82.0\% & 87.0\% & 86.3\% & 80.8\%  & 90.0\%   \\ \cline{2-11}
& \makecell[c]{$m_{N}$=\\  2.0 TeV} & 77.7\% & 75.2\% & 80.0\% & 82.6\% & 83.0\%  & 88.0\% & 90.0\% & 85.0\% & 66.0\% \\ \cline{2-11}
& $\chi$ & 98.4\% & 98.0\% & 98.8\% &  98.5\% & 98.7\% & 98.8\% & 98.8\% & 98.4\%  & 98.9\%  \\ \hline
		\end{tabular}
        }
\end{table}

\begin{figure}[h!]
	\centering
    \includegraphics[width=0.6\linewidth]{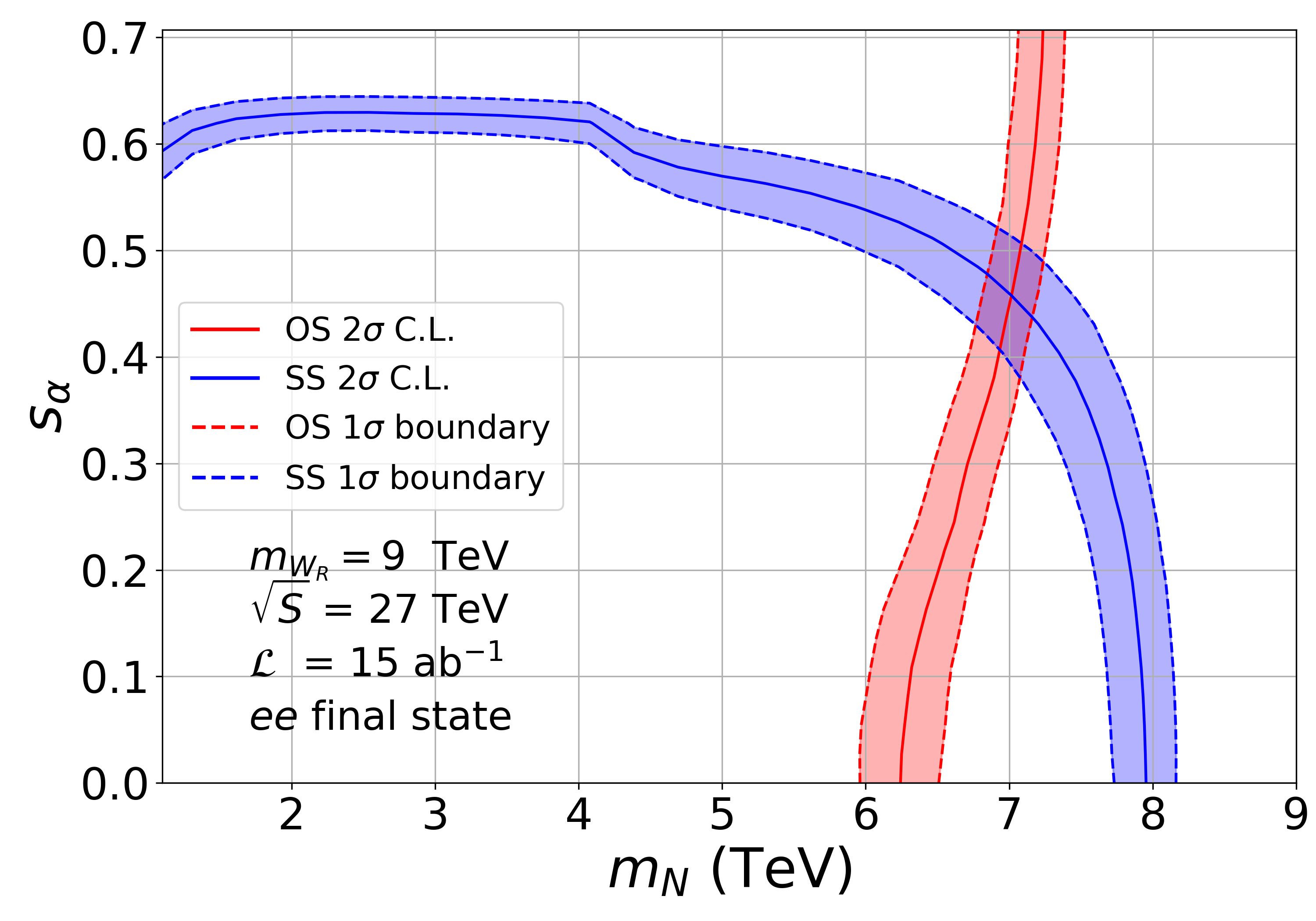}
	\includegraphics[width=0.6\linewidth]{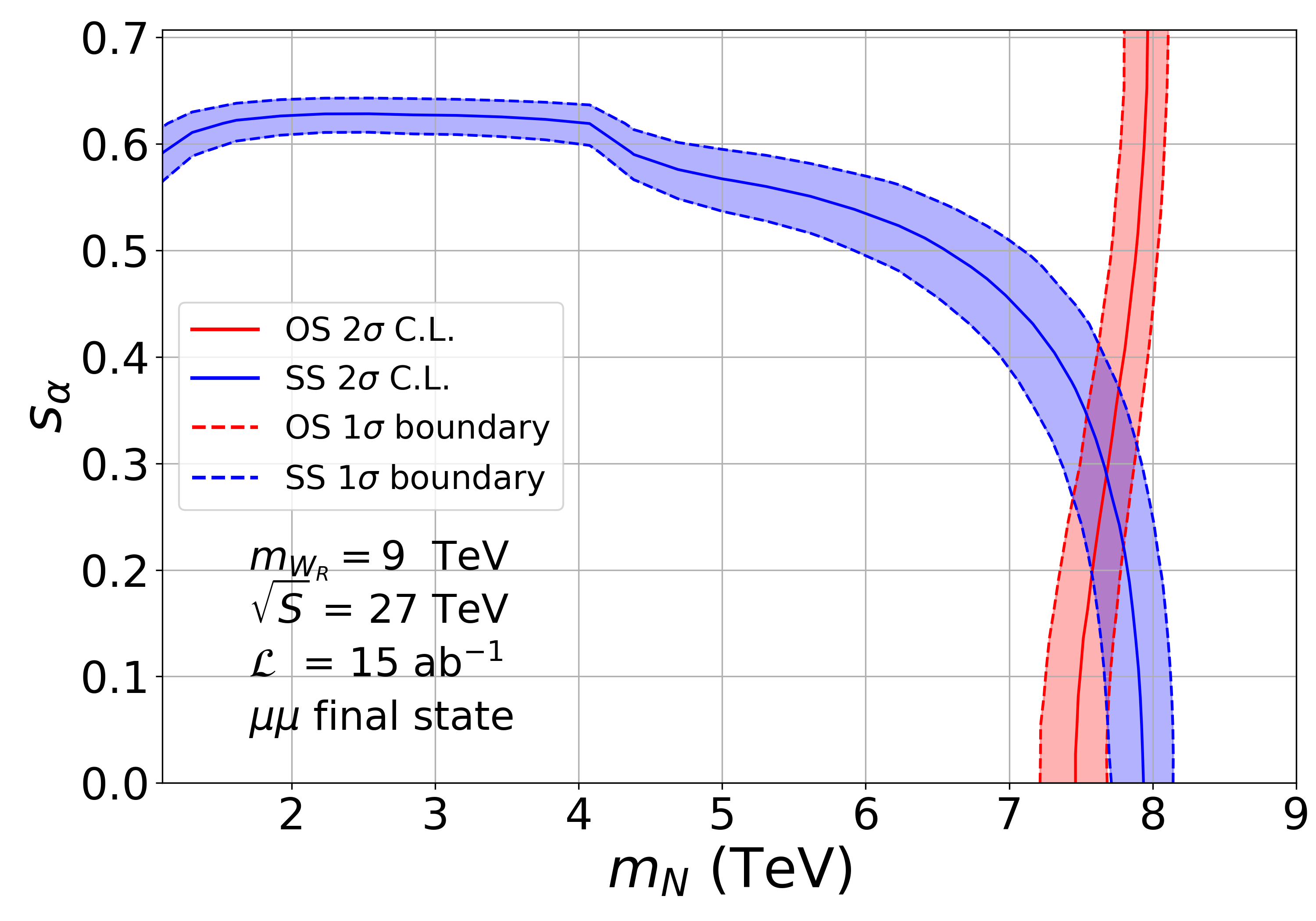}
    \includegraphics[width=0.6\linewidth]{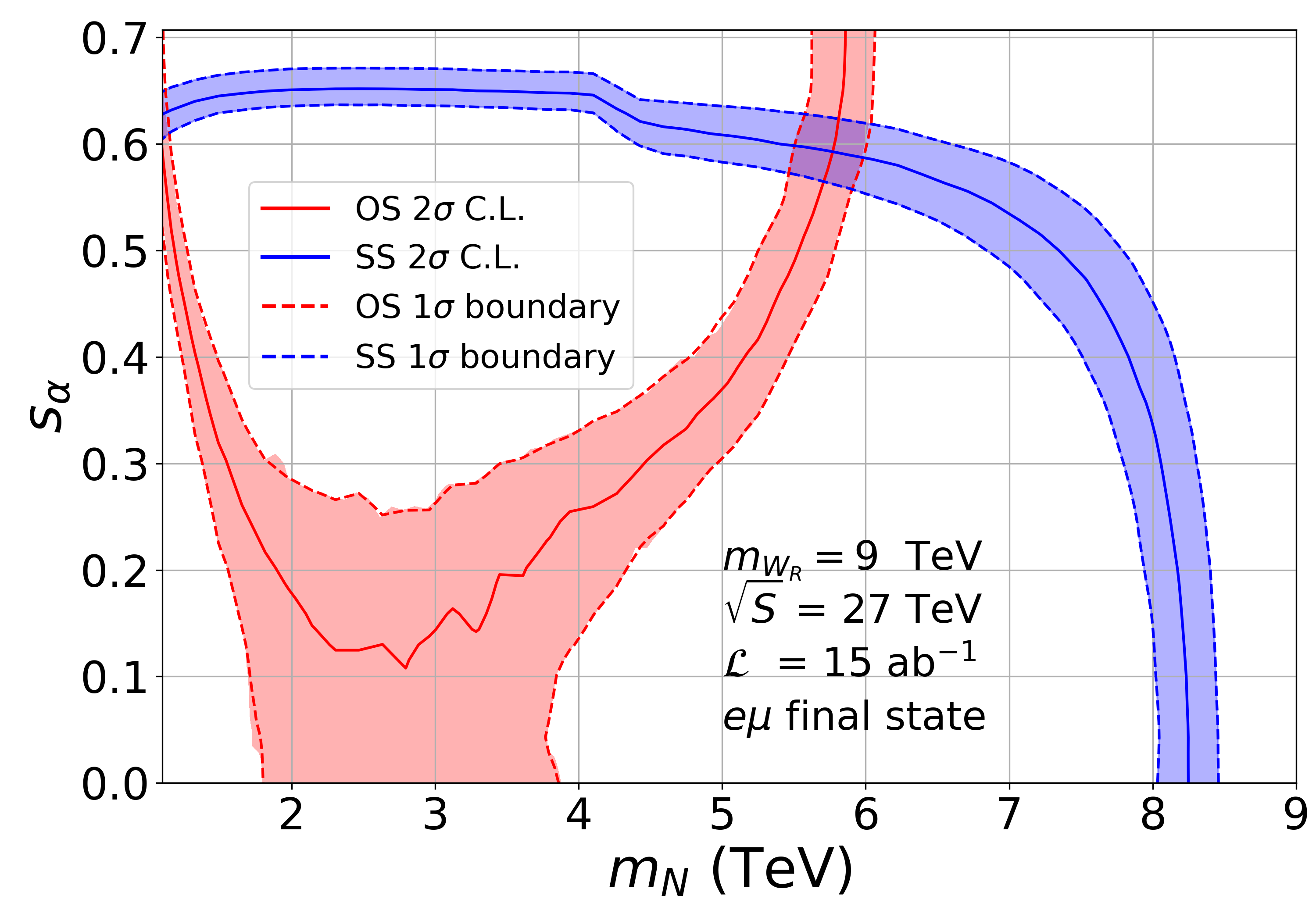}
	\caption{The same as Fig.~\ref{fig:sensitivity:14}, but the solid red and blue lines represent the sensitivities at the $2\sigma$ C.L. for the OS and SS processes, respectively at $\sqrt{s} = 27$ TeV with an integrated luminosity of 15 ab$^{-1}$.
    In the bottom panel, the OS dilepton signal can only probe the region above the red line. See text for more details.  }
	\label{fig:sensitivity:27}
\end{figure}

\begin{figure}[h!]
	\centering
	\includegraphics[width=0.58\linewidth]{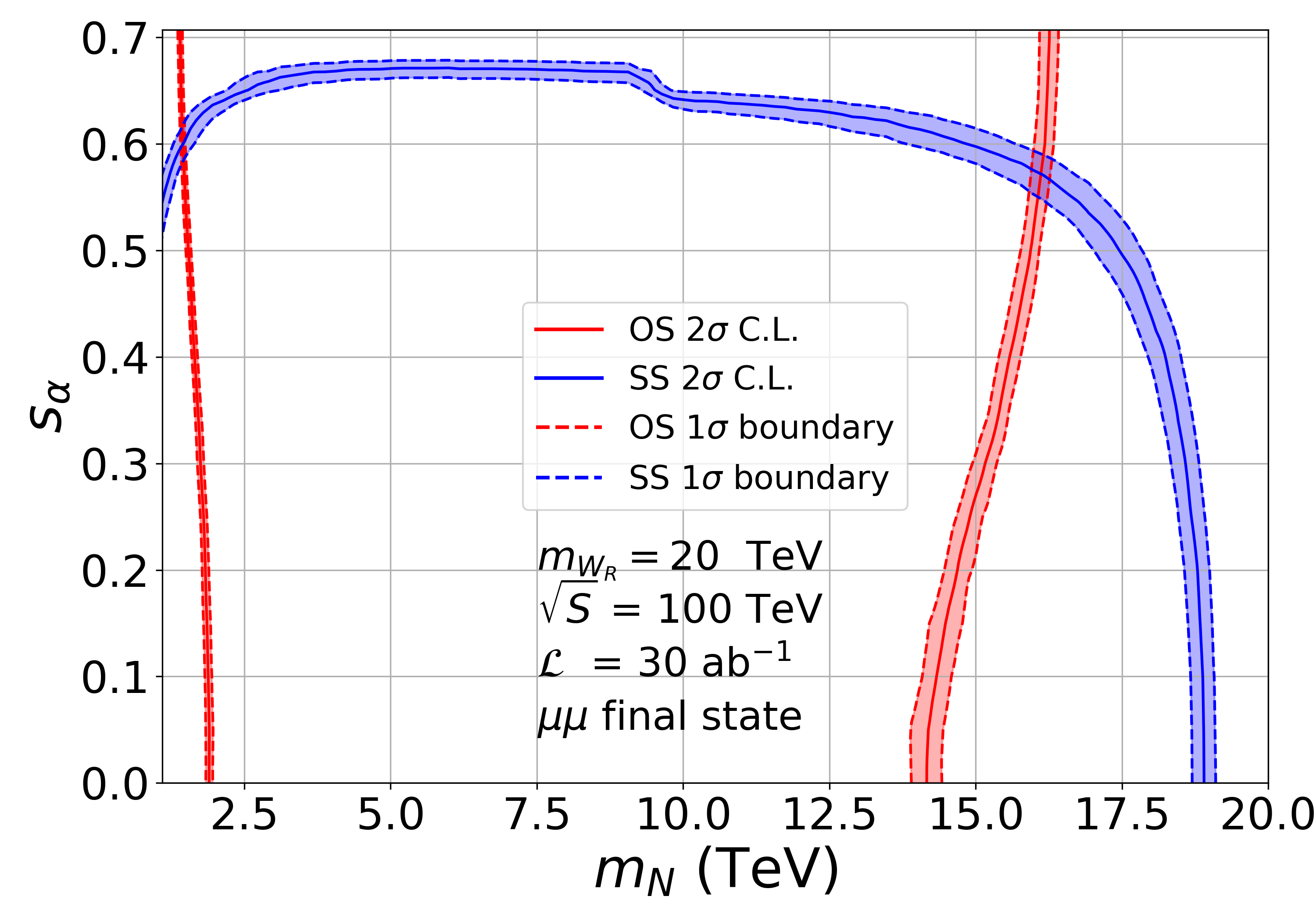}	
    	\includegraphics[width=0.58\linewidth]{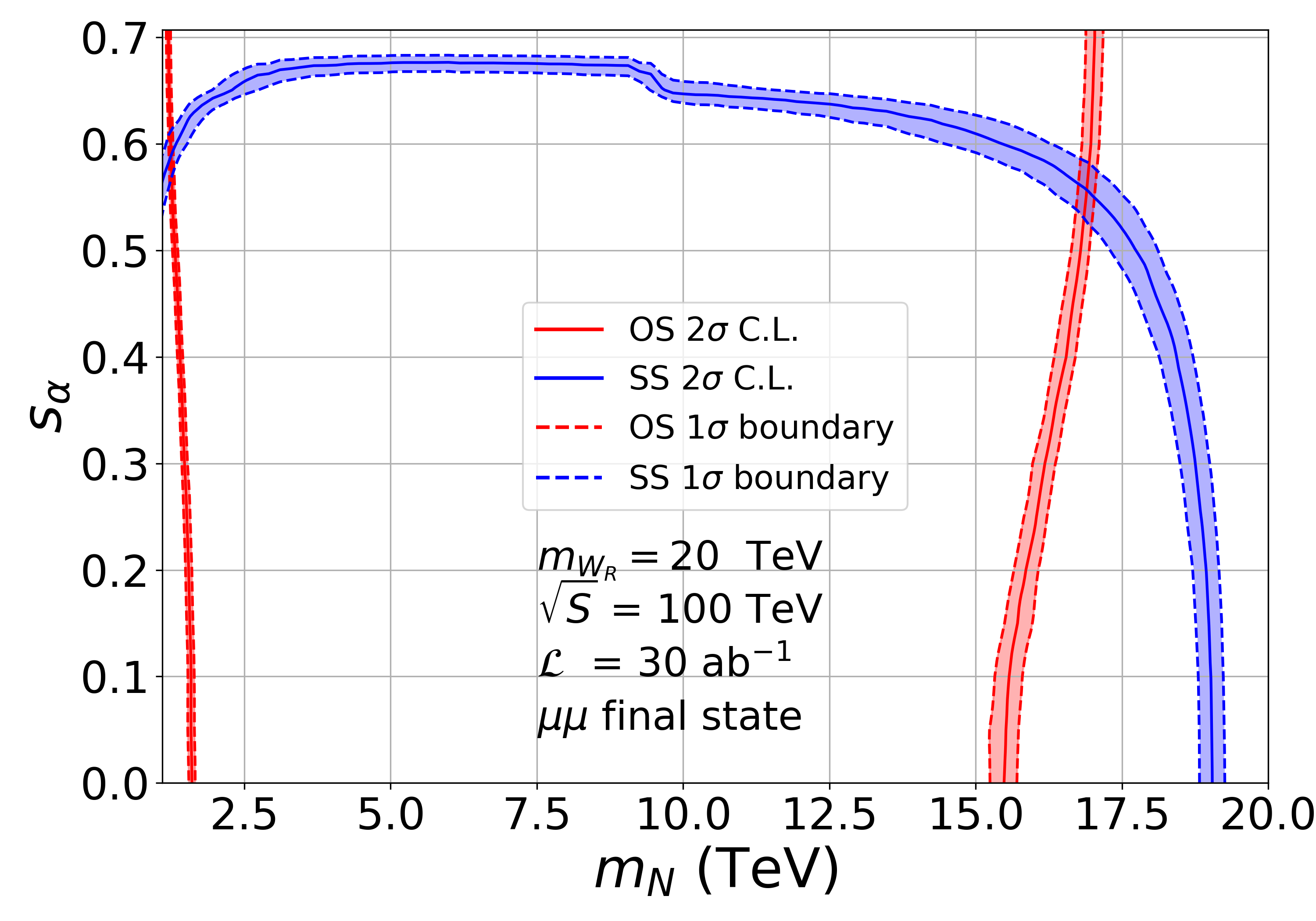}
	\includegraphics[width=0.58\linewidth]{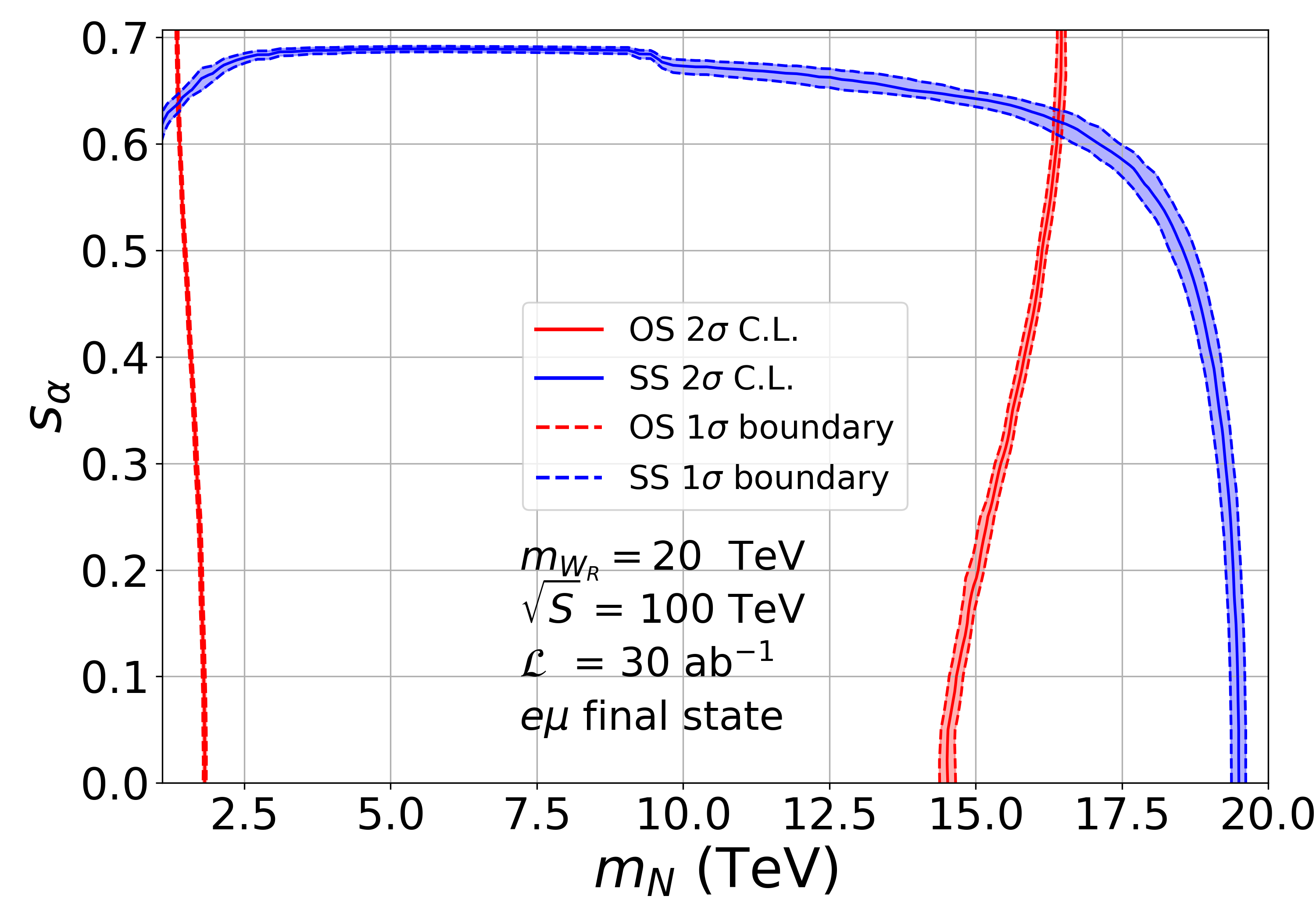}
	\caption{The same as Fig.~\ref{fig:sensitivity:27}, but at $\sqrt{s} = 100$ TeV with an integrated luminosity of 30 ab$^{-1}$. In the three panels, the OS dilepton signal can only probe the region between the two red lines. See text for more details.
    }
\label{fig:sensitivity:100}
\end{figure}

\begin{table}[t!]
	\centering
 	\caption{Main results for the sensitivities of $m_N$ and and $s_\alpha$ at the 95\% in the OS and SS signals, with the final states of $ee$, $\mu\mu$ and $e\mu$ at $\sqrt{s}=14$ TeV, 27 TeV and 100 TeV.  
    See Figs.~\ref{fig:sensitivity:14} to \ref{fig:sensitivity:100} and the text for more details. 
    }
	\label{tab:results}
	\vspace{5pt}
	\begin{tabular}{|c|c|c|c|c|c|}
		\hline
  \multicolumn{3}{|c|}{$\sqrt{s}$ [TeV]}  & 14 & 27 & 100 \\ \hline
  \multicolumn{3}{|c|}{luminosity [ab$^{-1}$]}  & 3 & 15 & 30 \\ \hline
  \multicolumn{3}{|c|}{$m_{W_R}$ [TeV]}  & 6.5 & 9 & 20 \\ \hline\hline
  \multirow{6}{*}{OS} &	\multirow{3}{*}{$m_N$ [TeV]} & $ee$  & [1.0,~3.57] & [1.0,~7.2] & [1.3,~16.2] \\ \cline{3-6} 
    && $\mu\mu$ & [1.0,~3.78] & [1.0,~7.9] & [1.2,~17.1] \\ \cline{3-6}
    && $e\mu$ & [1.3,~2.56] & [1.0,~5.8] & [1.3,~16.5] \\ \cline{2-6}
	&	\multirow{3}{*}{$s_\alpha$} & $ee$ & [0.36,~0.707] & [0,~0.707] & [0,~0.707] \\ \cline{3-6}
    && $\mu\mu$ & [0.30,~0.707] & [0,~0.707] & [0,~0.707] \\ \cline{3-6}
    && $e\mu$ & [0.58,~0.707] & [0.11,~0.707] & [0,~0.707] \\ \hline\hline
	\multirow{6}{*}{SS} &	\multirow{3}{*}{$m_N$ [TeV]} & $ee$  & [1.0,~4.16] & [1.0,~7.9] & [1.0,~19.0] \\ \cline{3-6} 
    && $\mu\mu$ & [1.0,~4.22] & [1.0,~7.9] & [1.0,~19.1] \\ \cline{3-6}
    && $e\mu$ & [1.0,~4.19] & [1.0,~8.2] & [1.0,~19.5] \\ \cline{2-6}
	&	\multirow{3}{*}{$s_\alpha$} & $ee$ & [0,~0.48] & [0,~0.63] & [0,~0.67]\\ \cline{3-6}
    && $\mu\mu$ & [0,~0.49] & [0,~0.63] & [0,~0.67] \\ \cline{3-6}
    && $e\mu$ & [0,~0.49] & [0,~0.65] & [0,~0.69] \\ \hline
	\end{tabular}
\end{table}

Although the statistical uncertainties are very large in Fig.~\ref{fig:sensitivity:14}, we can still extract the following information from the plots. For the SS signal, when the mixing angle \( \alpha \) is small, the sensitivity is better than that of the OS process. It is easy to be understood: the SS process has cleaner SM backgrounds than the OS signals. 
Furthermore, as seen in Eq.~(\ref{casetwo}), a large mixing angle $\alpha$ will enhance the cross section for the OS signals. In contrast, as implied by Eq.~(\ref{casetwo}), when the mixing angle $\alpha$ approaches $\pi/4$, the interference of the two heavy neutrinos $N_{1,\,2}$ can lead to a reduction in the cross section for the SS dilepton process, which diminishes its sensitivities at the high-energy hadron colliders. 
In all the three panels of Fig.~\ref{fig:sensitivity:14}, the regions above the solid red lines can be probed at the 95\% C.L. in the OS channels, while in the regions below the solid blue lines we can achieve the 95\% C.L. sensitivity in the SS channels.



The corresponding sensitivities at the center-of-mass energies of 27 TeV and 100 TeV are presented in Figs.~\ref{fig:sensitivity:27} and \ref{fig:sensitivity:100}, respectively. As expected, at higher energy colliders with an enlarged luminosity, the heavy neutrinos can be probed to a larger mass, and the statistic uncertainties shrinks significantly. 
As in Fig.~\ref{fig:sensitivity:14}, the regions below the solid blue lines can be probed at the 95\% C.L. in the SS channels, while for the OS signals it is to some extent different. 
In the lower panel of Fig.~\ref{fig:sensitivity:27}, the region above the solid red line can be probed at the 95\% C.L. in the $e^\pm \mu^\mp$ channel; in the all the three panels of Fig.~\ref{fig:sensitivity:100}, the 2$\sigma$ sensitivity region in the OS processes lies between the two solid red lines. 
In other words, if the neutrino $N$ is relatively light, i.e. $m_N \ll m_{W_R}$, the decay products from the heavy neutrino $N \to \ell q \bar{q}'$ tend to be highly boosted. Then the signal events will be highly suppressed by the angular separation requirement of $\Delta R > 0.4$. 
For the sake of convenience, 
the probable ranges of $m_N$ and $s_\alpha$ at the 95\% C.L. in the OS and SS signals at future high-energy hadron colliders at $\sqrt{s}=14$ TeV, 27 TeV and 100 TeV are collected in Table~\ref{tab:results}. 

\section{Conclusion}
\label{con}

Exploring the origins of the tiny masses of active neutrinos and their mixing remains a key issue in particle physics. One of the well-motivated models beyond the SM is the inverse seesaw mechanism, incorporating the heavy neutrinos \(N\) and \(S\) to impart masses to the active neutrinos. 
In this work, we delve into the framework of LRSM with the inverse seesaw mechanism, and focus on improving the sensitivities of heavy neutrinos and the mixing angle between $N$ and $S$ at future high-energy hadron colliders using machine learning.
In particular, we study both the OS and SS dilepton processes $pp \to \ell_\alpha^\pm \ell_\beta^\mp jj,\; \ell_\alpha^\pm \ell_\beta^\pm jj$ with $\alpha,\, \beta = e,\, \mu$, which are mediated by the heavy $W_R$ boson and the heavy neutrino mass eigenstates $N_{1,2}$. The corresponding dominant SM backgrounds for the OS and SS dilepton signals are taken into account in our analysis (cf. Table~\ref{tab:process}). It should be noted that the backgrounds are different for the same-flavor OS dilepton signals $ee$ and $\mu\mu$, different-flavor OS signal $e\mu$, and the SS dilepton signals $\ell_\alpha^\pm \ell_\beta^\pm$.

The distributions of the transverse momenta of the charged leptons and jets, the missing transverse energy, the invariant mass of the dilepton pair, and the angular separations of the charged leptons and jets can be found in Figs.~\ref{fig:HTandMET} through \ref{fig:14os_emu_kin_4}. The correlations between these observables are shown in Fig.~\ref{cos}. We utilize XGBoost for machine learning to handle all the information from the distributions and correlations. The machine learning results for the $ee$, $\mu\mu$ and $e\mu$ final states at $\sqrt{s}=14$ TeV, 27 TeV and 100 TeV are collected in Figs.~\ref{fig:14Events} through \ref{fig:100os_emu}. It is clear in these figures that machine learning can effectively differentiate between signals and backgrounds. The sensitivity regions of heavy neutrino mass $m_N$ and the sine of mixing angle $s_\alpha$ of heavy neutrinos in the OS and SS $ee$, $\mu\mu$ and $e\mu$ signals at $\sqrt{s}=14$ TeV, 27 TeV and 100 TeV are shown in Figs.~\ref{fig:sensitivity:14} to \ref{fig:sensitivity:100}. The main results of $m_N$ and $s_\alpha$ at these colliders are summarized in Table~\ref{tab:results}. It turns out that the statistical uncertainties are quite large at the HL-LHC in all the OS and SS channels, which can be improved significantly at the higher energy colliders. 
It is remarkable that the heavy neutrinos can be probed up to roughly 17.1 TeV and 19.5 TeV in the OS and SS dilepton signals at the future 100 TeV collider, respectively; the sine of the mixing angle $s_\alpha$ for heavy neutrinos can be probed up to the maximal value of $\sqrt2/2 \simeq 0.707$ and $0.69$ in the OS and SS channels, respectively (cf Table~\ref{tab:results}). The results in this paper not only provide the theoretical predictions of heavy neutrino mixing at future high-energy hadron colliders, but also offer a new perspective for particle physics experiments in searching for the LNV processes.




It is crucial to note that these conclusions in this paper are valid only under conditions in which interference of heavy neutrino states occurs. When the neutrino mass splitting is very significant, the signals at the high-energy colliders could be very different. Furthermore, in this paper we have focused only on relatively simple cases for the machine learning analysis. It should also be very beneficial if machine learning methods are used in more general cases, for instance:
(i) the case with $m_N > m_{W_R}$; (ii) the signals from the tau flavor heavy neutrino; (iii) the case with CP violating phases in the heavy neutrino sector; (iv) the inclusion of the heavy-light neutrino mixing in the LRSM framework; and (v) the prospects of heavy neutrino mixing at future high-energy lepton and lepton-hadron colliders.


\begin{acknowledgments}
Qi-Shu Yan is supported by the Natural Science Foundation of China (NSFC) under Grant No. 12275143. Hong-Hao Zhang is supported by the NSFC under Grant No. 12275367. Yongchao Zhang is supported by the NSFC under Grant No. 12175039. 
This work is also supported by the Fundamental Research Funds for the Central Universities, and the Sun Yat-Sen University Science Foundation. 
\end{acknowledgments}

\bibliographystyle{JHEP}
\bibliography{ref}

\end{document}